\documentclass[oneside,12pt]{classes/Ricethesis_PDF_new}
\usepackage[toc,page]{appendix}
\usepackage{amsmath}
\usepackage{amsthm}
\usepackage{amssymb}

\usepackage{float}
\usepackage{algpseudocode}
\usepackage{algorithm}
\usepackage{algorithmicx}
\usepackage{amsmath}
\usepackage{amssymb}
\usepackage{graphicx,caption}
\usepackage{subcaption}
\usepackage{color}
\usepackage{xspace}
\usepackage{float}
\usepackage{url}
\usepackage{lipsum}
\usepackage{multicol}
\usepackage[printonlyused,withpage]{acronym}
\usepackage[acronym, nopostdot, nonumberlist]{glossaries}

\def\bibsecname{References}

\author{Peshal Nayak}
\title{AP-side WLAN Analytics}
\degree{Doctor of Philosophy}
\degreedate{October 2019}

\newcommand{\BAP}{B_{\textrm{\tiny AP}}}
\newcommand{\BSTA}{B_{\textrm{\tiny STA}}}
\newcommand{\NAP}{N_{\textrm{\tiny AP}}}
\newcommand{\NSTA}{N_{\textrm{\tiny STA}}}
\newcommand{\SDOWN}{S_{\textrm{\tiny down}}}
\newcommand{\SUP}{S_{\textrm{\tiny up}}}
\newcommand{\SSSTA}{S_{\textrm{\tiny sta}}}
\newcommand{\TF}{T_{\textrm{\tiny F}}}
\newcommand{\TDOWN}{T_{\textrm{\tiny down}}}
\newcommand{\TUP}{T_{\textrm{\tiny up}}}
\newcommand{\equaref}[1]{(\ref{eq:#1})}
\renewcommand{\P}{\mathbb{P}}
\newcommand{\diff}{{\rm\,d}}

\newcommand{\E}{\mathbb{E}}

\newcommand{\SSTA}{S_{\textrm{\tiny STA}}}
\newcommand{\FAP}{F_{\textrm{\tiny AP}}}
\newcommand{\FSTA}{F_{\textrm{\tiny STA}}}

\newcommand{\sdl}{\theta_{\textrm{\tiny dl}}}
\newcommand{\sul}{\theta_{\textrm{\tiny ul}}}

\newcommand{\esdl}{\theta_{\textrm{\tiny dl}}}
\newcommand{\esul}{\theta_{\textrm{\tiny ul}}}
\newcommand{\wmax}{W_{\textrm{\tiny m}}}

\newcommand{\dldelay}{d_{\textrm{\tiny access}}}
\newcommand{\dltx}{d_{\textrm{\tiny tx}}}
\newcommand{\uldelay}{u_{\textrm{\tiny access}}}
\newcommand{\ultx}{u_{\textrm{\tiny tx}}}

\newcommand{\usage}{V}

\newcommand{\technique}{\textit{uScope} }
\newcommand{\tenq}{t_{\textrm{enq}}}
\newcommand{\thead}{t_{\textrm{head}}}
\newcommand{\tsuccess}{t_{\textrm{rx\_start}}}
\newcommand{\tend}{t_{\textrm{rx\_end}}}
\newcommand{\ttxend}{t_{\textrm{tx\_end}}}
\newcommand{\segend}{t_{\textrm{tx\_end}}^{\textrm{seg}}}
\newcommand{\ackend}{t_{\textrm{rx\_end}}^{\textrm{ack}}}
\newcommand{\ackstart}{t_{\textrm{rx\_start}}^{\textrm{ack}}}
\newcommand{\ackenq}{t_{\textrm{enq}}^{\textrm{ack}}}
\newcommand{\ackhead}{t_{\textrm{head}}^{\textrm{ack}}}
\newcommand{\intkend}{t_{\textrm{rx\_end}}^{\textrm{int(N)}}}
\newcommand{\intkstart}{t_{\textrm{rx\_start}}^{\textrm{int(N)}}}
\newcommand{\qdelay}{\bar{\Phi}_{\textrm{queuing}}}
\newcommand{\adelay}{\bar{\Phi}_{\textrm{access}}}
\newcommand{\tdelay}{\bar{\Phi}_{\textrm{tx}}}
\newcommand{\ulatency}{\bar{L}_{\textrm{uplink}}}
\newcommand{\retrans}{\bar{R}}
\newcommand{\tdefer}{\bar{\Psi}_{\textrm{defer}}}
\newcommand{\deferk}{\bar{\theta}_{\textrm{defer,k}}}\newcommand{\contnk}{\bar{\theta}_{\textrm{contn,k}}}
\newcommand{\txk}{\bar{\theta}_{\textrm{occ,k}}}
\newcommand{\deferone}{\bar{\theta}_{\textrm{defer,1}}}

\begin{document}

\begin{romanpages}

\maketitle

\doublespacing
\begin{abstract}
Monitoring the network performance experienced by the end-user is crucial for managers of wireless networks as it can enable them to remotely modify the network parameters to improve the end-user experience. Unfortunately, for performance monitoring, managers are typically limited to the logs of the Access Points (APs) that they manage. This information does not directly capture factors that can hinder station (STA) side transmissions. While the AP-observable measurements do indeed help to characterize the PHY performance for downlink and uplink, managers today lack models and tools to translate them into user experience metrics (such as TCP throughput). Consequently, state-of-the-art methods to measure such metrics primarily involve active measurements. For instance, typically to measure achievable download and upload TCP throughputs, users use internet speed tests that perform 10s of MB of TCP uploads and downloads. Unfortunately, such active measurements increase traffic load and if used regularly and for all the STAs can potentially disrupt user traffic, thereby worsening performance for other users in the network and draining the battery of mobile devices.

This thesis enables passive AP-side network analytics. Therefore, for performance monitoring, I consider that a monitoring framework will have access only to the logs of the AP that the manager controls. Further, I consider that there is no STA side co-operation and no access to STA side logs. As a result, the framework is constrained to make an estimate solely based on passive AP-side observables.

In the first part of the thesis, I present virtual speed test, a measurement based framework that enables an AP to estimate speed test results for any of its associated clients solely based on AP-side observables. Virtual speed test employs a novel L2 edge TCP model to perform throughput estimation. We implemented virtual speed test using commodity hardware, deployed it in office and residential environments, and conducted measurements spanning multiple days having different network loads and channel conditions. Overall, virtual speed test has mean estimation error less than 10\% compared to ground truth speed tests, yet with zero overhead, and outcomes available at the AP.

Next, I present Uplink Latency Microscope (uScope), an AP-side framework for estimation of WLAN uplink latency for any of the associated STAs and decomposition into its constituent components. Similar to virtual speed test, uScope makes estimations solely based on passive AP-side observations. The key idea in uScope is to leverage the layer-4 handshake as a virtual probe to estimate and decompose layer-2 latency. We implement uScope on a commodity hardware platform and conduct extensive field trials on a university campus and in a residential apartment complex. In over 1 million tests, uScope demonstrates high estimation accuracy with mean estimation errors under 10\% for all the estimated parameters.	
\end{abstract}

\begin{acknowledgements}
This thesis has been possible due to two sets of entities. The first set of entities lies in the spiritual world and the second in the human. I'll start with the spiritual world. First and foremost, I'd like to thank God without whose secret and invisible support, none of this would have been possible. God always responded to my prayers and sent help whenever I needed it.   

In the human world, a large number of people have knowingly and unknowingly helped me in direct and indirect ways during my time at Rice. Unfortunately, due to limitations of space, I can only thank a handful of those people. However, it is worth noting that if I were to thank them all one by one, this section would be far bigger than the rest of the thesis!

First and foremost, I would like to thank my advisor Dr. Edward Knightly for being both a mentor and a friend. Without his excellent guidance, constant motivation, support, and help, this thesis would have been impossible. He provided me with the best research environment anyone could ask for which played a vital role in sharpening my problem-solving skills and helped me develop a rigorous and systematic approach to identifying and solving research problems. 

Next, I would like to thank my thesis committee members: Dr. Ashutosh Sabharwal, Dr. Lin Zhong, and Dr. Ang Chen for their inputs and constructive feedback. I would also like to thank my research collaborator Dr. Santosh Pandey (previously with Cisco Systems and now with Prosimo.io) for his feedback on my work. I also thank Dr. Carlos Cordeiro (Intel labs) for his periodic feedback and Dr. Seongwon Kim (SK Telecom) for his valuable tips on tweaking ns 3. 

I'd like to give special thanks to Dr. Michele Garetto (Universit\`{a} di Torino, Italy) for being like an elder brother to me. His guidance and motivation played a crucial role in helping me complete my research work. 

I would like to express my gratitude towards all the members of Rice Network Group who have contributed their precious time and effort to help me with my research. A special thanks to Yasaman Ghasempour, Chia-Yi Yeh and Dr. Kumail Haider for listening to my frustrations and providing constant motivation. I thank Dr. Ryan Guerra, Dr. Naren Anand and Dr. Clayton Shepherd for answering my questions and helping me when I stuck with conceptual problems. I would also like to thank Dr. Sharan Naribole, Dr. Xu Zhang, Dr. Adriana Flores, Dr. Oscar Bejarano, Dr. Riccardo Petrolo, Furqan Ahmad, Keerthi Dasala, Vinicius Da Silva Goncalves, Maryam Khalid, Zhambyl Shaikhanov and Jonghun Park for their critical feedback on my presentations. I would also like to thank all my friends at Rice who have helped me get through some of the toughest times in my life. 

A special thanks also to the ns-3, OpenWrt and Linux community for responding to my questions on their respective online forums and to all my anonymous paper reviewers who helped me look at my research work from alternative directions.  

I owe this success to my parents late Dr. Balakrishna K. Nayak and Mrs. Seema B. Nayak who always motivated and supported me in every way possible. My mom patiently listened to my research frustrations \textit{everyday for the last five years} and provided valuable guidance and motivation. Without my parents' love, encouragement and assistance I would not be what I am today. 
	
\end{acknowledgements}

\begin{dedication}
	I dedicate this work to my parents late Dr. Balakrishna K. Nayak and Mrs. Seema B. Nayak who made great sacrifices to help me succeed in life. 
\end{dedication}

\setcounter{secnumdepth}{3}
\setcounter{tocdepth}{2}
 
\onehalfspacing
\tableofcontents
\listoffigures
\listoftables

\end{romanpages}
 
\doublespacing



\chapter{Introduction}
\label{sec:intro}

Managers of wireless networks often want to understand the performance that the end users experience over their wireless connection. Network analytics or in other words the analysis of data collected from the network enables the manager to gain a deeper understanding of how users perceive their current wireless connectivity and the quality of service provided by the manager. Further, monitoring the performance experienced by each end-user can enable the network manager to remotely make changes to the network to improve the end-user experience. 

\begin{figure}
    \centering
    \includegraphics[width=0.5\linewidth]{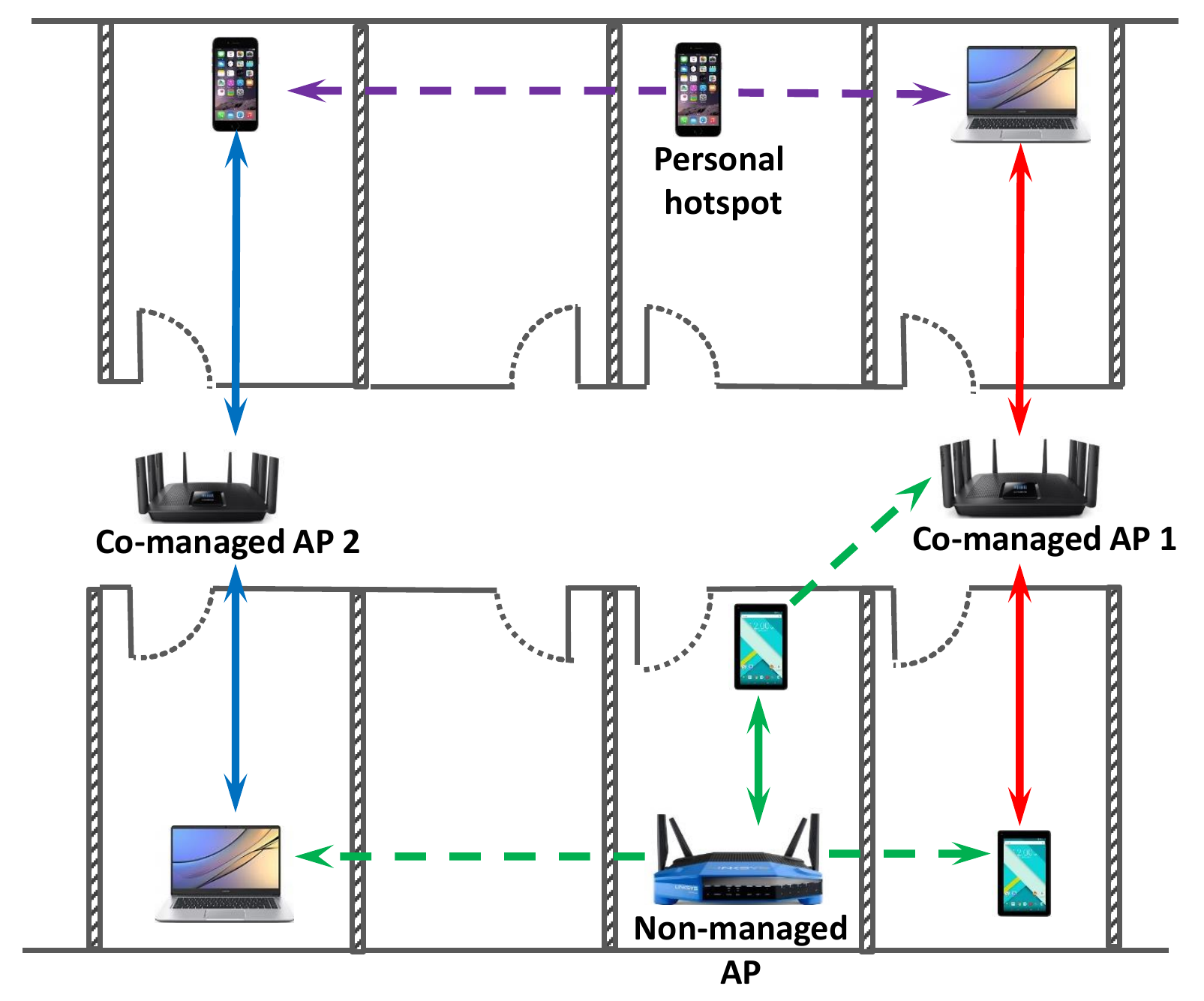}
    \caption{A typical WLAN today comprises of a number of co-existing APs. Some of these APs could fall under a different network manager. This co-existence results in a complex internode connectivity among devices.}
    \label{net_scene}
\end{figure}

Unfortunately, monitoring the performance experienced by the end-users is challenging. This is because wireless LANs are extremely complex today. A typical WLAN as shown in Fig.~\ref{net_scene} comprises a managed infrastructure deployed by a network manager to serve his clients. In addition to this managed infrastructure, there may be one or more non-managed WLANs that may be interfering. Such WLANs could correspond to a personal hotspot or a WLAN under a different network manager resulting in complex inter-node connectivity among these devices. The problem of performance monitoring is exacerbated by the fact that managers are typically limited to only the information that can be collected via the logs of the AP that they control. While these logs are rich with information pertaining to layer-2 and layer-1 performance of the Access Point (AP), managers lack the tools and models to translate the AP-side information to higher-layer performance metrics (\textit{e.g.}, TCP throughput). Further, while this information does capture layer-1 statistics of the end user's device, it does not contain the layer-2 statistics (\textit{e.g.}, WLAN latency, retransmission rates, etc.) of the end-user. 

Possible workarounds to this problem are to either seek end-user co-operation (\textit{e.g.}, third party software installations for periodic reporting) and perform active measurements to measure the required metrics. Unfortunately, such techniques are not feasible in practice for two key reasons. First users typically do not like to install third-party software on their devices and consequently user side information may not always be available. Secondly, active tests impose an additional traffic load on the network. Consequently, if frequently used for monitoring and all the STAs, they can potentially disrupt user traffic thereby worsening performance for other users and draining the battery of mobile devices. 

The goal of this thesis is to enable and AP-side network analytics. Therefore, for performance monitoring we are constrained by the fact that we cannot perform any active measurement, cannot seek any STA-side co-operation and cannot install any additional hardware infrastructure (\textit{e.g.}, a group of sniffers) to collect more information. Therefore, we are limited to information that can be directly observed via the logs of the AP that the manager controls. 

To this end, we design, implement and evaluate two novel frameworks that enable AP-side performance monitoring. We specifically focus on download and upload speeds and WLAN latency as these are two key performance metrics that determine end-user experience. The key idea in these frameworks is to exploit information that can be observed via a layer-4 handshake and design analytical models that exploit this information to estimate the performance metrics under consideration. Since TCP ACK is transmitted as a layer-2 frame, the duration between the transmission of a TCP segment on the downlink to the reception of TCP ACK on the uplink exposes valuable information (as described later) that enable estimation of metrics that are generally not observable at the AP. We evaluate our frameworks using commodity hardware and perform extensive field trials on a university campus and in a residential apartment complex to validate our models. An overview of the two frameworks developed as a part of this thesis is as follows and their details are provided in the following chapters.

\section{Download and Upload Speed Estimation Based on AP-side Observables}
TCP speed tests are end-to-end tests of network health and are available via a plethora of online apps~\cite{ooklaspeedtest,attspeedtest,xfinity}. As part of the measurement process, a client performs an active TCP download and an active TCP upload to a server to measure the download and upload TCP throughput respectively. Since more than 80\% of current Internet traffic is transmitted via TCP \cite{murray2017analysis}, the performance of numerous online applications is crucially dependent on the maximum TCP throughput achievable over an underlying network path. 

If a client's speed test uses a nearby server (\emph{i.e.}, a server with minimum possible latency to the AP), the WLAN becomes the key part of the end-to-end path. Consequently, the results would be valuable to the network manager to assess WLAN performance and make decisions on network infrastructure alterations to improve the quality of service experienced by the end-user. However, the results can only be seen by the end-user and are unavailable to the administrator without seeking end-user co-operation. Moreover, regularly performing such speed tests imposes additional traffic load on the network and hence doing so can potentially disrupt user traffic and drain the battery of mobile devices. 

To this end, we make the following contributions.
First, we present a framework that enables an AP to estimate the outcome of a speed test, \emph{i.e.}, the upload and download TCP  throughputs that any of its associated STAs should obtain from a nearby server, yet, without any special-purpose probing, with zero co-operation of endpoints (\emph{i.e.}, the server and the client), and solely based on measurements that are passively observable at the AP. We call our measurement-based framework \emph{virtual speed test}. The speed test results obtained by a STA can vary over time depending on numerous factors such as the number of active STAs, interference level, etc. Likewise, virtual speed test can enable the network manager to dynamically track any given STA's speed test results based on its unique characteristics (\textit{e.g.}, via a dynamic dashboard).

Virtual speed test employs a novel L2 edge TCP model to perform throughput estimation. The key challenge for the AP to estimate these inherently bi-directional, end-to-end and layer-4 throughputs, is that the AP only has a limited view of the network. Since the AP is unaware of the presence of hidden terminals, interference from neighboring BSS to the STAs, etc. (which affect the STA's queuing delays, NAV timers, and packet retransmissions), the AP cannot estimate how long it takes a STA to successfully transmit after it starts to attempt. Our design is motivated by the fact that since the WLAN is the final hop for any TCP segment directed towards a STA, this duration can also be estimated by measuring the delay incurred between the transmission of a TCP segment on the downlink to the reception of the corresponding TCP ACK on the uplink from the STA. This TCP segment, therefore, can belong to any TCP flow (\textit{e.g.}, a Netflix video stream) and need not be a part of a flow from a nearby server. To carry out these measurements, the AP must identify TCP flows. To this end, we leverage TCP's inherent bi-directionality and packet size signatures to spot TCP flows. Specifically the fact that TCP flows involve TCP segment traversing on the forward path and small-sized TCP ACKs on the reverse path enables the AP to identify these flows and perform its measurements.  

Second, we develop a virtual speed test enabled AP (VST AP) by using commodity hardware. We build APIs that enable the VST AP to passively collect several per-packet statistics and feed them into the L2 edge TCP model to obtain throughput estimates. While virtual speed test does not require the collection of STA-side statistics, for validation purposes, we also implement APIs for data collection from STAs associated with the VST AP for characterizing the operating environment and for ground truth measurement. We deploy VST AP in two environments: an office located inside a university building and an apartment in a residential complex. The VST AP is deployed in the office for 2 days and in the apartment for 7 days. Both deployment settings are characterized by interference from non-BSS devices co-existing in the same frequency band, human mobility and link diversity with respect to signal propagation (\emph{i.e.}, LoS vs non-LoS paths) and supported PHY rates. The office and the residential scenario cover a total of 36 and 49 topologies respectively with a varying number of STAs. Overall, the VST AP observes a total of 113,047 TCP flows across both deployments. These TCP flows result from multiple applications running on end devices such as video streaming, music streaming, pdf downloads, and email activities. For validation, actual client-based speed tests are employed as ground truth. Virtual speed test demonstrates a high level of estimation accuracy compared with ground truth, with average estimation error under 6\% for both upload and download speed estimation. 

Finally, we implement virtual speed test into ns-3's source code and perform extensive simulations to investigate operating conditions beyond those encountered in our field trials. The simulation results concur with field trial conclusions demonstrating estimation errors below 5\%.

To the best of our knowledge, virtual speed test is the first to estimate both upload and download TCP throughputs of STAs in the network by using passive measurement metrics at only the access point, \emph{i.e.}, without any active probing, additional hardware infrastructure or user participation.

\section{WLAN Latency and Constituent Component Estimation Based on AP-side Observables}
Wireless LAN latency is a key metric for network managers to understand user experience. WLAN latency comprises of three key components - channel access delays (which is further composed of 802.11 contention, retransmissions and defer delays), queuing delays and transmission delays (as determined by chosen data rates, transmission modes, overhead, etc.). Remotely monitoring WLAN latency for each device in the network and decomposing it into its constituent components can enable the network manager to take timely diagnostic actions to improve the quality of experience of the end-user. 

WLAN latency and its components are determined by the joint effect of several factors such as the number of conflicting nodes, their traffic load, air time utilization, etc., whose impact can be directly observed at the transmitter. Therefore, remotely computing WLAN \textit{downlink} latency and decomposing it into its constituent components is trivial based on direct observations obtained from the AP logs. However, an AP-side estimation of uplink latency is challenging for two reasons - (i) the factors affecting uplink latency are only known at the STA (\textit{i.e.}, at the transmitter) (ii) the factors and magnitude of their impact can be different for different STAs in the network.  

State-of-the-art techniques to estimate WLAN uplink latency and its components involve active probing~\cite{huang2011mobiperf1, speedtest, ooklaspeedtest, attspeedtest, xfinity}. However, probing increases traffic load and if used regularly and for all the STAs can potentially disrupt user traffic, thereby worsening latency for other users in the network and draining the battery of mobile devices. 

To this end, we make the following contributions. 

First, we present \technique (\textbf{\underline{u}}plink latency micro\textbf{\underline{scope}}), an AP-side framework for passive monitoring and analysis of WLAN uplink latency. While management and inference of WLANs and ad hoc network parameters have been the focus of intense research for decades, \technique is the first to enable estimation of WLAN uplink latency and breakdown into its constituent components. While doing so, \technique does not require any active measurements, special-purpose software installation on the STAs (and hence no additional messages between the AP and STAs), nor any additional hardware infrastructure to collect more information. \technique estimates and decomposes uplink latency solely based on passive observations made from a single AP. 

\technique employs virtual probing to enable a measurement-based analysis of uplink latency. The key idea in virtual probing is to employ layer-4 handshakes of the STA as virtual probes. Since the WLAN is the final hop for any TCP segment intended for a STA, the duration between transmission of a TCP segment on the downlink to the reception of the TCP ACK on the uplink exposes the total WLAN uplink latency for that STA. Because virtual probing leverages the fundamental closed-loop property of TCP, it can employ the layer-4 handshake of any TCP download (\textit{e.g.}, a Netflix video stream) to estimate WLAN uplink latency. Further, virtual probing does not impose any additional traffic load on the network as it uses the layer-4 handshakes that occur due to TCP. 

However, the virtual probe only reveals the total uplink latency. To further decompose it into its constituent components \technique leverages the transmissions received from the STA during the handshake. \technique performs timing analysis on these transmissions to decompose the total uplink latency. The timing analysis leverages the fact that the STA is guaranteed to be backlogged with at least one packet (the TCP ACK) in the duration between the end time of transmission of the TCP segment to the reception of the TCP ACK. Consequently, any intermediate transmission that occurs from the STA in this duration exposes the time when the TCP ACK reaches the head of the queue. Thus, by leveraging the timestamps of reception start and reception end of packets from the STA, \technique can estimate its queuing and access delays. Finally, \technique uses a novel estimation technique that couples virtual probing with knowledge of 802.11 protocol rules, to estimate the average number of retransmissions and the average defer delays faced by the STA. 
 
Next, we implement \technique on an 802.11ac compliant off-the-shelf Access Point. The implemented framework comprises of over \textit{7,000 lines} of code in Python to process the AP log and implement \technique. While \technique does not require any STA side information, for the purpose of evaluation, we also build APIs to collect STA side observations from portable laptops for measurement of ground truth values. 

Finally, we deploy our commodity hardware-based testbed on a university campus and in a residential apartment and perform a total of \textit{1,296,000 tests} to validate \technique. Both of these trials are characterized by interference from co-existing BSSs, light user and environmental mobility, diversity with respect to links (\textit{i.e.}, LoS and non-LoS paths), supported PHY rates, etc. In these field trials, the STAs run various internet applications performing video streaming, music streaming, pdf downloads, email activities, etc. Our field trials reveal that the estimation accuracy of \technique is dependent on the number of TCP handshakes that the AP observes. However, even with as few as 1,000 observed handshakes representing typically 1 MB of traffic received by any application of the STA, \technique demonstrates an estimation error under 10\% across all the parameters.

To the best of our knowledge, \technique is the first AP-side framework that passively estimates and decomposes WLAN uplink latency for any STA in the network. While doing so, \technique does not require any active measurements, special-purpose software installations on the STA, additional hardware infrastructure and can make estimates solely based on passive AP side observations.

The organization of this thesis is as follows. Chapter 2 provides a detailed description of our technique for passive AP-side estimation of download and upload speeds. In Chapter 3, we explore traditional modeling approaches and present novel analytical models for 802.11ac MU-MIMO systems operating under TCP traffic. This is followed by the description of our latency estimation and decomposition framework in Chapter 4. The experimental testbeds developed as a part of this thesis are described in Chapter 5. We present the evaluation of our frameworks in Chapter 6 and 7. Chapter 8 provides description and comparison with prior work. We conclude in Chapter 9. 
\chapter{Passive AP-side Estimation of Download and Upload Speeds}
This chapter presents the design of virtual speed test, a framework that enables a passive AP-side estimation of download and upload speed test throughputs for any of the associated STAs. 

First, we describe the network scenario in Sec.~\ref{scenario} that we target in this thesis. Next, we present a detailed description of the operation of internet speed tests in Sec.~\ref{background} followed by a high level problem description in Sec.~\ref{problem}. We then present the design of virtual speed test framework. Virtual speed test employs a novel L2 Edge TCP model which captures the packet dynamics involved into a tandem server closed queuing network and provides analytical expressions for throughput. The L2 Edge TCP model is described in Sec.~\ref{model} followed by how virtual speed test estimates the model parameters in Sec.~\ref{AP_side_measurement}.\footnote{This work has been previously published in IEEE INFOCOM 2019 \cite{nsk}.}

\section{Network Scenario}\label{scenario}
\begin{figure}
    \centering
    \includegraphics[width=0.6\linewidth]{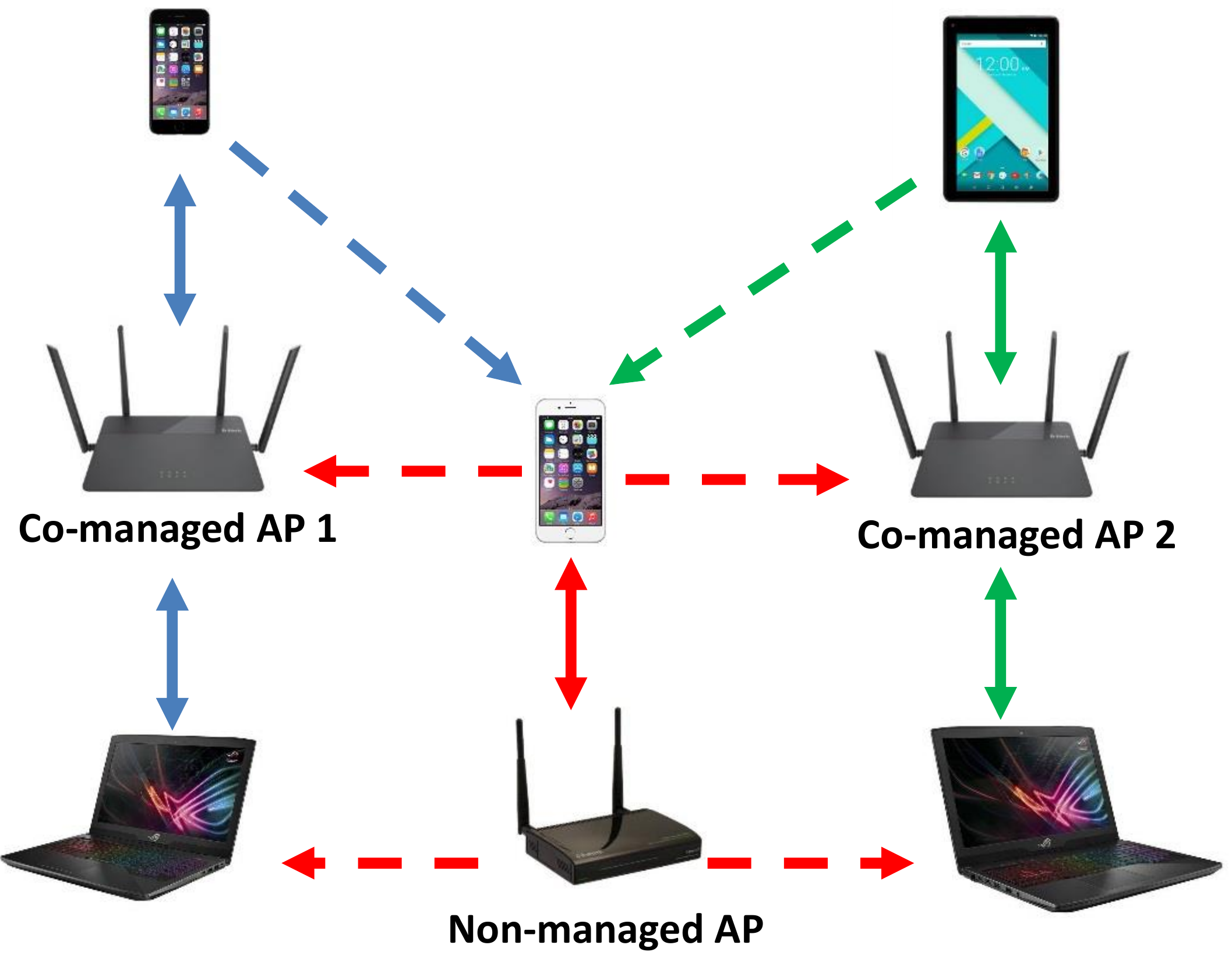}
    \caption{Enterprise WLAN scenario: bold lines indicate connectivity while dotted lines indicate interference.}
    \label{scenario}
\end{figure}

We consider an enterprise WLAN environment such as illustrated in Fig.~\ref{scenario}. As depicted, the network comprises of multiple APs. While the network may use channelization, for ease of exposition we consider only APs with at least partially overlapping channels such that they can potentially interfere with each other. Moreover, we consider that in addition to the managed infrastructure, there may be one or more non-managed WLANs that may be interfering. Such WLANs can correspond to an LTE hot spot or a neighboring WLAN under different administrative control. 

Ideally, all such networks should have sufficient physical separation to enable full spatial reuse for each AP (\emph{i.e.}, simultaneous transmission for each network). However, as depicted, the unwanted interconnectivity creates interference and contention among nodes. Moreover, inter-node connectivity can form a complex relationship: while all STAs are necessarily connected to the APs that they associate with, a particular STA may or may not be in range of other APs. Likewise, STAs may be “hidden” from each other or mutually in range. It is further possible that a STA is in range of other APs which are not in range of the AP that is serving it. The interference and contention possibilities are further compounded by the need to consider both downlink transmissions (AP to STA), uplink transmissions (STA to AP), and mixes. 

We do not make any assumptions about the PHY layer capabilities of the AP or the STAs. For instance, the AP may have advanced physical layer capabilities such as multi-user MIMO. Likewise, the AP can have any channelization strategy, \textit{e.g.}, dynamically bonding channels to 80 MHz as available.

\section{Packet Dynamics of Internet Speed Tests}\label{background}
Speed tests measure the upload and download TCP throughput that a client would get from a  server on the internet. If the speed test happens from a nearby server, the WLAN becomes a key part of this end-to-end path and the network manager can use these results to assess WLAN  performance. For the remainder of this chapter, we will only focus on speed tests that happen from a nearby server. A speed test is user initiated and the results are visible to the user at the end of the measurement. Speed tests primarily consist of two phases: a setup phase during which the speed test parameters are configured and a measurement phase which involves an active TCP upload and download. 

\textbf{Setup phase.} The setup phase begins with a server selection process which can either be manual or app driven. If this is app driven, a server is selected by probing a pool of available servers and a download or an upload session is established with it. Typically the server is selected such that the backbone delay between the server and the AP is as minimum as possible to ensure a maximum TCP throughput \cite{startbutton}. An ideal case would be one in which the selected server is in the same LAN as the AP since this would completely eliminate the effect of backbone delay on the measurements. Since the goal is to measure the maximum TCP throughput, while running a speed test, a STA is recommended to turn off other applications. Next, the client and server side TCP parameters are configured. The exact mechanism used for performing this configuration differs from one speed test application to another. A commonly used mechanism is to conduct a test download and a test upload from the STA. For instance, in the case of Ookla speed test, the STA initially downloads or uploads a small sized file to estimate an initial throughput. Following this initial phase, the STA adjusts the file size, buffer size and number of parallel TCP flows (limited to maximum of 8) to maximize the network connection usage while preventing congestion during the measurement phase \cite{ooklaoperation}.  

\begin{figure}
    \centering
    \includegraphics[width=1\linewidth]{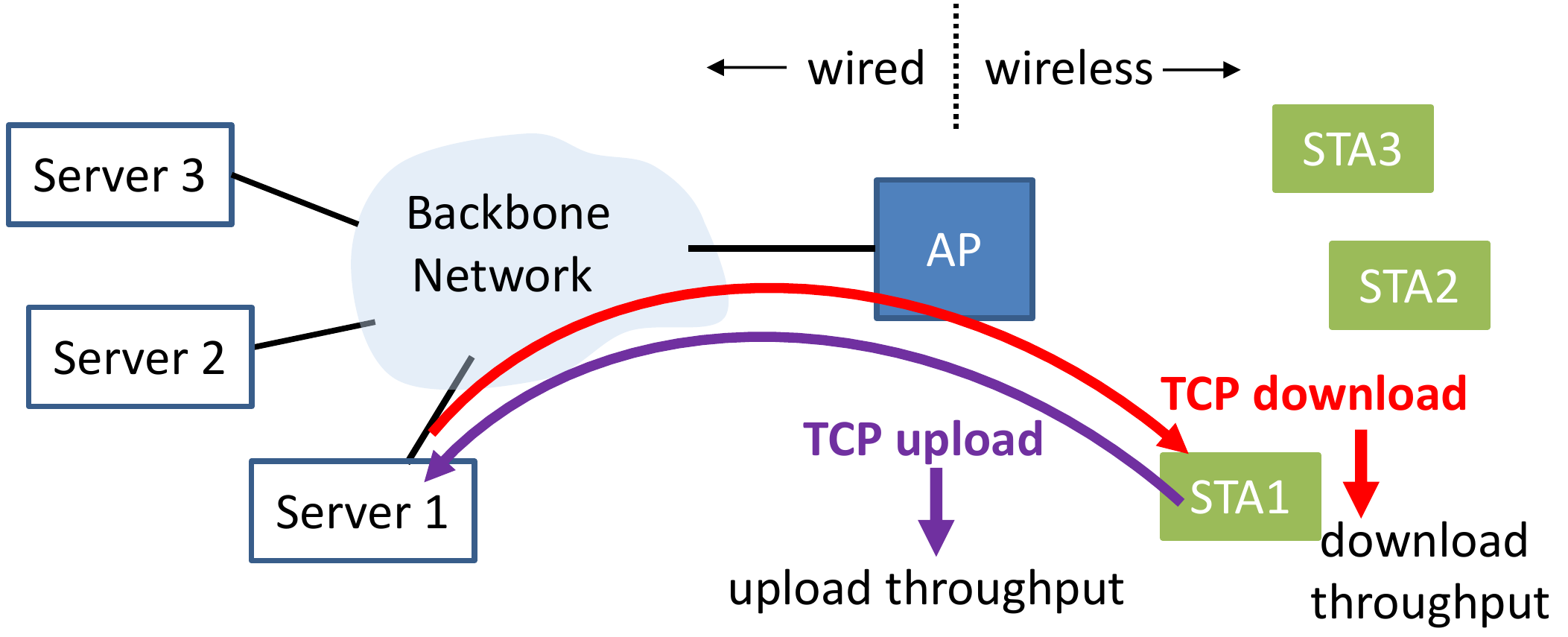}
    \caption{Illustration of an upload and download speed test: here STA 1 performs a speed test. Server 1 is selected from a pool of servers consisting of server 1, 2 and 3. Here server 1 has minimum backbone delay to the AP.}
    \label{tcp_speed_test}
\end{figure}

\textbf{Measurement phase.} As shown in Fig.~\ref{tcp_speed_test}, the measurement phase consists of two sessions: an upload session and a download session. A vast majority of the speed test apps available online follow a flooding based mechanism in the upload and download sessions \cite{goga2012speed}. A flooding mechanism involves establishment of several parallel TCP flows between the server and a STA with a calculation of aggregate throughput across all the flows. This ensures that the results obtained are robust to factors such as a small TCP window size  \cite{altman2006parallel} (due to, for instance, loss of a TCP segment) or any bounds on the maximum window size  \cite{bauer2010understanding} (for instance, due to a small receive window size advertised by the receiver) which could potentially make the total number of circulating TCP segments the prime bottleneck. The number of parallel flows to be established is determined in the setup phase. During the upload phase, a STA performs an active upload to the selected server and measures the TCP throughput by averaging the total data transmitted end-to-end over the total time taken. During the download phase, the STA performs an active download and measures throughput in a similar fashion.

\section{High Level Problem Formulation}\label{problem}
Analogous to online speed tests, our goal is to realize a \emph{virtual} speed test that enables an AP to estimate the TCP download and upload throughput that a STA can achieve from a nearby server. As described in our network scenario, an AP can have an arbitrary number of STAs associated with it and the AP should be able to estimate the throughput for any of the associated STAs. The speed test results of a STA can vary as driven by factors such as number of active STAs, interference level, etc. Likewise, we target that virtual speed test also tracks the speed test results for a given STA based on its own unique characteristics. Note that the STA does not perform the actual speed test. The AP is required to make the prediction using only passively collected information available on the AP side whereas no reports are available from STAs and out-of-network APs. Further, we consider that no additional commands can be required of STAs, \textit{e.g.}, STAs cannot be requested to send packets for testing purposes. Moreover, STAs cannot be requested to download special purpose software or report STA-side measurements. Instead, we consider that by leveraging AP-side observables, the AP can estimate the following metrics \cite{pgb}.

\textbf{Aggregate AP metrics.} We consider that the AP can measure the  airtime usage due to transmission and reception, defer time, contention time, idle time (no backlogged downlink traffic) as well as byte counts for downlink and uplink frames. 

\textbf{Per-STA metrics.} Likewise, while the STAs do not report STA-side statistics, the AP can observe some per-STA metrics at the AP such as uplink RSSI and SNR, downlink and uplink MCS and PHY parameters including use of advanced PHY features such as channel bonding, spatial multiplexing, multi-user transmission and downlink retransmission statistics. 

\textbf{Non-associated device metrics.} Lastly, the AP may be in range of a number of non-associated 802.11 devices that are transmitting on a different BSS. When the AP is forced to defer to a non-BSS device, it can record  interferer air time consumption. 

While the above may appear to be an exhaustive set of information for performance characterization, there are a number of STA-side metrics that remain unobservable by the AP. For instance, the AP does not know the STA's idle times or the STA's defer times due to NAV especially when the STA is deferring to a non-BSS device. Since the network scenario considers a complex inter-node connectivity which may lead to inter-cell interference, hidden terminals, etc., these parameters cannot be directly calculated based on the metrics mentioned above. However, the throughputs that we want the AP to estimate are  inherently bi-directional, end-to-end and layer-4 and can indeed be degraded by the above factors. To this end, we infer the impact of these unknowns using the above AP-side observables with the help of techniques described in Section~\ref{AP_side_measurement}.

\section{L2 Edge TCP Model}\label{model}
To enable an AP to estimate the upload and download throughputs that a STA would obtain if it performs a speed test, we develop a novel L2 edge TCP model 
that uses AP-side observables as inputs. 

\subsection{Assumptions for Mathematical Analysis}

Here, we state the key assumptions that we make to capture important aspects of the aforementioned  speed test setup and measurement phases in our model. 

In the measurement phase of an actual speed test, multiple TCP flows are initiated between the server and the client so that the measurement phase is not bottlenecked by the number of circulating TCP segments. Instead, we model this  by representing the packet flow dynamics by a single long lived TCP flow with a maximum congestion window size of $\wmax$ which is large enough so that there are a sufficient number of TCP segments circulating in the network. We further assume that this flow does not experience any permanent packet losses. This is not to say that collisions or packet errors do not occur on the wireless channel. Rather, packets lost on the wireless channel are locally retransmitted by the MAC layer and we do account for these collisions and retransmissions in our analysis. We hereby refer to this modeled flow as the speed test flow, the STA under consideration as the target STA and the remaining STAs as non-target STAs. 

In an actual speed test, the server selection process in the setup phase selects a server with minimum latency to the AP to reduce the impact of backbone elements on the measured results. Consequently, we consider backbone congestion and delays as factors that do not impact the throughput. Also, recall that the parameters of the TCP flow used during the speed test are adjusted by the STA based on an initial measurement performed to ensure that TCP does not drive the network into congestion.

\subsection{Virtual End Point Representation}
\begin{figure}
    \centering
    \includegraphics[width=0.8\linewidth]{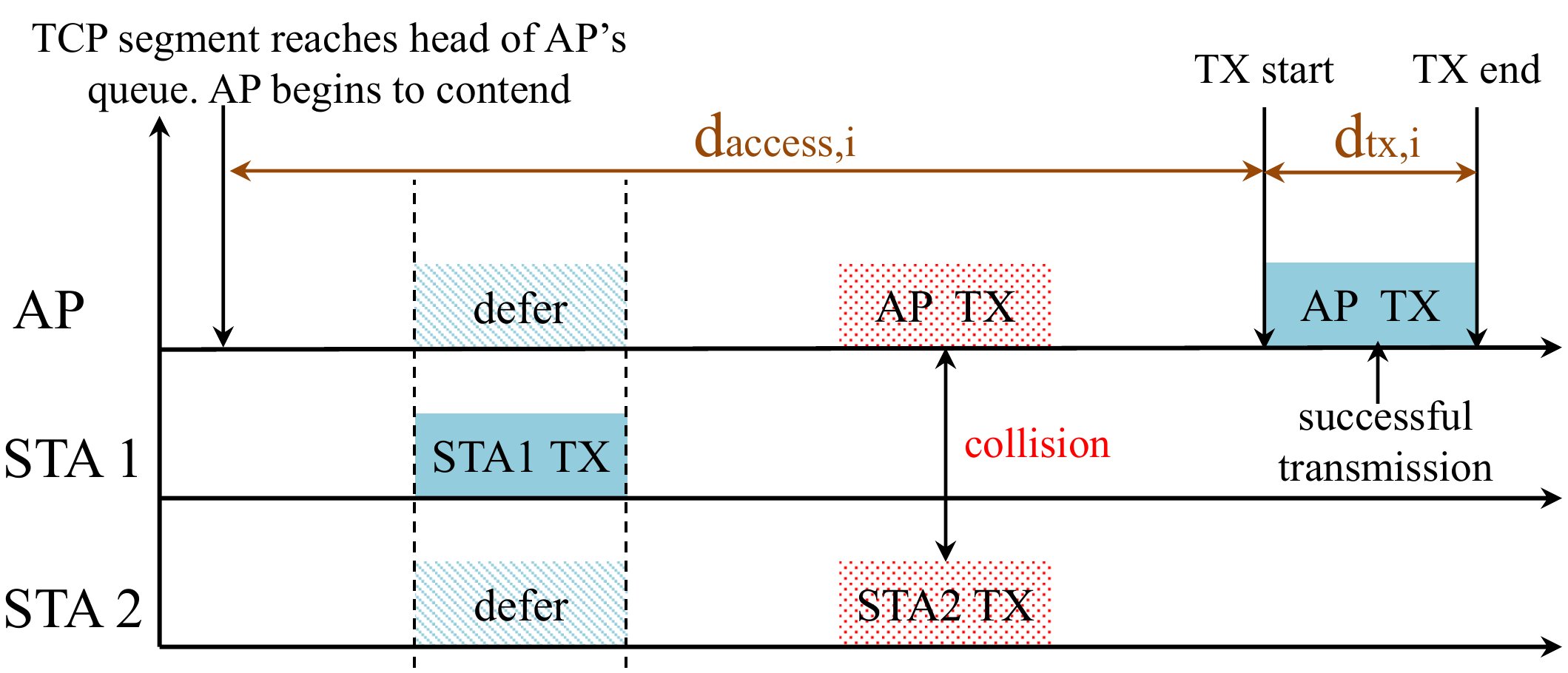}
    \caption{Example timeline of a downlink transmission to depict $d_{\textrm{\tiny access,i}}$ and $d_{\textrm{\tiny tx,i}}$. $d_{\textrm{\tiny access,i}}$ and $d_{\textrm{\tiny tx,i}}$ denote the `access' and `tx' values respectively for the $i^{\textrm{th}}$ downlink transmission. $\dldelay$ and $\dltx$ denote their mean values.}
    \label{downlink_tx}
\end{figure}

The discussions in this sub-section are mainly in the context of a download speed test. However, the arguments and explanation are applicable to upload speed tests as well and will be generalized later. 

In the network scenario of Fig.~\ref{scenario}, there are no restrictions on the traffic flows of non-target STAs and STAs in neighboring BSSs and they may have UDP and/or TCP traffic going on the downlink and/or the uplink. Further, the number of these flows per device can also be variable and differ from STA to STA. Since we make no assumptions about the network topology, interfering links or the type or number of flows, it is not possible to state precisely the inputs for a queuing model. We remark that a majority of TCP models for Wi-Fi require AP-side knowledge of network topology, interfering nodes including those from neighboring BSSs, their traffic patterns, PHY capabilities, data rates, etc. Removing this requirement is vital to the realization of virtual speed test.

The modeled speed test flow comprises both its TCP segments and TCP ACKs. First, we analyze the speed test flow by considering the journey of a speed test flow segment from the server to the target STA. On the forward path, a TCP segment experiences delays on the queues of devices on the backbone.\footnote{A example cause of these delays is that due to cross traffic sharing a common queue on the backbone with the TCP segment.} When the packet  enters the queue at the AP, it encounters another delay before reaching the head of the queue, part of which arises from the AP serving non-speed test flow packets. We denote the average amount of time the AP spends on non-speed test flow packets prior to serving a speed test flow packet by $\usage$. Upon reaching the head of the queue, the AP begins to contend to access the channel. It is possible that as the AP counts down, the target STA or a non-target STA or another AP wins the channel, causing the AP to defer. It is also possible that a transmission from the AP fails either due to collision or poor channel quality, forcing the AP to double its contention window size and re-contend and transmit (with the same or adapted data rates\footnote{The exact rate adaptation policy is vendor implementation dependent.}). We denote the mean time the AP takes to win the channel prior to a successful transmission by $\dldelay$ as shown in Fig.~\ref{downlink_tx}. Notice that the value of this parameter can vary depending on the STA being considered as the target STA. The average amount of time to transmit the TCP segment is represented by $\dltx$. This includes any MAC and physical layer overhead, MAC frame transmission time, all interframe spacings and the MAC layer acknowledgement. Just like the TCP segment, the TCP ACK also faces a similar journey back to the server. The terms $\uldelay$ and $\ultx$ are defined in a similar manner for the target STA.  

\begin{figure}
    \centering
    \includegraphics[width=0.8\linewidth]{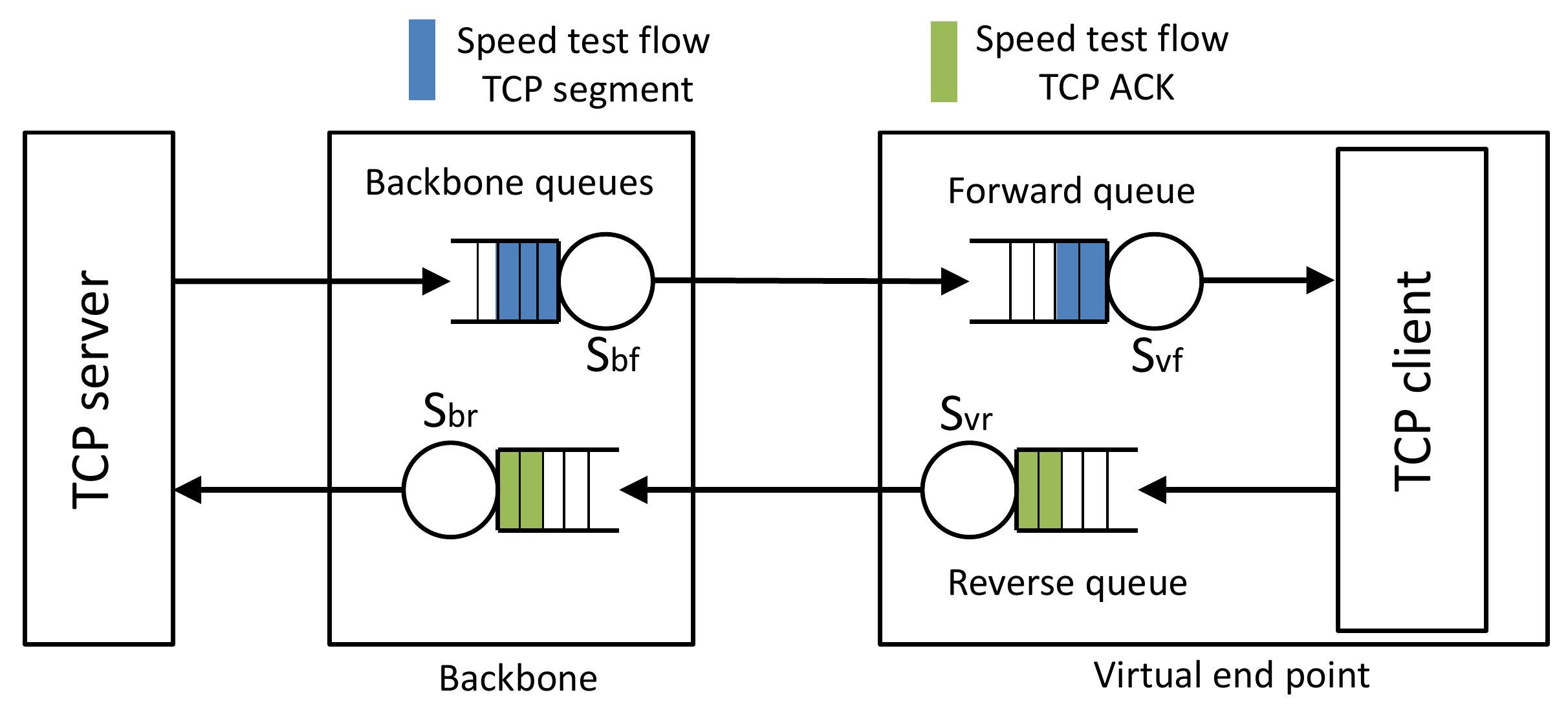}
    \caption{WLAN representation as a virtual end point consisting of a forward and a reverse queue. The TCP client here refers to the socket level client and is not to be confused with the physical STA itself.}
    \label{quemod}
\end{figure}

For our analysis, we represent the WLAN (AP and STAs) as a virtual end-point consisting of two queues: a forward queue and a reverse queue. For now, let us assume that the non-speed test flows are non-existent and that only the speed test flow packets exist in the WLAN (we subsume the impact of non-speed test flow packets into the model parameters later). With this consideration, we can treat the virtual end point as a black box replacing the WLAN that runs a speed test. The socket level TCP client (not to be confused with the physical STA) runs on the virtual end point itself as shown in Fig.~\ref{quemod}. We can think of TCP segments and TCP ACKs as jobs circulating in the network. The service time of each job is a sum of its `access' term and its `tx' term. \textit{E.g.}, for jobs in the forward queue, the service time is a sum of $\dldelay$ and $\dltx$. Since we account for the `tx' term in the service time itself, the jobs themselves become indistinguishable. As we have not yet subsumed the effect of non-speed test flows, the throughput of the virtual end point is not the same as that of the target STA in our WLAN. 

In our second step, we  account for the impact of non-speed test flow packets on the throughput of this system by inflating the service times of each queue to account for the non-speed test flow packets. In essence, this inflation makes the effective speed of each server as seen by the speed test flow packet in the virtual end point system the same as that in the original system where some server time should have been consumed by non-speed test flow packets as well. Consequently, on the forward queue, the service time is inflated by $\usage$. However, since the target STA has no other uplink traffic while performing a speed test, the reverse queue service time requires no inflation. Similarly, we can subsume the impact of cross traffic on the backbone queues into their respective service times. 

\subsection{Throughput Analysis}
To analyze the throughput of the network shown in Fig.~\ref{quemod}, we consider two cases. First we consider a case wherein TCP performs no ACK thinning. Consequently, in this case, each TCP segment received by the STA results in the generation of a TCP ACK. Next, we  generalize this to account for the case of ACK thinning with an ACK thinning ratio of $n$. In this case, the client generates a TCP ACK following the receipt of every $n^{th}$ TCP segment.
 
\subsubsection{No TCP ACK Thinning}
Ignoring the initial transient stage during which TCP's window size grows,  the speed test flow will reach a steady state wherein TCP operates at $\wmax$. Consequently, the number of packets that are contained in the speed test flow, which can either be TCP segments or TCP ACKs, remain constant and the system behaves as a closed queuing network with tandem servers and a constant number of jobs circulating inside it. 

Based on the aforementioned notations, the mean service time for the forward and the reverse queue in the virtual end point (Fig.~\ref{quemod}) is given by:
\vspace{-1pt}
\begin{equation}
S_{vf} = \dldelay + \dltx + \usage
\label{s1}
\vspace{-18pt}
\end{equation}

\begin{equation}
S_{vr} = \uldelay + \ultx  
\label{s2}
\vspace{-1pt}
\end{equation}

\noindent Let $S = S_{bf}+S_{br}+S_{vf}+S_{vr}$, $S_{max} = max(S_{bf}, S_{br}, S_{vf}, S_{vr})$ and $\theta$ denote the throughput in terms of jobs per second. It can be shown  \cite{harchol2013performance} that 
\vspace{-1pt}
\begin{equation}
    \theta \leq min \Bigg(\frac{\wmax}{S} , \frac{1}{S_{max}} \Bigg) 
    \label{bounds}
\end{equation}

\noindent where $\frac{\wmax}{S}$ is an asymptotic bound for small values of $\wmax$ and $\frac{1}{S_{max}}$ acts as an asymptotic bound for large values of $\wmax$. The cases of  small and large here are relative to a critical value $\wmax^{*}$ which is the point at which the asymptotes cross each other. Consequently, 
\vspace{-1pt}
\begin{equation}
\wmax^{*} = \frac{S}{S_{max}}  
\label{eqwmax}
\end{equation}

\noindent To understand the physical relevance of the two components of Eq.  (\ref{bounds}), let us consider two extreme case scenarios. Let us assume that $\wmax = 1$ which makes the number of jobs circulating in Fig.~\ref{quemod} the botteneck. The throughput, therefore, is given by $\frac{\wmax}{S}$. On the other extreme, if $\wmax$ is sufficiently large (again large as compared to $\wmax^{*}$) to not bottleneck the system, then the slowest queue acts as a bottleneck. In this case the slowest queue always remains busy and in accordance with the utilization law, $\theta = \frac{1}{S_{max}}$. 

Recall that due to the server selection process, $S_{br}$ and $S_{bf}$ are not the bottleneck in the system. To understand the typical values that $\wmax^{*}$ can take, let us consider the critical point wherein $S_{br} = S_{bf} \sim max(S_{vf}, S_{vr})$. Substituting in Eq. (\ref{eqwmax}), we will get $\wmax^{*} = \frac{2*(S_{vf} + S_{vr})}{max(S_{vf}, S_{vr})}$. The maximum value of $\wmax^{*}$ occurs when $S_{vf} = S_{vr}$ and thus $\max (\wmax^{*}) = 4$. In practice, $\wmax \gg 4$ and consequently, we can see that $\theta \leq \frac{1}{S_{max}}$ will act as a asymptotic bound on the values of $\theta$. In fact, we find in our experimental evaluation that for a typical speed test, the values of $\wmax$  is extremely large as compared to 4 and $\theta$ will tend to the bound yielding 
\vspace{-1pt}
\begin{equation}
\theta \sim \frac{1}{S_{max}}.   
\label{throughput}
\end{equation}

\subsubsection{TCP ACK Thinning} Now, we extend the above to the more general case of TCP ACK thinning. For an ACK thinning ratio of $n$, we can view a maximum of only $\frac{\wmax}{n}$ jobs circulating in the system and the remaining jobs can again be accounted for by further inflating the service times of each of the queues (just as for non-speed test flows). Consequently, when the wireless nodes transmit only one frame per transmission, the service times of both the forward and reverse queue in the virtual end point stretch by an amount equal to ($n - 1$)$\times$($\dldelay + \dltx + \usage$) for the case of the download speed test. Here we inflate the service time of the reverse queue to account for the fact that the TCP ACK is not generated until the $n^{th}$ TCP segment is received. The numerator of Eq. (\ref{throughput}) should also be multiplied by $n$ to compensate for the shrinking of the total number of TCP segments. For the upload speed test, the service times stretch by ($n - 1$)$\times$($\uldelay + \ultx$). However, when the nodes transmit multiple frames per transmission, such an inflation is not necessary since the STA receives multiple TCP segments in a single downlink transmission and there is no  additional delay in the generation of a TCP ACK. These multiple frames may be transmitted using frame aggregation in single stream transmissions (\textit{e.g.}, SISO) or by using multi-stream transmissions (\textit{e.g.}, MIMO) or a combination of both frame aggregation and multi-stream transmissions. We emphasize that this is possible since typical ACK thinning ratios of TCP are much smaller than the number of frames that can be transmitted in a single transmission via the above mentioned policies under 802.11 \cite{dmurray,braden1989rfc,std11ac,std11n}. 

In summary, the throughput in bits/sec is given by

\vspace{-1pt}
\begin{equation}
\esdl = \frac{\E[\textrm{TCP segment size}] \times \FAP}{max(S_{vf},S_{vr})} 
\label{dlst}
\vspace{-5 pt}    
\end{equation}

\begin{equation}
\esul = \frac{\E[\textrm{TCP segment size}] \times \FSTA}{max(S_{vf},S_{vr})}
\label{ulst}
\end{equation}

\noindent where we denote $\esdl$ and $\esul$ as the download and upload TCP throughputs respectively. $\FAP$ denotes the average number of frames transmitted by the AP in a single downlink transmission to the target STA.  For the case of the upload speed test, we use $\FSTA$ instead.

Note that while calculating $S_{vf}$ and $S_{vr}$ for Eq. (\ref{dlst}), $\dltx$ is the average time to transmit $\FAP$ number of TCP segments at the AP's data rate and $\ultx$ is the average time to transmit $\FSTA$ number of TCP ACKs at the target STA's data rate. In Eq. (\ref{ulst}), this is reversed since the target STA is now the one transmitting TCP segments and the AP is the one transmitting the TCP ACKs. $S_{vf}$ and $S_{vr}$ further vary depending on which STA is chosen as the target STA. Consequently, the AP has to estimate these two parameters with respect to the particular STA that is chosen as the target STA.

We remark that while the L2 edge TCP model needs to be supplemented with AP-side measurements, it is not restricted by a requirement for AP-side knowledge of inter-node connectivity or an assumption on network traffic characteristics. Next we show how the model parameters are estimated. 

\section{AP-side Passive Estimation of Model Parameters}
\label{AP_side_measurement}
In this section, we show how the AP can measure all of the parameters required for the above model, thereby enabling a dynamic AP-side speed test estimate for each STA.

\subsection{AP-side Estimation Problem}
We observe that Eq. (\ref{dlst}) and (\ref{ulst}) are independent of $S_{br}$ and $S_{bf}$. To estimate $\esdl$ and $\esul$ at the AP, the key challenge is computation of $S_{vr}$, as the remaining parameters are based on common AP side observables described in Sec~\ref{problem}. Recall from Eq. (\ref{s2}) that $S_{vr}$ is composed of $\ultx$ and $\uldelay$. While the average uplink transmission time $\ultx$ is known to the AP via per-STA metrics, the uplink access time $\uldelay$ is known only at the STA side. Let $t_{\textrm{\tiny hq}}^{U,i}$ denote the time at which the $i^{th}$ \emph{uplink} packet reaches the head of the STA's queue,  $t_{\textrm{\tiny ts}}^{ U,i}$ denote the start time corresponding to the successful transmission of this packet and $t_{\textrm{\tiny te}}^{U,i}$ denote the end time of this packet transmission. By definition, $\uldelay = \E[(t_{\textrm{\tiny ts}}^{U,i} - t_{\textrm{\tiny hq}}^{U,i})]$. While the AP can observe $t_{\textrm{\tiny ts}}^{ U,i}$ for any uplink transmission, $t_{\textrm{\tiny hq}}^{U,i}$ remains unknown. If the STA is assumed to be fully backlogged, the end time of the previous transmission can be approximated to be the time when the next packet reached the head of the queue. However, STA backlog is user activity dependent and is not known to the AP. As a result, the AP cannot estimate $\uldelay$ by a simple observation of packets received on the uplink.

\subsection{Snooped Handshakes for Model Parameter Estimation}
Suppose that the client is performing a TCP download from a server (\textit{e.g.}, streaming a Netflix video). This can be any server on the internet with any backbone delay to the AP. The client will attempt to return a TCP ACK as fast as possible after reception of the corresponding TCP segment. This TCP ACK is ``data" at layer 2. For now, consider a case where there are no other flows on the uplink from the target STA and no ACK thinning. Since the WLAN is the final hop for the TCP segment, upon reception of a TCP segment, \emph{i.e.}, at the end of the AP's successful downlink transmission (denoted by $t_{\textrm{\tiny te}}^{D,i}$), the STA has the corresponding TCP ACK and begins to contend. Consequently, in this case, $t_{\textrm{\tiny hq}}^{U,i} = t_{\textrm{\tiny te}}^{D,i}$ and thus the AP will have inferred a parameter that is not directly observable. In essence, the delay incurred between the transmission of the segment to the reception of the TCP ACK enables the AP measure how long it takes the STA to successfully transmit after it starts to attempt. Thus, our general approach is to selectively sample TCP data-ACK handshakes from any TCP download performed by the target STA and use them to drive a measurement based prediction of $\esdl$ and $\esul$. We refer to such TCP flows as \emph{snooped flows}.

This can be generalized under a flow hypothesis (\emph{i.e.}, knowing that a given flow on the downlink is a TCP flow) by the following two cases.

\textbf{ACK queuing. } This case occurs when the target STA has other uplink flows whose packets get queued prior to the TCP ACK. Consequently, in such scenarios, $t_{\textrm{\tiny hq}}^{U, i} = t_{\textrm{\tiny te}}^{U, i - 1}$. In such cases, we abuse the term  $t_{\textrm{\tiny te}}^{U, i - 1}$ to refer to the end time of transmission of the immediately preceding uplink packet.

\textbf{ACK immediate. } However, if the target STA has no other uplink flow, it begins to contend as soon as the TCP ACK is queued. Consequently,  $t_{\textrm{\tiny hq}}^{U,i} = t_{\textrm{\tiny te}}^{D,i*n}$ where the superscript `$D$' refers to a downlink transmission. 

\subsection{TCP Flow Inference}
Because  the layer four handshake  is needed to estimate  $\uldelay$, it is crucial to identify this handshake at the AP, which does not have layer four visibility.  To this end, we employ IP addresses and size signatures as follows. 

\textbf{IP address signature. } Due to the inherent bi-directionality of TCP,  the source and destination addresses for TCP segments traversing on the forward path are swapped for the corresponding TCP ACKs on the reverse path. This key factor enables us to distinguish individual TCP flows and separate them from the remainder of the downlink and uplink traffic.

\textbf{Packet size signature. } Although the above signature enables identification of a bidirectional flow, it does not aid in spotting the forward and reverse paths distinctly. While the size of TCP segments on the forward path may fluctuate during the course of a download, the reverse path is characterized by small TCP ACKs whose size remains fixed during the entire duration of the flow. Typically a TCP ACK is 20 bytes long \cite{jon1981transmission}. Having distinctly identified the forward and reverse paths, the AP can employ the $\uldelay$ estimation process described in the previous sub-section. 

\chapter{Analytical Model for TCP Throughput Estimation}

The L2 Edge TCP model is a measurement based model that enables a data-driven estimation of download and upload speeds achievable by a target STA over its current wireless connection. In this chapter, we explore traditional analytical modeling approaches to estimate TCP throughput. We present the first analytical model that captures the performance of 802.11ac MU-MIMO under the impact of closed loop TCP dynamics. Our model reveals interesting bottleneck regimes that can be used as guidelines for designing next generation WLANs.\footnote{This work has been previously published in IEEE INFOCOM 2017 \cite{ngk}.} 

Downlink multi-user MIMO (DL MU-MIMO) is a promising physical-layer
technology to boost the capacity of wireless LANs by transmitting data
streams to multiple stations (STAs) concurrently, thus scaling up the
achievable data rate by a factor equal to the number of antennas on
the Access Point (AP).  This approach is different from traditional
single-user (SU) networks where only one STA gets served at a time.
With inclusion in the IEEE 802.11ac standard \cite{bejarano2013ieee, ieee2013ieee}, DL MU-MIMO has moved from theoretical research into the real world.

In this chapter, we show that DL MU-MIMO alone, without UL MU-MIMO, does
not necessarily correspond to an equivalent gain in terms of
throughput perceived by users at the transport layer, even if the vast
majority of bytes are transmitted in the downlink direction, \textit{e.g.}, via
download of large files via TCP.  Specifically, severe
performance degradation can occur, {\color{black} in some scenarios,} when DL MU-MIMO is coupled with a single-user uplink under closed-loop traffic such as that generated by TCP, which still carries more than 80\% \cite{dmurray,garcia2012characterization,marnerides2018internet} of Internet traffic today. {\color{black} In particular, we show that a key performance factor is the amount of frame aggregation performed during each transmission in the downlink or in the uplink.}\footnote{This work has been previously published in IEEE/ACM Transactions on Networking \cite{nayak2019modeling}.}

\section{Network Scenario} \label{sec:scenario}
\subsection{Cross-layer Setup}
To investigate the performance of DL MU-MIMO under closed-loop traffic, 
we consider a simple network scenario and adopt some simplifying assumptions
to analyze it. We emphasize that our goal is not to develop a comprehensive model
to predict TCP throughput over MU-MIMO WLANs under very general and realistic conditions,
but to identify crucial performance factors that can offset the gains achievable
by MU-MIMO. Such factors, which are more easily understood and quantitatively
analyzed in a simple (but not unrealistic) scenario, are expected to 
affect likewise the performance of MU-MIMO WLANs in more realistic and complex 
conditions.\footnote{Further, it would be extremely interesting to experimentally
verify our findings in a real network testbed, however this effort goes beyond
the modeling purposes of this work.}
 
\begin{figure}
    \centering
    \includegraphics[width=0.9\linewidth]{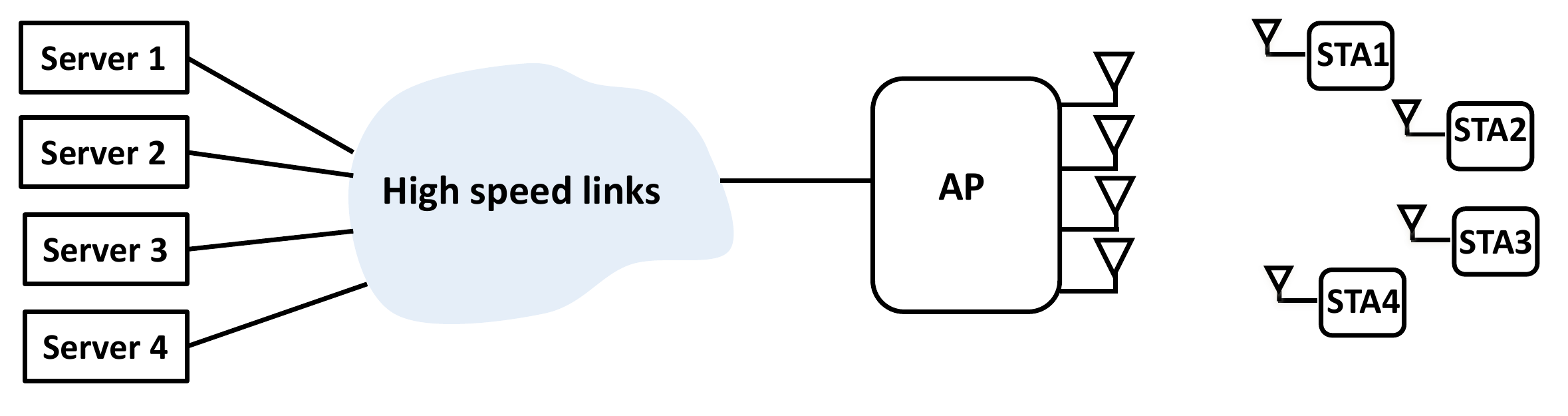}
    \vspace{-0.2cm}
    \caption{Network topology for studying impact of closed loop traffic.}
    \label{cltop}
    \vspace{-0.5cm}    
\end{figure}

We consider the network scenario illustrated in Fig. \ref{cltop}. 
A set of users (or stations\footnote{In this Chapter, we use the term user and station interchangeably.}) 
attached to a wireless LAN establish long-lived TCP flows to download bulk data 
from a set of servers located in the wired network. To isolate the targeted factors,
we assume that data 
is sent only on the downlink, so that just TCP ACKs are sent in the uplink direction.   
Servers are connected to the AP over high speed links which ensures
absence of congestion and queueing delays in the wired portion of the network.

In this scenario,  there are no losses in the backbone, therefore
each TCP flow (discarding an initial transient) operates at the maximum TCP congestion window size.  
As a consequence, TCP dynamics related to specific versions of the TCP protocol 
do not come into play in our scenario. Essentially, the only TCP feature
that matters is the fact that data (ACK) packets are transmitted by TCP senders (receivers) 
in response to ACK (data) packets received in the opposite direction.
This captures the closed-loop nature of the traffic generated by almost 
all versions of TCP.

Note that, while operating at the maximum congestion window size, 
TCP senders transmit one data packet in response to each TCP ACK
(or two data packets, if the delayed ACK option is enabled \cite{braden1989rfc}).  
We assume that all TCP flows traverse the same AP, which is equipped 
with multiple antennas and performs MU-MIMO transmissions on the wireless channel
whenever possible, \textit{i.e.}, when the AP has backlogged traffic for more than one user.

As is the case with IEEE 802.11ac, uplink transmissions by the stations 
are instead single-user, \textit{i.e.}, the STAs
transmit on the uplink one at a time as dictated by random access. In general, the STAs
could also be equipped with multiple-antennas, and thus perform SU-MIMO
by transmitting multiple streams to the AP simultaneously (we account for 
this in our analysis).

We will be especially interested in analysing the standard case  
in which channel access is governed by the fair 802.11 
contention mechanism, which provides equal probability
of contention victory to all nodes competing for transmission:
each node that intends to transmit generates a random value for the backoff timer chosen 
uniformly from $[0,W_0-1]$ where $W_0 = 16$ is the minimum contention window size. 
While the channel is sensed idle, the node counts down with a slot duration of $\sigma$, 
and transmits when the backoff timer becomes zero.  

Since the random channel access protocol of 802.11 can be responsible
for severe throughput degradation of MU-MIMO under conditions that we will
uncover in this work, alternative channel access strategies
will be considered later in Sec. \ref{sec:comparison}.

\subsection{Background on 802.11ac Compliant MU-MIMO}\label{subsec:ac}
Here, we review the key components of the 802.11ac timeline
for our analysis. When the AP obtains access to the channel by winning contention, it performs a transmission including three main phases:

\textbf{Channel Sounding and feedback phase.} The AP requires channel
state information at the transmitter (CSIT) to limit interference among users.
Consequently, it initiates a sounding process 
by transmitting a Null Data Packet Announcement (NDPA) which contains information 
that identifies the STAs that the AP intends to transmit data to on the downlink. Following this, the AP 
transmits a Null Data Packet (NDP) which contains the pilot sequence that the STAs use to estimate the CSI. 
The STAs  process the CSI to calculate the angles $\phi$ and $\psi$ that are used to build the transmit 
weight matrix at the AP \cite{perahia2013next}. The STAs transmit these in 
a compressed beamforming report (CBR), as polled by AP. 

\textbf{Data transmission phase.} Data is transmitted simultaneously to the
users, typically via zero-forcing beamforming using the collected CSIT.
To amortize overhead and improve performance,
the AP aggregates multiple frames destined to the same STA into the same data bundle.
We emphasize that 802.11ac allows up to 1 MB to be aggregated per STA.

\textbf{Acknowledgement phase.} After the AP transmits data, 
the first STA responds with a Block Acknowledge (BA). Following this, the AP 
subsequently transmits a block acknowledgement request (BAR) to other STAs, which 
then transmits their BA.   

\begin{figure}
    \centering
    \includegraphics[width=0.9\linewidth]{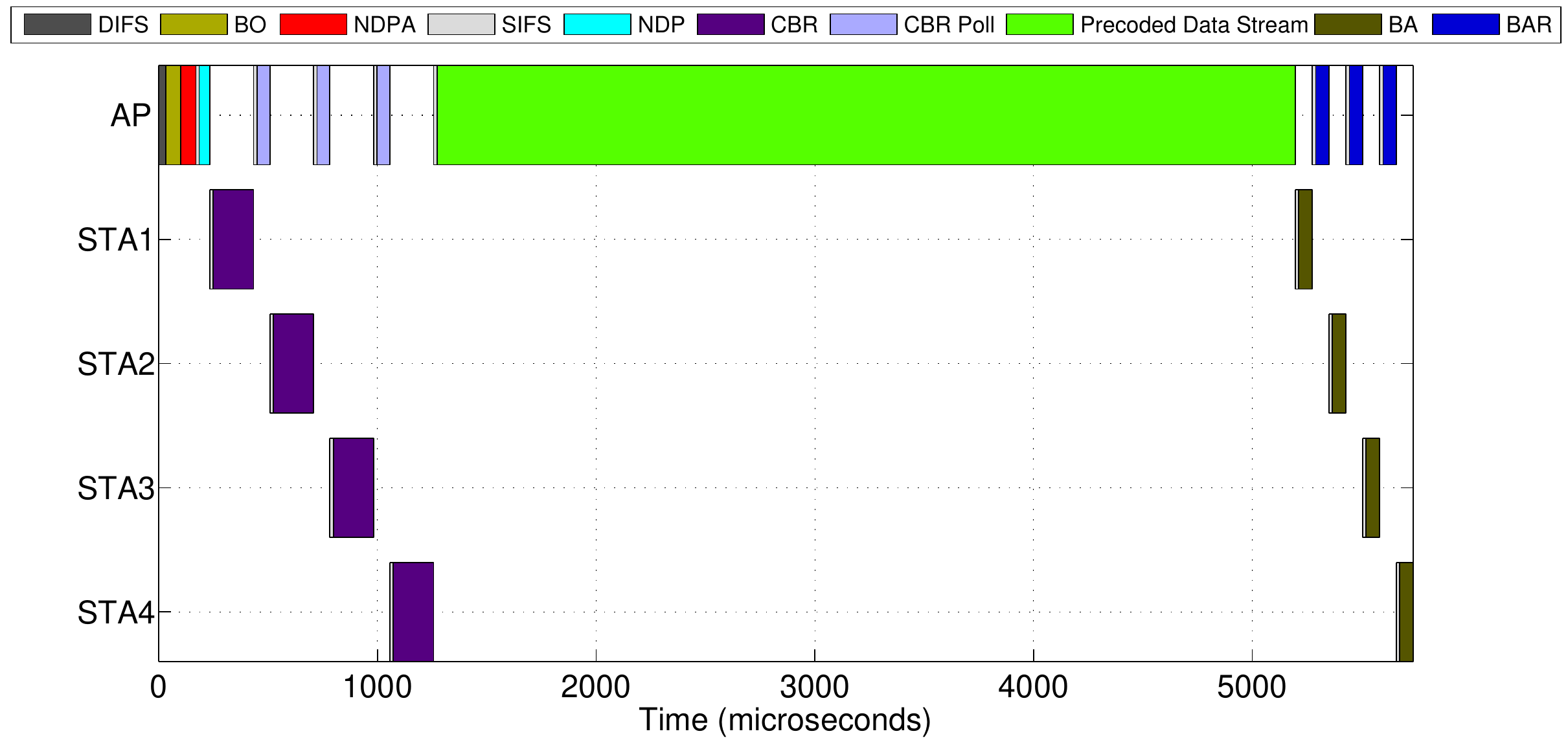}
    \vspace{-0.2cm}
    \caption{An example of 802.11ac downlink transmission timeline in the case of an AP with 4 transmit antennas  serving 4 single-antenna STAs.}
    \label{timeline}
    \vspace{-0.5cm}
\end{figure}

Fig.~\ref{timeline} shows an example 802.11ac downlink transmission for an AP with four 
transmit antennas serving four single-antenna STAs, in the case of channel bandwidth 20MHz, 
sub-carrier grouping of 4 and quantization bits for $\phi$ and $\psi$ being 7 and 5 respectively. 
These values result in the minimum possible sounding and feedback phase duration at this bandwidth. 
Note that, even in this case, the total overhead due to channel sounding
and feedback phases is about 1.5 milliseconds.
During this interval, roughly 10 data packets of size 1 KB could be transmitted using
standard SISO. Therefore, aggregation of at least a few tens of frames (among all stations)
is necessary to get any performance gain from MU-MIMO with respect to traditional SISO. 

To validate the results obtained in this work, we extended the simulator ns3 \cite{ns3rng} to incorporate detailed behavior of 802.11ac compliant MU-MIMO WLANs. 

\section{System Model}\label{sec:system}
\begin{table}
\begin{center}
\vspace{0mm}
\caption{Descriptions of symbols used for analytical modeling\label{tab:notation}}{%
\scriptsize
\begin{tabular}{||c|l||}
\hline
\hline
\rule{0pt}{3ex}
$K$  & number of stations \\[1ex]
\hline
\rule{0pt}{3ex}
$F_s$  & Number of TCP flows for each station \\[1ex]
\hline
\rule{0pt}{3ex}
$W_{\max}$ & TCP maximum window size \\[1ex]
\hline
\rule{0pt}{3ex}
$T_F$  & TCP ACK thinning factor \\[1ex]
\hline
\rule{0pt}{3ex}
$D$ & two-way propagation delay of each flow \\[1ex]
\hline
\rule{0pt}{3ex}
$\NAP$  & number of antennas in the AP \\[1ex]
\hline
\rule{0pt}{3ex}
$\NSTA$ & number of antennas in a station \\[1ex]
\hline
\rule{0pt}{3ex}
$\BAP$ & maximum frame aggregation by AP \\[1ex]
\hline
\rule{0pt}{3ex}
$\BSTA$ & maximum frame aggregation by a station \\[1ex]
\hline
\rule{0pt}{3ex}
$A(h, b)$ & channel holding time of the AP \\[1ex]
\hline
\rule{0pt}{3ex}
$\Lambda$ & aggregate system throughput \\[1ex]
\hline
\hline
\end{tabular}}
\end{center} 
\vspace{-1cm}
\end{table} 

\subsection{Assumptions and Notation}\label{subsec:assume}
The main notation used to describe the considered system is summarized in Table \ref{tab:notation}.
Let $K$ be the number of stations attached to the AP, each of which is a 
destination of at least one long-lived TCP flow.
Our goal is to compute the aggregate steady-state  
throughput $\Lambda$ achieved by the set of all TCP flows.
 
In some of the scenarios that we will consider, the aggregate throughput
will be limited by the TCP maximum window size $W_{\max}$ (expressed in number of
segments). In those cases, we will assume for simplicity a symmetric
traffic scenario: stations establish an equal number $F_s$ of TCP flows,
and all flows experience the same two-way propagation delay 
$D$ in the fixed network.   

To simplify the analysis, we further assume a perfect wireless channel 
(without errors) and a collision-free MAC protocol.\footnote{Under the 802.11 MAC protocol, 
the absence of collisions can be obtained (\textit{i.e.}, simulated with ns3) by 
assuming that the backoff extracted by a node is continuous, rather than 
discrete, and that nodes instantaneously freeze their backoff as soon as another
node starts transmitting.}    
While these assumptions are simplifications of the real system, they
enable us to capture macroscopic effects into a parsimonious analytical model. Channel errors and/or
collisions could be incorporated in the analysis using well-established techniques \cite{bianchi2000performance,kumar2005new},     
but we do not do so here to keep the analysis focused on the joint
impact of a closed-loop transport layer with a multi- and single-user 
MAC. Further, collisions typically produce only a second-order effect, 
while they do not lead to closed-form expressions (\textit{i.e.}, they require numerical fixed-point solutions).

We consider an AP implementing a work-conserving policy: when it has
at least one packet to transmit, the AP starts contending for channel access.
When it wins the channel, the AP employs multi-user MIMO whenever
it has packets queued for at least two different stations (if it has packets destined 
to only a single station, the AP employs single-user MIMO).
Note that the AP maintains a separate queue to store the packets destined to each
attached station. Let $\NAP$ be the number of antennas in the AP. 
Let $\NSTA$ be the number of antennas in each of the stations.
If $\NAP < K$, it is possible that the number of stations for which the AP has a non-zero backlog is larger than the number of antennas at the AP. In this case, we assume
that the AP will pick $\NAP$ different stations with non-zero backlog
uniformly at random.         
Let $A(h, b)$ be the channel holding time of the AP, which depends on two 
parameters: the number of non-empty queues $h$, and the largest backlog $b$
of these queues. Note that $A(h, b)$ is a known deterministic function
of $h$ and $b$, given physical system parameters.

Let $\BAP$ be the maximum number of frames destined to the same station that can be
aggregated and sent by the AP in the same channel access.
Note that $\BAP$ will never constrain performance when $\BAP > F_s W_{\max}$, since 
in any case the AP cannot store a number of frames destined to the same station larger than the 
product of the TCP maximum window size times the number of flows per station.

Let $\BSTA$ be the maximum number of frames (TCP ACKs, in our case) destined to the AP 
that can be aggregated and sent by a station in the same channel access.  

We emphasize that the vast majority of existing performance evaluation studies
of 802.11, focused on early versions of the standard, only consider the case
$\BSTA = \BAP = 1$. The impact of aggregation (in particular, possibly different
levels of aggregation performed by the AP and by the stations) is instead fundamental
to understand the performance of MU/SU MIMO systems.        

\subsection{High-Level Packet Dynamics}\label{subsec:highlevel}

\begin{figure}
    \centering
    \includegraphics[width=1\linewidth]{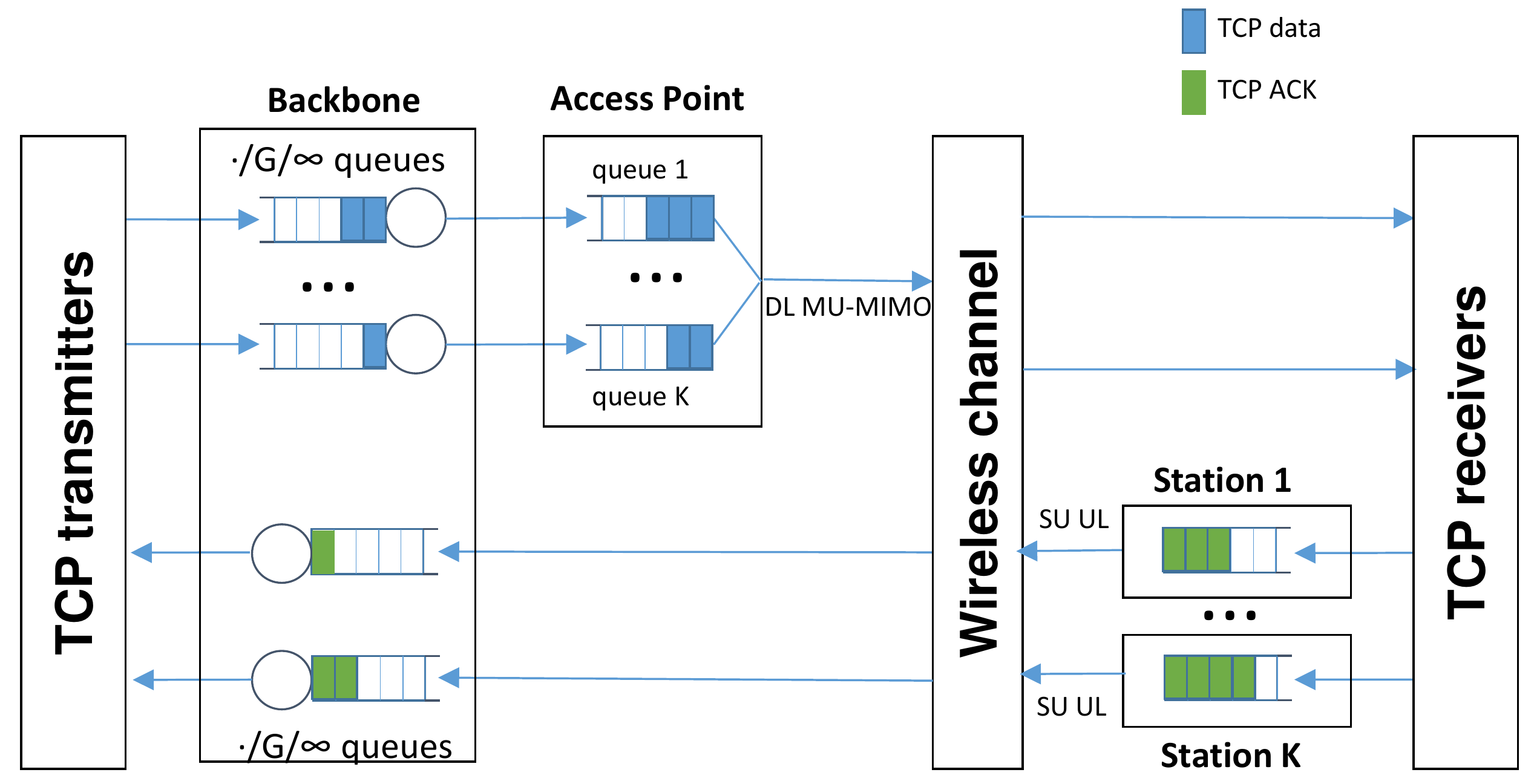}
    \caption{Representation of the system as a closed queueing-network.}
    \label{modeling}
    \vspace{-0.5cm}
\end{figure}

The diagram in Fig. \ref{modeling} illustrates the high-level
dynamics of the system represented as a closed queueing network.
The top part represents TCP data packets flowing downlink from servers to clients.
The bottom part represents TCP ACKs flowing uplink from clients to servers.
Recall that the AP maintains a separate queue for each of the attached stations.
Based on our assumptions, one-way delays in the backbone can be
modeled as infinite-server queues with deterministic service time.
However, for greater generality, we describe
propagation delays incurred by individual packets as i.i.d. random variables
with general distribution. Consequently, we model one-way delays 
in the backbone as $\cdot/G/\infty$ queues.

Consider, for now, the case in which TCP receivers send one ACK for each data packet
(we relax this assumption later). Then TCP receivers essentially \lq transform' data 
packets into ACKs, whereas TCP transmitters transform ACKs into data packets. 
Except for their different sizes, data packets and ACKs
can be both considered as individual customers circulating around the network.  

Since we assume that packets are never lost, each long-lived TCP flow reaches a steady-state condition with $W_{\max}$ outstanding packets in the network. As a consequence, the system indeed behaves as a multi-class closed queueing network
with a constant number of \lq customers', where it is not really important
to distinguish whether customers are data packets or TCP ACKs\footnote{System customers are 
classified only by the ID of the station acting as source/destination.}.

Unfortunately, batch arrival/services, and more importantly the
fact that wireless channel contention correlates the dynamics of  
MAC queues in the AP with those in the stations, do not allow us
to actually solve the queueing network model with traditional techniques
(such as product-form solution). Nevertheless, bottleneck analysis
can still be applied to compute the long-term throughput of the system.
Indeed, the aggregate system throughput $\Lambda$ essentially depends on how fast the customers
of the closed queueing network depicted in Fig. \ref{modeling} circulate
around the network. 

We can view the population of customers as a fluid
pushed forward by three main \lq pumps': the downlink pump (the AP), the uplink
pump (the stations), and the backbone. Correspondingly, each of the above
pump has a reservoir where the fluid gets accumulated waiting to be
drained. Note that the three pumps are in a specific circular order, each
pushing fluid into the reservoir of the next pump in the sequence.
Since the power of the three pumps is, in general, different,
we can expect the fluid to be found most of the time in the reservoir of 
the slowest of them, which will act as the system bottleneck.
The difficulty of the analysis lies in the fact that the power
of the pumps depends on the amount of fluid (belonging to each flow)
found on the associated reservoir. However, the maximum power of either the downlink
or the uplink pump quickly reaches a saturation level as soon as enough fluid 
is found in their buffers. Hence we can easily determine which one of them is the strongest 
under the assumption that a large enough amount of fluid (for each flow) is present
in their reservoirs (by comparing the saturation throughput in downlink/uplink).
This will lead us to make a first distinction
between two fundamental regimes: the {\em downlink bottleneck} case (when the maximum
power of the downlink pump is smaller than the maximum power of the uplink pump)
and the {\em uplink bottleneck} case (viceversa).

Note, however, that we will also consider cases in which this distinction
is not possible, because the total amount of fluid in the system is not
large enough to steadily operate at the saturation level neither in downlink nor in uplink.
In these cases, the system does not have a well-defined bottleneck.
  
The backbone pump is somehow different because its capacity does not saturate to any value
(the data rate on the backbone is supposed to be infinite). Indeed, the impact of the
backbone is just to delay the fluid in transit from the uplink pump to the downlink pump.
Nevertheless, there are cases (large propagation delays) in which almost all fluid is found in 
the reservoir associated with the backbone (\textit{i.e.}, packets flying in the backbone), 
rather than in the other buffers of the system.  
Therefore, by increasing the propagation delays, we eventually reach a third regime in which 
the backbone becomes the main system bottleneck, despite the fact that its capacity is infinite,
because of the limitation in the total amount of fluid in the system. 

Consider, initially, the case in which the number of packets
flying in the backbone reaches its maximum value. This case always 
occurs when $D$ is very small (possibly zero), or when $W_{\max}$ is large enough
that TCP flows completely \lq fill the pipe'.
Then a simple saturation throughput analysis, to be described next,  
allows us to understand where the rest of customers are primarily to be found
(\textit{i.e.}, either in the AP or in the stations).
          
\subsection{Saturation Throughput Analysis}\label{subsec:satur}        
Suppose we start from a condition in which the MAC queues of the AP, and the MAC queue
of each station, have a large backlog. The AP moves packets down into the stations,
while stations push up packets back into the AP (through the backbone). Who wins?

The key observation here is that contention for the wireless 
channel is fair among all nodes trying to transmit on it. Therefore, on average,
for one downlink transmission performed by the AP, we will have $K$ uplink
transmissions performed by the set of all stations.
Now, under the assumption that the AP employs multi-user MIMO (if $\NAP > 1$), whereas
stations employ single-user MIMO, the AP will push down on average
$$ \SDOWN = \BAP \cdot \min\{\NAP,K \cdot \NSTA\}$$ 
in each cycle of $K+1$ transmissions.
Indeed, the number of concurrent streams is given by the minimum between the number of 
antennas on the transmitting and receiving sides, and we can assume that
the maximum allowed number of packets (equal to $\BAP$) is transmitted on each stream.
During the same cycle of $K+1$ transmissions, the stations
will send up on average 
$$ \SUP = K \cdot \BSTA \cdot \min\{\NAP,\NSTA\} \cdot \TF$$ 
{\em effective} TCP ACKs. Indeed, each station will have (on average)
one opportunity to transmit $\BSTA$ packets using single-user MIMO, and we 
have accounted for the fact that TCP receivers might thin the
feedback traffic to improve performance \cite{miorandi2006queueing}, by transmitting
only one out of $\TF$ (Thinning Factor) ACKs. 
For example, the standard delayed ACK option of TCP \cite{braden1989rfc} corresponds to $\TF = 2$.
For later purposes, let $\SSSTA = \BSTA \cdot \min\{\NAP,\NSTA\} \cdot \TF$
be the maximum number of (effective) TCP ACKs sent by a station in one access,
so that $\SUP = K \SSSTA$. 

If $\SDOWN > \SUP$, the AP will eventually be able to move its backlog
into the stations, maintaining its queues almost empty from that time on.
If $\SDOWN < \SUP$, the stations will instead be able to drain their backlog, 
and most of the packets will be found in the AP.
If $\SDOWN = \SUP$, the AP and the set of all stations will maintain on average 
an equal backlog.

We emphasize that existing analytical models of IEEE 802.11
have focused only on the case $\SDOWN \leq \SUP$. 
This can be explained by the fact that, prior to the introduction of 
multi-user technique, it was reasonable to assume $\BSTA = \BAP$ (and in many models $\BSTA = \BAP = 1$),
and \mbox{$\NAP \leq K \cdot \NSTA$}. 
Note that earlier versions of 802.11 (without MIMO) correspond to $\NAP = \NSTA = 1$.
In all cases above, the AP becomes the performance bottleneck under 
closed-loop (\textit{e.g.}, TCP) traffic.

Multi-user MIMO has changed the picture by making the AP much more
powerful than the typical station. Not only can the AP be equipped
with many more antennas than its attached stations (which by itself would not
be enough to move the bottleneck to the uplink), but more importantly,
the AP must employ significant frame aggregation ($\BAP \gg 1$) to amortize
the overhead necessary to set up multi-user transmissions.
As a consequence, the performance bottleneck can shift
to the uplink, which is one novel scenario analysed in our work. 

\subsection{Fundamental Regimes}\label{subsec:regimes}
When there are enough packets flowing in the system 
to \lq fill the backbone pipe', \textit{i.e.}, when the propagation delay $D$
is small enough and, jointly, the average window size of TCP 
transmitters is not too small,\footnote{So far we have assumed 
for simplicity a loss-free network bringing TCP sources to steadily operate at the 
maximum window size. However,  analogous considerations can be done when 
the congestion window size of each flow oscillates around 
some (large) value due to a (small) packet loss probability.}
previous discussion leads us to distinguish the following three fundamental regimes:

\begin{itemize}
  \item {\bf {\em  downlink bottleneck} regime}. This regime occurs
  when both $\SDOWN \leq \SUP$ and $K F_s W_{\max} \gg \SDOWN$.
  Under the above conditions, the AP can be assumed to operate in
  saturation conditions, \textit{i.e.}, to be always fully backlogged. This is actually
  a desirable property to achieve the capacity gain of DL MU-MIMO.
  \item {\bf {\em  uplink bottleneck} regime}. This regime occurs
  when both $\SDOWN > \SUP$ and $F_s W_{\max} \gg \SSSTA$.
  Under the above conditions, each station can be assumed to operate in
  saturation conditions, \textit{i.e.}, to be always fully backlogged.  
  \item {\bf {\em  full aggregation} regime}. This regime occurs
  when both $\SDOWN \geq K F_s W_{\max}$ and $\SSSTA \geq F_s W_{\max}$.
  Under the above conditions both the AP and the stations perform a large enough 
  packet aggregation to completely empty their buffers at each channel access.
  This regime is different from the others because no node transmitting on the channel 
  operates in saturation conditions. 
\end{itemize}
  
Note that the {\em  full aggregation} regime is a limiting case of the downlink (uplink)
bottleneck regime as we increase the aggregation level performed by the AP (the stations).   

As we increase the backbone delay $D$, or reduce the average window
size of TCP flows, the system performance will eventually
be limited by the wired network delay, rather than the wireless channel dynamics.
In our analysis, we will also (partially) explore the impact of the backbone delay 
$D$ in the regimes described above. In Sec. \ref{sec:losses},
we will also explore by simulation what happens when TCP flows experience
non-zero loss probability, due to buffer overflows or other reasons. 

{\bf Remark.} One crucial observation that we can already make at this point is the following: 
the size of data packets, and that of TCP ACKs, plays no role in determining
the regime in which the system operates, as one can check by inspecting the conditions
listed above for each regime. Specifically, the fact 
that TCP ACKs are much smaller in size than a TCP data packet 
does not modify in any way the system bottleneck. 
This fact is in sharp contrast to a common misconception,
according to which the impact of uplink traffic is 
negligible because TCP ACKs are ``small'' (in size).
As we will see, instead, the uplink feedback process can determine
the overall system performance, although the large majority of traffic volume
(in terms of bytes) flows downstream.

\subsection{Reference System and Basic Throughput Bounds}\label{subsec:reference}
To validate our analysis we will consider a reference system  
closely following the network topology illustrated in Fig. \ref{cltop} and the 
physical-layer parameters described in Sec. \ref{subsec:ac}.
Specifically, we will always assume an AP equipped with 4 antennas 
(equal to the maximum number of concurrent streams considered in 802.11ac), 
operating at 54 Mb/s physical data rate per stream.

Stations are instead assumed to have a single antenna\footnote{Due to size and cost constraints on 
mobile hand held devices, STAs tend to have fewer antennas than the AP.},
thus performing single-user SIMO transmissions in the uplink.
Unless otherwise specified, we assume 4 stations in the network,
so that all of them can potentially be served concurrently by the AP.

We further assume that each station establishes a single long-lived
TCP flow with a server ($F_s = 1$). Unless otherwise specified, 
the maximum TCP congestion window size is $W_{\max} = 200$.
The TCP segment size is 1024 bytes, and we enable the 
delayed ACK option ($T_F = 2$). 

In the next section, we will compare analytical results
(for each of the regimes in Sec. \ref{subsec:regimes})
with detailed ns3 simulations obtained in our reference system. 
To put our throughput figures under the right perspective,
it is important to keep in mind the following simple upper bounds on $\Lambda$.

Given a physical data rate of 54 Mb/s, and 4 antennas, clearly we cannot exceed the 
trivial upper bound $\Lambda^{(1)} = 54 \cdot 4 = 216$ Mb/s, corresponding to the unrealistic
case of zero overhead everywhere.
Under the constraint of adopting the best 802.11ac-compliant MU-MIMO in the downlink, 
we obtain a better (tighter) bound as $\Lambda^{(2)} = K F_s W_{\max}/A(K,F_s W_{\max})$,
by assuming zero overhead in the uplink: after the AP sends down the aggregate of all system packets, 
all data is acknowledged in zero time by the TCP receivers. 
In our reference system with $K=4$, $F_s=1$, $W_{\max} = 200$, we obtain $\Lambda^{(2)} = 192.5$ Mb/s.
At last, assuming that all system packets, after been dumped by the AP, 
are sequentially acked by the stations (actually, one ACK every 2 packets, since $T_F = 2$), we obtain
$\Lambda^{(3)} = K F_s W_{\max}/[A(K,F_s W_{\max}) + \TUP(K F_s W_{\max}/2)] = 172.5$ Mb/s,
where $\TUP(K F_s W_{\max}/2)$ is the channel time to send 400 TCP ACKs, in our case. 

\section{Analysis}\label{sec:analysis}   
\subsection{Downlink Bottleneck Regime}\label{subsec:downbottle}   
\begin{figure}
    \centering
    \includegraphics[width=1.0\linewidth]{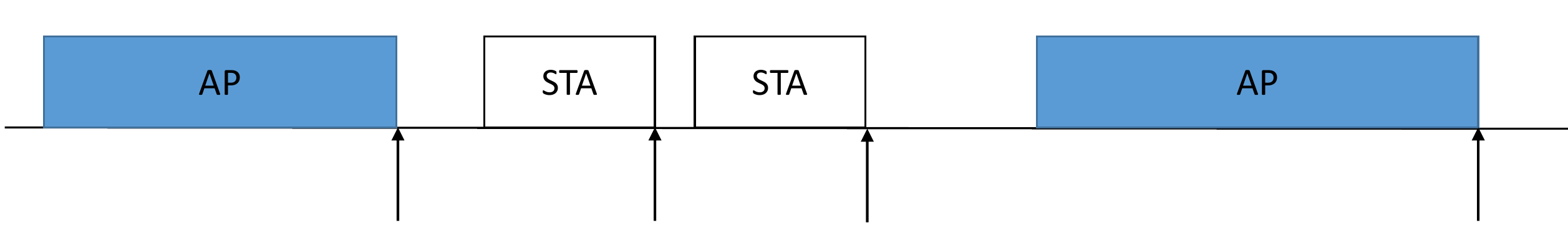}
    \vspace{-0.5cm}
    \caption{Embedded discrete-time Markov Chain to analyse the downlink bottleneck regime.}
    \vspace{-0.5cm}
    \label{embed}
\end{figure}

Recall that in this regime we assume the AP to be always fully backlogged.
We consider a discrete-time Markov Chain embedded at the time instants
at which the wireless channel becomes idle (\textit{i.e.}, at the end of a transmission) -- see Fig. \ref{embed}.
The state of this Markov Chain is the set of queue lengths of the stations
at the beginning of a cycle. 

Standard renewal theory allows us to write the aggregate throughput $\Lambda$ (in packets per seconds) as
\begin{equation}\label{eq:ren}
\Lambda = \frac{\textrm{average number of packets sent in a cycle}}{\textrm{average cycle duration (s)}} 
\end{equation}
where packets can be {\em either} TCP data packets or (effective) TCP ACKs. Indeed, flow conservation 
(closed-loop traffic) implies that throughput in terms of data packets must 
be equal to throughput in terms of (effective) ACKs.
  
Any cycle is divided into two parts: a contention phase
and a packet transmission phase. Let $\hat{K}$ be the random variable denoting the number of contending stations 
at the beginning of a cycle. 
To simplify the analysis, we assume that random backoffs are chosen according to an exponential
distribution of mean $1/\mu$, instead of a uniform distribution in $[0,W_0-1]$ (in number of slots of duration $\sigma$).
To match the first moment of the backoff distribution, we correspondingly set $\mu = 2/(W_0 \sigma)$.
This way we can exploit the memoryless property of the exponential distribution
and ignore the backoff time spent by a node in previous cycles. Note that this is 
a standard technique to simplify the analysis of 802.11, and it is known to introduce
negligible errors (see \cite{cali2000dynamic} and Fig. \ref{plot5}).     

It follows that the average duration of the contention phase, conditioned on having $\hat{K} = k$ contending stations ($k=0,1,\dots,K$), is\footnote{Recall that the (saturated) AP is always contending for channel access.} 
$\frac{1}{(k+1) \mu}$. If the AP wins the contention, which occurs with probability $\frac{1}{k+1}$, we have a downlink
transmission of a data bundle by the AP consisting of $\SDOWN$ TCP data packets, occupying the channel
for a duration \mbox{$\TDOWN = A(K,\BAP)$}.
Instead, with probability $\frac{k}{k+1}$ the contention is won by a station, that will 
occupy the channel for a duration $\TUP$.

An exact analysis of the system requires to track the queue lengths of the stations. 
However, following this approach would be an 
overkill, given that the system obeys flow conservation in the downlink and
uplink directions. Actually, the only advantage of performing the above exact analysis  
would be to perfectly characterize the duration of the contention phase at the beginning of a cycle, 
which has however negligible impact on the overall throughput. Therefore, we adopt the following simplifying
assumptions: i) a station always transmits $\min(\BAP,\SSSTA)$ packets when it gets access on the channel; ii) the number
$\hat{K}$ of contending stations, which is a random variable, is replaced by a constant
value $k^*$ obtained by flow conservation:
$$ \frac{1}{k^*+1} \SDOWN = \frac{k^*}{k^*+1} \min(\BAP,\SSSTA) $$
which provides\footnote{The value of $k^*$ computed in this way is, in general, not an
integer, but we do not have to worry about this.} $k^* = \frac{\SDOWN}{\min(\BAP,\SSSTA)}$.
These might appear to be rough approximations but, to say it again, 
they only impact the computation of the average contention time at the beginning of a cycle,
which has negligible impact on the throughput.     

The above considerations allows us to derive the throughput according to \equaref{ren}:
\begin{eqnarray}\label{eq:ren2}
\!\! \Lambda \!=\! \frac{\frac{1}{k^*+1} \SDOWN}{\frac{1}{(k^*+1)\mu} +\! \frac{1}{k^*+1} \TDOWN +\! \frac{k^*}{k^*+1}\TUP} \!=\!
\frac{\SDOWN}{1/\mu + \TDOWN + k^* \TUP} 
\end{eqnarray}

At last, we account for the fact that, as we increase the backbone two-way delay $D$,
we will enter at some point the regime in which the backbone becomes the performance 
bottleneck. To do so, we adopt a simple approach based on the assumption that the queues
of the AP are in one of two states: they are either empty, or they have sufficient backlog
to send $\SDOWN$ packets in one channel access.

Let 
$$\bar{C} = \frac{1}{\mu} + \TDOWN + k^* \TUP$$
be the average time to send $\SDOWN$ packets
downlink (the denominator of \equaref{ren2}). 
Suppose that we start from a condition in which all $K F_s W_{\max}$ packets in the system are stored in the AP.
If the backbone delay is too large, the queues of the AP will not get refilled
in time to maintain it constantly backlogged. In particular, the AP will run out of packets
if $D > {\bar{C}} \frac{K F_s W_{\max}}{\SDOWN}$, \textit{i.e.},
if the backbone delay is larger than the (average) time to completely drain the AP queues.
Moreover, to be sure that the AP sends $\SDOWN$ packets in each channel access,
we assume that at least $\SDOWN$ packets have to be stored in its buffers:
if there are not enough packets in the system to fill the pipe and guarantee
enough backlog in the AP, we simply assume that the AP remains completely idle for some time.
Specifically, we consider the AP to be fully backlogged for a fraction 
of time given by $\frac{K F_s W_{\max}}{\left(1 + \frac{D}{\bar{C}} \right) \SDOWN }$, if 
this fraction is smaller than one.

The final formula for the throughput, valid whenever \mbox{$\SDOWN \leq \SUP$}, $K F_s W_{\max} \gg \SDOWN$, becomes:
\begin{equation}\label{eq:ren3}
\Lambda = \frac{\SDOWN}{1/\mu + \TDOWN + k^* \TUP} \cdot 
\min\left(1,\frac{K F_s W_{\max}}{\left(1 + \frac{D}{\bar{C}} \right) \SDOWN }\right)
\end{equation}

\begin{figure}
    \centering
    \includegraphics[width=0.9\linewidth]{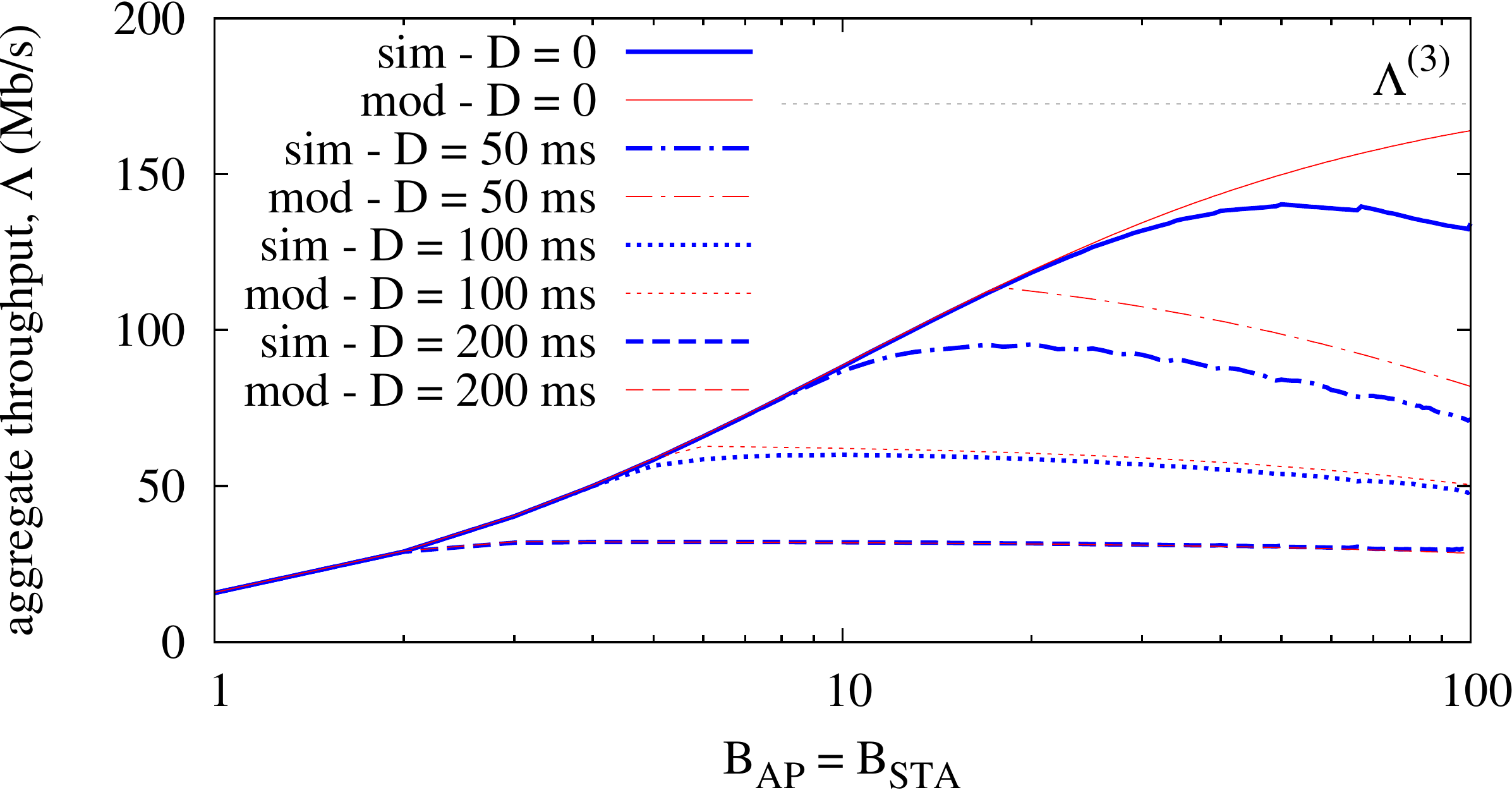}   
    \caption{Throughput comparison (model vs simulation) in the reference system, with $\BAP = \BSTA$.}
    \label{plot1}
    \vspace{-0.2cm}
\end{figure}

Fig. \ref{plot1} compares simulation results (blue, thick lines) 
against analytical prediction \equaref{ren3} (red, thin lines) 
in our reference system, as we vary the aggregation level
employed by all nodes, for different values of backbone delay $D$. We do not show confidence intervals for simulation results 
since they are too narrow (at 95\% level) to be visible. 

Note that, with $\BAP = \BSTA$, we are 
in the {\em downlink bottleneck} regime. As expected, the model is less
accurate when $D$ comes into play, or (for $D=0$) when the assumption $K F_s W_{\max} \gg \SDOWN$ (which here reads
$800 \gg 4 \BAP$) does not hold. Interestingly, there is an optimal
aggregation level (strongly related to $F_s W_{\max}$) which maximizes
throughput. This can be explained by the fact that, as we push 
$\BAP$ close to $F_s W_{\max}$, we obtain diminishing
returns from amortizing the overhead of setting up MU-MIMO,
while increasing the probability that the AP completely
empties one of its MAC queues, resulting is lower multiplexing gain.
Unfortunately, such kind of optimization of the aggregation level requires
knowledge of $F_s W_{\max}$, and can hardly be done in practice.

We conclude that in the {\em downlink bottleneck} regime
the fundamental reason that can prevent achieving 
the theoretical performance gains of MU-MIMO is the 
limited amount of data packets in the queues of the AP, 
necessary to amortize the overhead of MU-MIMO transmissions.
Such limitation, in the absence of packet losses at the transport layer\footnote{Note that 
losses on the wireless channel due to
poor channel quality are automatically recovered by the MAC protocol.},
is essentially related to the maximum TCP congestion window size
and the number of concurrent flows for each station.      
In Sec. \ref{sec:losses} we will explore by simulation
scenarios in which TCP flows experience losses, for example due
to congestion. Of course in this case another 
performance factor coming into play is the packet loss probability,
which can make the window size of TCP flows oscillate
around too small values, causing inefficient frame 
aggregation by the AP.

\subsection{Uplink Bottleneck Regime}\label{subsec:upbottle}   
Recall that in this case we assume the stations to be always fully backlogged.
In this work, we will analyze this regime under two additional assumptions\footnote{Relaxing
either of these two assumptions is analytically challenging, and we leave it to future work.}:
i) the backbone delay $D=0$; ii) the AP completely empties its queues
when it gets access on the channel.
Assumption i) can represent the network scenario in which servers are located within 
the same LAN of the stations. Assumption ii) holds in the uplink bottleneck regime 
when $\NAP \geq K$. 

The main difficulty of the analysis lies in the fact that now the AP, differently from the downlink bottleneck regime,
is not fully backlogged, thus it typically aggregates only a limited number 
of packets, which can severely degrade the maximum theoretical throughput 
computed under saturation conditions. 

Recall that the channel holding time $A(h,b)$ of the AP depends 
on both the number of non-empty queues $h$ in the AP (hereinafter called user diversity) 
and their maximum backlog $b$. 
Let $H$ be the random variable denoting the user diversity, 
and $B$ the maximum queue length among the AP queues. Let $P(h,b) = \P[H = h, B = b]$  be the 
joint discrete distribution of the above two variables at the time instant
at which the AP gets access on the channel. Note that since the AP has contended
for channel access, we have $h \in \{1,\ldots,K\}$, $b \in \{1,\ldots,F_s W_{\max}\}$. 

\begin{figure}
    \centering
    \includegraphics[width=1\linewidth]{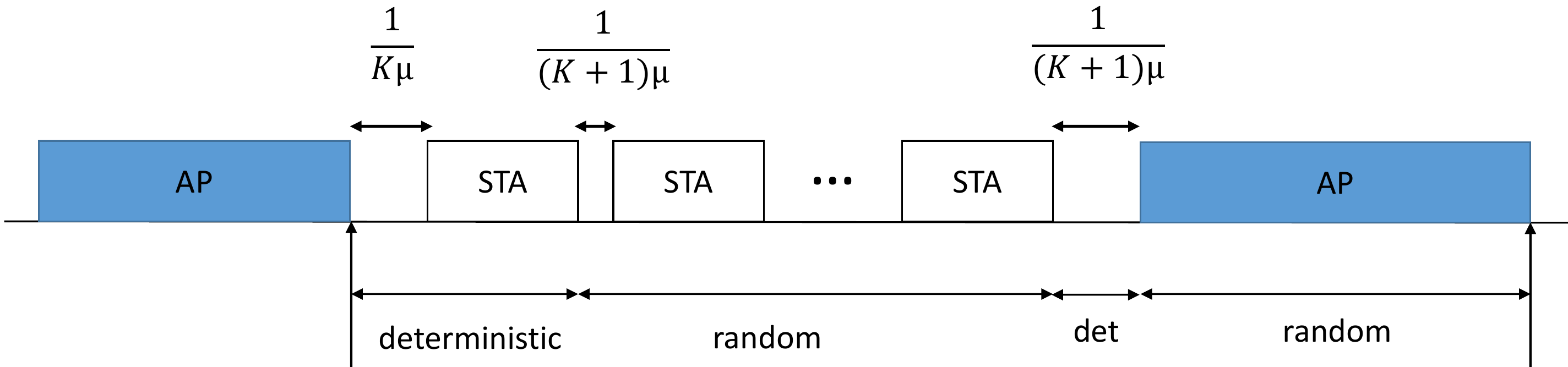}
    \vspace{-0.5cm}
    \caption{Cycle analysis for the uplink bottleneck regime with $D =0$.}
    \vspace{-0.5cm}
    \label{fig:embed2}
\end{figure}

Suppose, for now, that $P(h,b)$ is known. In Appendix \ref{app:joint}
we show how $P(h,b)$ can be analytically computed.  
The aggregate system throughput can then be derived by a simple cycle
analysis, illustrated in Fig. \ref{fig:embed2}.

This time we consider cycles delimited by time instants at which the AP
releases the channel. Since the AP flushes out all its backlog, any cycle
starts deterministically with a contention phase among $K$ backlogged
stations, of average duration $\frac{1}{K \mu}$, followed by the 
transmission of the winning station, of duration $\TUP$.
Now, since the backbone delay is zero, the ACKs sent up by this station 
will immediately create new data packet(s) in the AP, which will start
contending as well. Before the AP will eventually win the contention, 
a random number of stations will be able to transmit. Actually, on average
each station will be able to put one transmission on the channel 
before the AP wins. This result derives from the  assumption that
backoffs are exponential: by conditioning on the value $x$ extracted by the AP,
the number of transmissions made by a station is Poisson distributed    
of parameter $\mu x$. Deconditioning w.r.t. $x$, we obtain that on average
each station makes one transmission before the AP, of duration $\TUP$, preceded by
a contention period of average duration $\frac{1}{(K+1)\mu}$.

The cycle ends deterministically with another contention period
of average duration $\frac{1}{(K+1)\mu}$ (the one won by the AP) followed
by the channel holding time by the AP, whose average duration is $\sum_{h,b} P(h,b) A(h,b)$.
To compute the average number of packets sent in a cycle, it is convenient 
to express this number in ACKs, rather then data packets, since we
have already shown that on average we see $K$ transmissions by the set of all stations,
plus the deterministic transmission at the beginning of the cycle.

Putting everything together, the usual renewal formula \equaref{ren} provides
the throughput for this scenario:
\begin{equation}\label{eq:ren4}
\Lambda = \frac{(K+1) \SSSTA}{\frac{1}{K \mu} + (K+1)\left(\frac{1}{(K+1)\mu} + \TUP\right)  + \sum_{h,b} P(h,b) A(h,b)}
\end{equation}

To get insights into the resulting system performance, we 
compute here the marginal user diversity distribution $P(h) = \P[H = h]$
through an alternative method that does not require us to first derive 
the joint distribution  $P(h,b)$. This computation leads indeed
to a rather simple and instructive result that we will discuss later on. 

We first isolate the impact of the initial deterministic ACK, computing
the user diversity distribution $\hat{P}(h)$ produced by stations' transmissions
following the first one.
By conditioning on the backoff value $x$ extracted by the AP, we can write:
$$\hat{P}(h) = \int_{0}^{\infty} \binom{K}{h} \left(1-e^{-\mu x}\right)^h e^{-\mu x (K-h)} \mu e^{-\mu x} \diff x$$
Integrating by parts, we get
\begin{eqnarray*}
\!\!\hat{P}(h)\! = \!\! \!
\int_0^\infty \!\!\!\binom{K}{h} \frac{h}{K-h+1} (1-e^{-\mu x})^{h-1} e^{-(K-(h-1))\mu x}   \mu e^{-\mu x} \diff x 
\end{eqnarray*}

Noticing now that $\binom{K}{h} \frac{h}{K-h+1} = \binom{K}{h-1}$,
the above expression means that $\hat{P}(h) = \hat{P}(h-1)$. 
In other words, the distribution of $\hat{P}(h)$ is uniform
over $h = 0,1,\ldots,K$, hence $\hat{P}(h) = \frac{1}{K+1}$.

To compute the distribution $P(h)$, that includes the contribution of the first ACK,
we observe that $H = h$ occurs in two possible ways: i) either the
first ACK belongs to one of the $h$ queues which are 
already non-empty for effect of subsequent transmissions of the stations, 
with probability $\frac{h}{K}$, or it increases by one the number $h-1$ of non-empty 
queues produced by the other transmissions, with probability $\frac{K-(h-1)}{K}$.
We obtain:
$$ P(h) = \hat{P}(h) \frac{h}{K} + \hat{P}(h-1) \frac{K-(h-1)}{K} = \frac{1}{K}$$  
meaning that $P(h)$ is also uniform over the set of possible values $h = 1,2,\ldots,K$.

This result has striking consequences on the efficiency of MU-MIMO,
which strongly relies, in addition to the availability of large per-station backlog, on 
large user diversity (\textit{i.e.} multiplexing gain). Note that the optimal operating point of MU-MIMO is 
full diversity ($h \geq \NAP$), which naturally occurs in the downlink 
bottleneck regime.

In the uplink bottleneck regime, instead, wireless channel contention can 
result into a random user-diversity far from the optimal one.
Under the scenario considered in this section ($K \leq \NAP$), the average
user diversity is $(K+1)/2 \leq K$, which results roughly into a throughput 
reduction by factor $(K+1)/(2K)$, which in our reference system (with $K = 4$) 
equals 5/8 = 0.625. 
Note that the penalty introduced by such sub-optimal user-diversity
is intrinsic to the random access nature of the channel, and thus
unavoidable (in the uplink bottleneck regime).
Instead, the penalty due to small per-station backoff
can be eliminated by letting the stations perform packet aggregation 
in a way similar to what the AP does.

\begin{figure}
    \centering
    \includegraphics[width=0.9\linewidth]{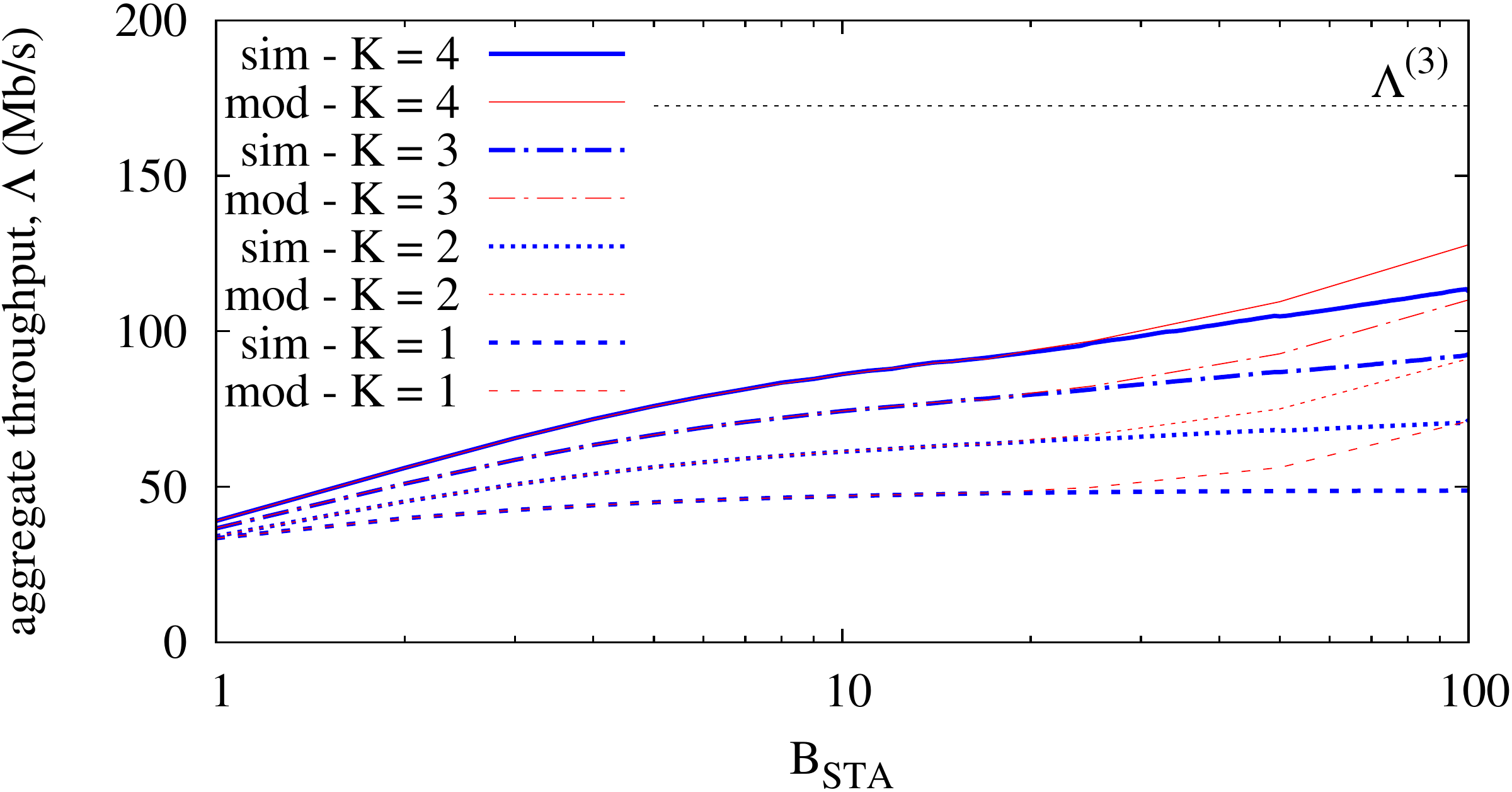}
    \vspace{-0.2cm}
    \caption{Throughput comparison (model vs simulation) in the reference system, with $\BAP = \infty$, $D = 0$, 
    as function of $\BSTA$, for different number of stations.}
    \label{plot3}
    \vspace{-0.2cm}
\end{figure}
                
Throughput formula \equaref{ren4} can be refined for the case in which
the number $\SSSTA$ of effective ACKs sent by a station in one channel access is so large that 
assumption $F_s W_{\max} \gg \SSSTA$ no longer holds (but notice that we still assume 
$\SDOWN > \SUP$, so that the system operates in the uplink bottleneck regime).
In particular, we can refine the analysis under the assumption that $F_s W_{\max} = \bar{m} \SSSTA$, where $\bar{m} \geq 1$
is an integer. In words, we assume for simplicity that the 
total number of packets belonging to a station is a multiple
of $\SSSTA$. In this case, a station cannot clearly transmits more than $\bar{m}$ times
in a cycle.
To account for this fact, we need to derive the distribution 
$P(m) = \P[M = m]$ ($0 \leq m \leq \bar{m}$) of the r.v. $M$ denoting the number of transmissions performed by a station
while contending with the AP, before the AP wins the contention.

By conditioning on the backoff value $x$ extracted by the AP, we can write: 
$$P(m) = \int_{0}^{\infty} \frac{(\mu x)^m}{m!} e^{-\mu x} \mu e^{-\mu x} \diff x = \frac{1}{2^{m+1}}$$
As expected, if the station could make an arbitrarily large
number of transmissions ($\bar{m} = \infty$), we would get the average value:
$$ \E[M] = \sum_{m=0}^\infty  \frac{m}{2^{m+1}} = 1 $$
When $M$ cannot exceed the maximum value $\bar{m}$ we get instead:
$$ \E[M] = \sum_{m=0}^{\bar{m}-1} \frac{m}{2^{m+1}} + \sum_{m=\bar{m}}^{\infty} \frac{\bar{m}}{2^{m+1}} = 
\frac{2^{\bar{m}}-1}{2^{\bar{m}}}$$               
which provides the improved throughput formula:
\begin{equation}\label{eq:ren4bis}
\Lambda = \frac{\left(1+\frac{2^{\bar{m}}-1}{2^{\bar{m}}}\right) \SSSTA}{\frac{1}{K \mu} + (K+1)\left(\frac{1}{(K+1)\mu} + \TUP\right)  + \sum_{h,b} P(h,b) \bar{A}(h,b)}
\end{equation}
where $\bar{A}(h,b) = A(h,\min\{b,F_s W_{\max}\})$ takes again into account the fact that
the number of packets sent by the AP, belonging to the same station, cannot exceed
$F_s W_{\max}$. 
                
Fig. \ref{plot3} compares the analytical prediction \equaref{ren4bis} against
simulation in our reference system (with $D = 0$), as we vary the aggregation level $\BSTA$
employed by stations. Here we assume unlimited aggregation by the AP (actually, $\BAP \geq F_s W_{\max}$), 
bringing to system to operate in the {\em uplink bottleneck} regime. As expected, the model is less
accurate when the assumption $F_s W_{\max} \gg \SSSTA$ (which here reads
$200 \gg 2 \BSTA$) does not hold. 

Focusing on the basic case $K = 4$, we observe severe 
throughput loss when $\BSTA$ is small, due to poor frame aggregation by 
the AP.\footnote{without the 
delayed ACK option, with $\BSTA = 1$ we would get $\Lambda = 23.9$ Mb/s, smaller than that
of a DL SU system! (see Sec. \ref{sec:comparison}).}
But even with unlimited aggregation by the stations (actually, the maximum level of aggregation 
by stations is already achieved with $\BSTA = 100$) 
the throughput is only about 113 Mb/s\footnote{This value requires
exactly $D=0$. Under more realistic conditions of small but not null delay,
we would get $\Lambda = 86$ Mb/s, see Sec. \ref{subsec:fullaggr}.}, which is $0.65 \cdot \Lambda^{(3)}$ (see
Sec. \ref{subsec:reference}), close to our analytical prediction 
of a throughput reduction by factor $\frac{K+1}{2K} = 0.625$.

\vspace{4mm}
{\bf Impact of backoff distribution.} 
Our analysis is based on the simplifying assumption that the random backoff chosen by 
each station contending for channel access is taken from an exponential distribution,
rather than the actual uniform (discrete) distribution dictated by the 802.11 standard.
Indeed, the uniform distribution would not allows us to make the simple cycle
analysis illustrated in Fig. \ref{fig:embed2}, since cycles would no longer be
independent of each others. On the other hand, the error introduced
by such simplifying assumption is expected to be of secondary importance,
as already observed in many 802.11 models, under both saturated and unsaturated conditions 
\cite{cali2000dynamic,garetto2005performance}. For this reason, and to better highlight the impact of 
the other simplifying assumptions introduced in our analysis, we have preferred
to adopt an exponential backoff distribution also in our simulations.

\begin{figure}
    \centering
    \includegraphics[width=0.9\linewidth]{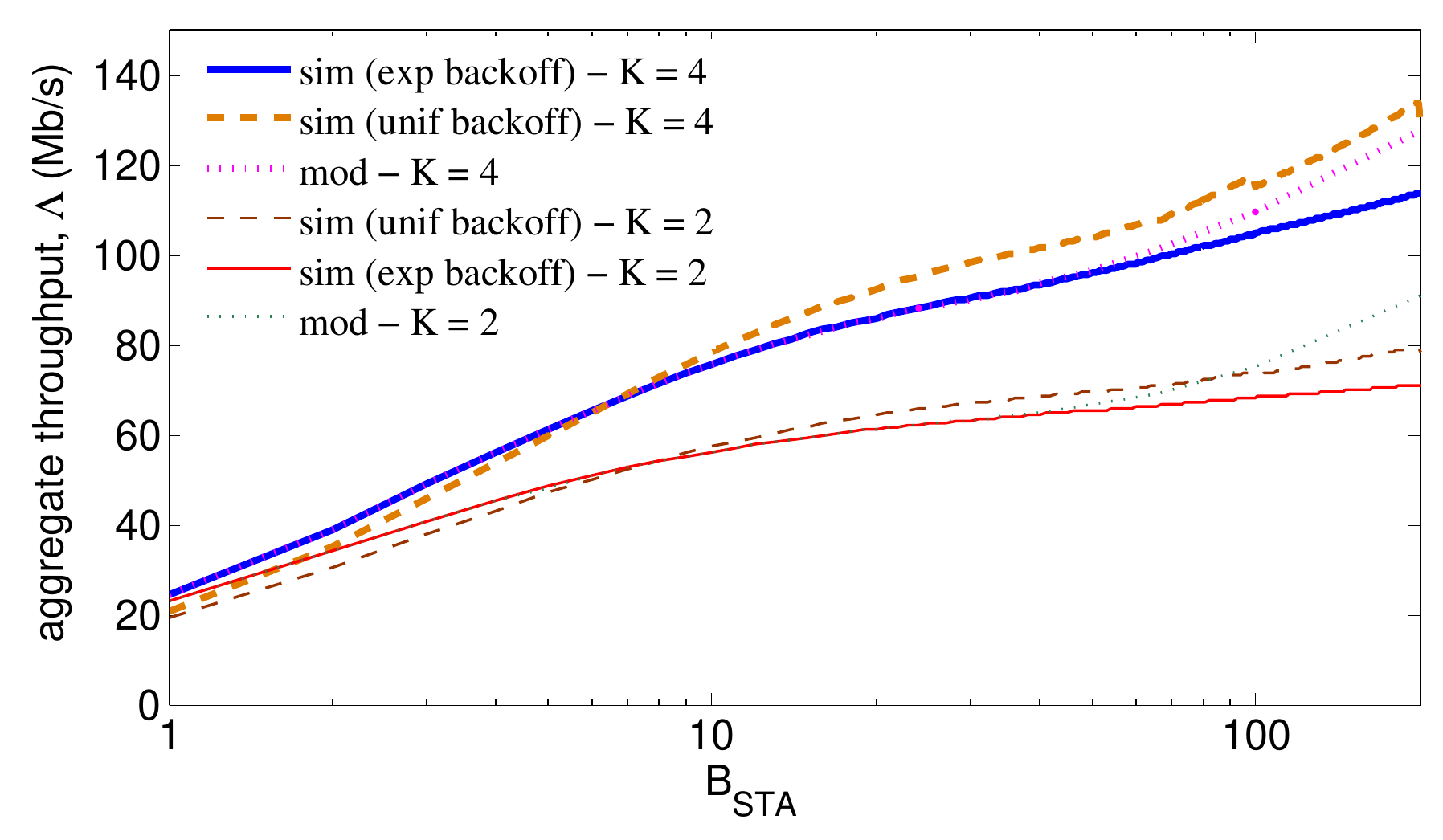}
    \vspace{-0.2cm}
    \caption{Throughput comparison (model vs simulation) in the reference system, with $\BAP = \infty$, $D = 0$, 
    as function of $\BSTA$, for $T_F = 1$, $K = 2$ or $K = 4$.}
    \label{plot5}
    \vspace{-0.2cm}
\end{figure}
            
For completeness, we report here a comparison of results obtained 
under the uniform vs exponential backoff distribution, for the scenario 
already considered in Fig. \ref{plot3}. However, to avoid repetition of results, this time 
we disable the delayed ACK TCP option, \textit{i.e.}, we consider $T_F = 1$.
Results shown in Fig. \ref{plot5} confirm the second-order effect
due to the shape of the backoff distribution. 
They also show that, by disabling the delayed ACK option, we obtain
worse performance, especially for small levels of aggregation performed by stations
(for $\BSTA = 1$, the throughput is only slightly larger than 20 Mb/s, whereas
it was around 40 Mb/s with $T_F = 2$, see Fig. \ref{plot3}). 

\subsection{The Case of Non-Zero Backbone Delay: Pump-And-Drain}\label{subsec:supercycle}  
\begin{figure}
    \centering
    \includegraphics[width=1\linewidth]{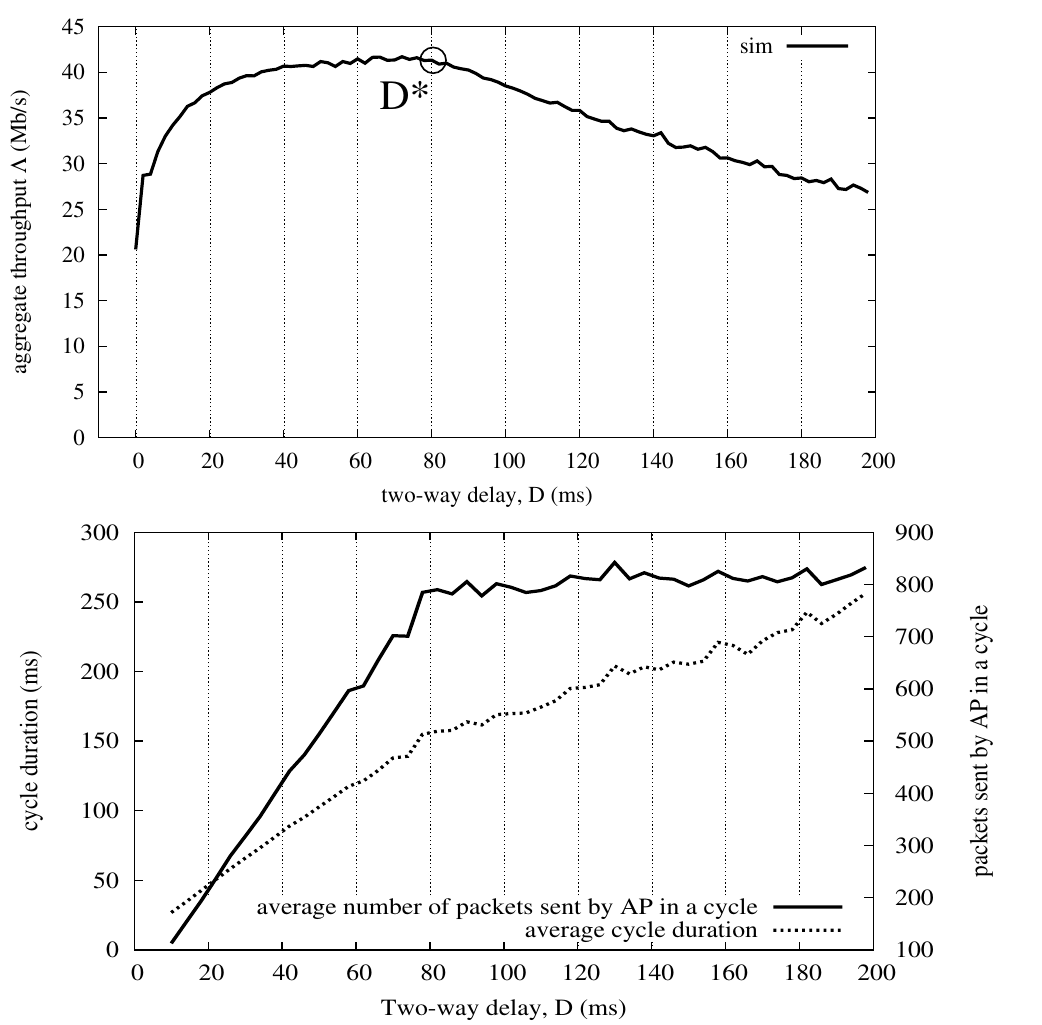}
    \caption{Simulation results obtained with $\NAP = K = 4$, $\NSTA = 1$, $\BAP = 256$, $\BSTA = 1$, $F_s = \TF = 1$, $W_{\max} = 200$.
    Aggregate throughput (top plot), and corresponding super-cycle statistics (bottom plot).}
    \label{fig:cycle200}
\end{figure}

Here we extend the analysis of previous section to the case of non-zero backbone
delay, while maintaining all of the other assumptions.
Quite surprisingly, we observed in simulation that backbone delay has a beneficial impact on the
aggregate throughput, which exhibits an intriguing non-monotonous behavior: it first increases
with the delay (though with diminishing returns, \textit{i.e.}, through a concave function),
up to a maximum value reached for $D = D^*$, after which it decreases 
to zero following a convex function, resulting into a \lq fin' shape -- see Fig. \ref{fig:cycle200} (top plot).

\begin{figure}
    \centering
    \includegraphics[width=1\linewidth]{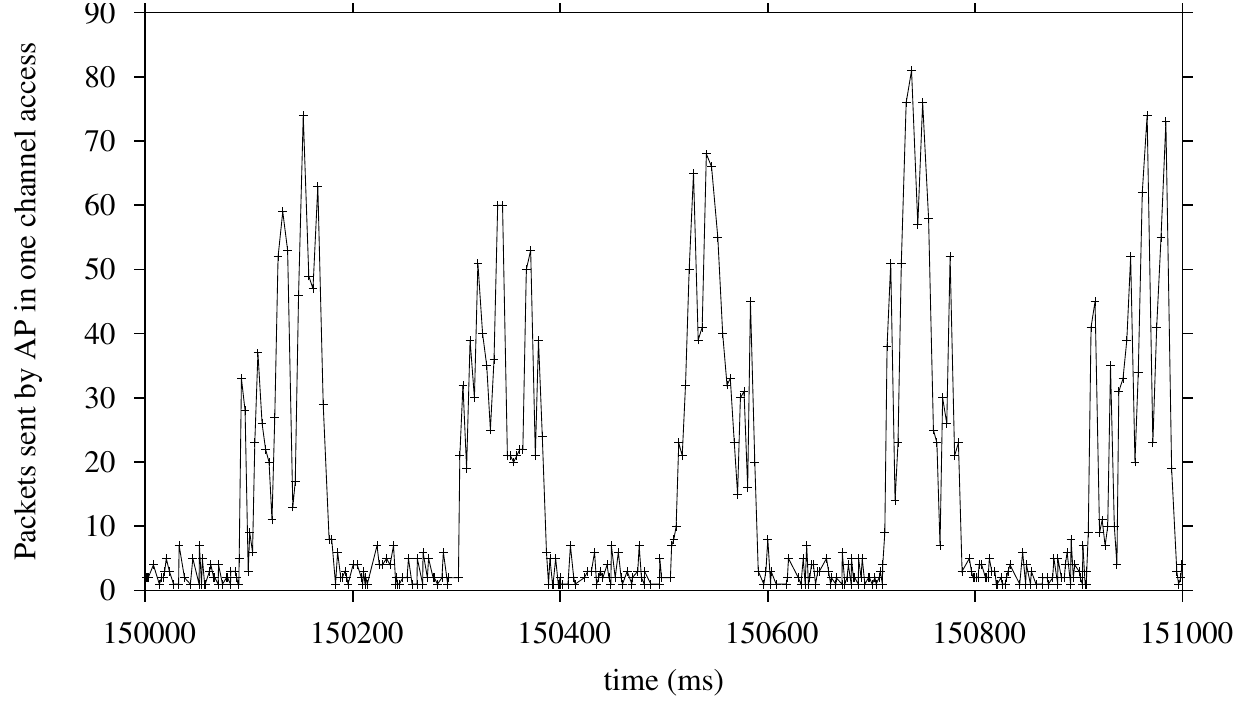}
    \caption{Simulation trace of the number of packets sent by the AP in one channel access. Parameters as in Fig. \ref{fig:cycle200}.}
    \label{fig:example200}
\end{figure}

Puzzled by this fact, we investigated simulation traces discovering the origin of this behavior.
We considered a scenario with $K = 4$ stations that do not perform any packet aggregation ($\BSTA = 1$),
each of them destination of a TCP flow with $W_{\max} = 200$.
The AP, equipped with $\NAP = 4$ antennas, can aggregate enough packets ($\BAP = 256$) for each station 
that we are guaranteed that all of its queues get emptied after each channel access (note that $\BAP > W_{\max}$).  

Fig. \ref{fig:example200} shows a simulation trace for the total number of packets sent by the AP 
in each channel access, for the particular delay value $D = 120$ ms.
It can be clearly recognized a cycle-stationary behavior
of duration slightly smaller than 200 ms. It the following, we will call \lq super-cycle'
each cycle of this roughly-periodic behavior. A \lq super-cycle'
is internally structured in two phases: a first phase in which the AP sends just a few packets ($<10$)
in each channel access (later called {\em pump} phase), and a second phase in which
the AP sends many more packets in each channel access (later called {\em drain} phase).
The roughly periodic behavior illustrated in Fig. \ref{fig:example200} shows up 
for any value of delay (except for very small delays, say smaller than 5 ms), though
with a different period of duration roughly proportional to $D$ itself (traces not shown here).     

We performed a post-processing of traces like that in Fig. \ref{fig:example200}, obtaining the super-cycle 
first-order statistics shown in the bottom plot of Fig. \ref{fig:cycle200} (note that the $x$ axes of the top and bottom plots
of Fig. \ref{fig:cycle200} are aligned). In particular, we obtained the (average) total number of packets
sent by the AP in each super-cycle (solid curve), and the (average) super-cycle duration (dotted curve).
We clearly see that the point $D^* \approx 80$ ms at which the aggregate throughput reaches its maximum
value corresponds to the point at which the total number of packets sent by the AP
in a super-cycle saturates to the total number of packets circulating in the system (4 TCP flows
with $W_{\max} = 200$ $\Rightarrow 800$ packets).

To convey the physical interpretation of this behavior, suppose to start from an ideal 
condition in which all packets circulating in the system are all initially stored in the stations.
The stations will then contend only among themselves, \lq pumping' their packets into the backbone
through a quite regular stream. If $D > D^*$, stations will actually exhaust at some point all of their 
backlog, and at this point all packets of the system will be \lq flying' in the backbone.
When the front of the stream generated by the stations arrives at the AP, the AP will start to contend,
sending initially a small number of packets in the first channel access (possibly just one). 
But than a multiplicative effect occurs: while the AP transmits on the channel, the stream arriving 
from the backbone (note that this stream is not interrupted by the fact that the channel is busy) refills the AP with
many more packets to send in the next channel access (note that the intensity of the backbone stream 
is large enough to produce the amplification effect, since ACKs are small and the backbone capacity is large -- a phenomenon
sometimes referred to in the literature as ACK compression), generating a sequence of elementary 
cycles with increasing duration and packet aggregation. As a consequence, the AP will be able to 
quickly drain the backlog arriving from the backbone down again into the stations, marking the end
of the super-cycle (and the beginning of a new super-cycle starting with a new pump phase). Interestingly, 
this pump-and-drain behavior occurs also when we do not start
from the condition in which all system packets are in the stations. Moreover, the same behavior
occurs when $D < D^*$. 

We observed that the pump-and-drain phenomenon is strongly non-linear and highly sensitive
to system parameters. For example, by reducing the DIFS parameter of 802.11 to the slot size (9 $\mu$s),
we observed a fin shape surprisingly sharper than the one shown in Fig. \ref{fig:cycle200} (top plot),
going up to 95 Mb/s! Moreover, the effect is so sensitive to system parameters to depend
also on the backoff process. To illustrate this, we compare in Fig. \ref{fig:cycle200} (top plot)
the throughput resulting from the (real) uniform backoff, with that resulting from
the (more model-friendly) exponential backoff (obtained hacking the ns simulation).  

\subsection{Full Aggregation Regime}\label{subsec:fullaggr}   
Recall that in this case both the AP and the stations perform a large enough 
packet aggregation to completely empty their buffers at each channel access.
Since the current 802.11 standards allow to adopt large levels of aggregation (around 1 MB), 
possibly larger than the TCP maximum window size, we believe this 
regime to be quite important in practice.
 
The main effect produced by large aggregation performed by both AP and stations is the following:
all packets circulating in the system, and associated to the same station (under our assumptions,
$F_s W_{\max}$ packets) cluster together and move as a single entity (a large batch)
across the network. Note that this phenomenon does not depend on initial conditions nor
on the value of backbone delay. 

The above behavior allows us to develop a simpler analytical model
than that in Sec. \ref{subsec:upbottle}, accounting also for backbone delay.    
We start analyzing the case of $D = 0$.
We adopt the same cycle analysis illustrated in Fig. \ref{fig:embed2}.
This time, however, we can have at most one transmission by each station
in between two consecutive transmissions by the AP.
Actually, we can directly exploit the computation of $P(h)$ done 
in Sec. \ref{subsec:upbottle}, and conclude that the number of transmissions performed 
by stations in a cycle has the uniform distribution over $1,\ldots,K$.
Let $\TUP$ be the time required by a station to send $F_s W_{\max}$ (effective) ACKs
in the uplink.
The usual renewal formula \equaref{ren} provides in this case:
\begin{equation}\label{eq:ren5}
\Lambda = \frac{\sum_{h=1}^K \frac{1}{K} h F_s W_{\max}}
{\frac{1}{\mu K} + \sum_{h=1}^{K}\frac{1}{K} \left(A(h,F_s W_{\max}) + h \TUP + \sum_{j=0}^{h-1} \frac{1}{\mu(K-j)} \right) }
\end{equation}

\begin{figure}
    \centering
    \includegraphics[width=0.9\linewidth]{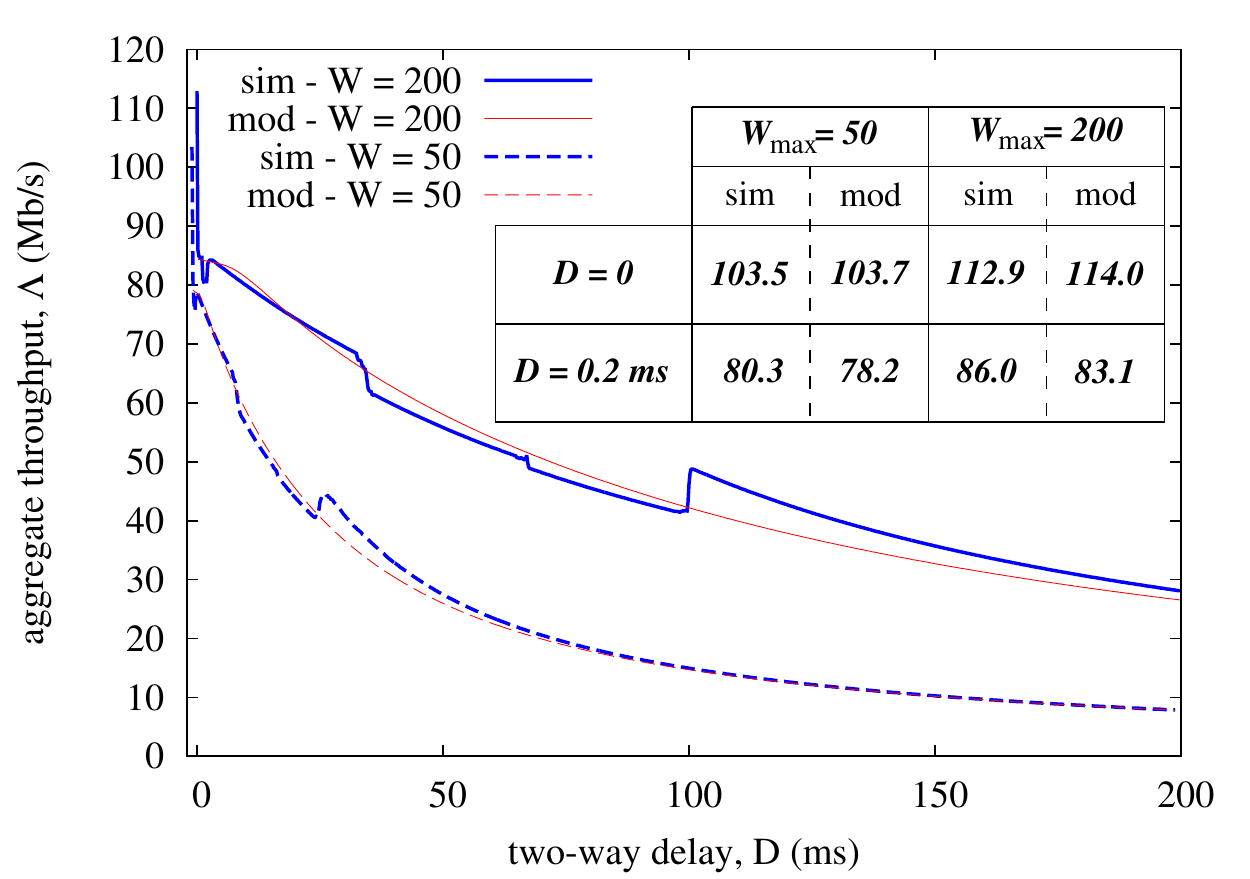}
    \caption{Throughput comparison (model vs simulation) in the reference system, 
    under the full aggregation regime ($\BAP \geq W_{\max}$, $\BSTA \geq W_{\max}$),
    for $W_{\max} = 50$ or $200$, as function of backbone delay $D$.}
    \label{plot4}
    \vspace{-0.2cm}
\end{figure}

Let us now consider a scenario in which the backbone delay is 
extremely small, but larger than the maximum channel contention time
(\textit{i.e.}, slightly larger than $W_0 \sigma$).   
It happens here that the last batch of ACKs sent up by a station in a cycle
cannot arrive at the AP in time to be immediately resent down
in the following AP transmission (marking the end of the current cycle). Therefore, this last 
batch will be aggregated with those sent by the AP at the end of the next cycle.
One important consequence of this fact is that, with non-zero delay,
we never see the maximum value ($h = K$) of user diversity.
This explains the sharp initial drop that we observe in the throughput
as we step out of $D = 0$ (see Fig. \ref{plot4}).

Specifically, with small delay the distribution of user diversity becomes:
\begin{equation}
  P(h) = \left\{\begin{array}{r@{}l@{\qquad}l}
    \frac{2}{K}  & \qquad h = 1 \\
    \frac{1}{K}  & \qquad 2 \leq h \leq K-1 
  \end{array}\right.
\end{equation}
whose average value is $\frac{K^2-K+2}{2 K}$.
Indeed, the AP only contends with $K-1$ stations during a cycle,
after receiving the last batch sent up in the previous cycle.
These $K-1$ stations send a number of batches uniformly distributed
in $[0,K-1]$, but the last of them (if any) cannot be transmitted
in the same cycle. On the other hand, we need to add the last
batch sent in the previous cycle.
Similarly, if stations do not send any batch during a cycle,
the AP will start contending after receiving 
one batch, again with $K-1$ stations.  

The corresponding throughput formula can be concisely written as: 
\begin{equation}\label{eq:ren6}
\!\!\Lambda \!=\! \frac{\sum_{h=0}^{K-1} \frac{1}{K} \max(1,h) F_s W_{\max}}
{\sum_{h=0}^{K-1} \!\frac{1}{K}\! \left(A(\max(1,h),F_s W_{\max}) \!+\! h \TUP \!+\! \sum_{j=0}^{h} \frac{1}{\mu(K-j)} \right) }
\end{equation}
leading to a throughput reduction roughly equal to $\frac{K^2-K+2}{2 K^2}$.
    
To compute the throughput in the case of larger delays, we 
adopt a useful approximation which consists of assuming that the 
network delay $D$ is exponentially distributed (instead of deterministic).
Such approximation greatly simplifies the analysis,
while providing an accurate throughput prediction.
Indeed, the memoryless property of the exponential distribution allows
us to embed a discrete-time Markov Chain at the boundaries
of the cycles as in Fig. \ref{fig:embed2}, with a bi-dimensional state $(m_1,m_2)$
denoting (assuming $K > 1$): the number $1 \leq m_1 < K$ of batches transmitted by 
the AP at the end of previous cycle (recall that this number cannot be 
equal to $K$, with non-zero delay);
the number $0 \leq m_2 \leq K - m_1$ of batches stored by the stations.
Then the remaining batches $m_3 = K - m_1 - m_2$ are  
still \lq flying' in the backbone, with remaining time
to arrive at the AP exponentially distributed with mean $D$.   
Note that the total number of states, equal to $\frac{K^2+K-2}{2}$,
is typically small (in the order of $K^2$).

We can easily express the transition probabilities among the above 
defined states, and use the stationary distribution of the Markov Chain
in \equaref{ren} (details can be found in App. \ref{app:markov}).

Fig. \ref{plot4} compares analytical predictions obtained by our Markov Chain model 
against simulation as we vary the backbone delay $D$, for two different values of
TCP maximum window size $W_{\max} = 50$ or $200$. We observe that the analytical predictions (based on the exponential delay assumption) nicely interpolate the rather complex curves obtained from simulation under deterministic delay.

The table inserted on the plot also shows the accuracy of \equaref{ren5} (for $D = 0$)
and \equaref{ren6} (for small but non null delay). Results for the latter 
(more realistic) case of non-null delay confirms that no more than 86 Mb/s can be achieved
by full aggregation in the reference system with $W_{\max} = 200$, which
is 50\% of bound $\Lambda^{(3)} = 172.5$ Mb/s, as roughly predicted by factor $\frac{K^2-K+2}{2 K^2}$,
equal to 44\%, with \mbox{$K=4$}.

\section{Comparison of Uplink Strategies} \label{sec:comparison}
Since the traditional random access mechanism of 802.11
does not allow us to fully exploit the capacity gain of downlink MU-MIMO
under closed-loop traffic, we may ask which alternative schemes (specifically intended 
for the uplink traffic) could be used to improve the throughput.

We will again focus on our reference system (under the best case $D = 0$), 
for which theoretical throughput bounds have been already computed
in Sec. \ref{subsec:reference}.

A simple solution to avoid the performance degradation inherent 
to random channel access is to make the uplink operate under 
the AP's coordination. Consider, for example, a simple polling mechanism
working as follows: right after transmitting down a data bundle, the AP
polls each station to which it has transmitted data
to send up a corresponding number of packets.
Clearly, this scheme allows to achieve bound $\Lambda^{(3)} = 172.5$ Mb/s.
  
Note that upper bound $\Lambda^{(2)} = 192.5$ could be approached
in a similar way, if stations were also able to send up a single (small) cumulative
ack for all data received from the AP. This could actually be obtained 
at the transport layer by increasing the thinning factor $T_F$.
Note, however, that massive use of delayed ACK techniques (beyond
the standard $T_F = 2$) has detrimental effects to TCP \cite{miorandi2006queueing}, and would 
require sophisticated cross-layer design to be implemented in a WLAN.
      
At last, we could employ multi-user transmissions 
also in the uplink (as is expected to be the case with the upcoming 802.11ax). In particular, consider a vanilla MU uplink with zero overhead\footnote{Similar 
to DL MU-MIMO, a multi-user uplink transmission also requires some overhead 
to set up communication \cite{tan2009sam}. 
While a multi-user uplink 
is yet to be standardized in the upcoming 802.11ax standards, prior works 
such as \cite{flores2016scalable} have demonstrated schemes to reduce this uplink overhead to 
as little as $100$ $\mu$s which is approximately 10 times less as compared to the sounding overhead 
for DL MU-MIMO.}, that allows backlogged stations to aggregate and concurrently 
send up many packets (TCP ACKs, in our case) in the uplink.
Even employing the standard delayed ACK option ($T_F = 2$), such scheme would achieve, with full
aggregation, throughput as high as $\Lambda^{(4)} = K F_s W_{\max}/[A(K,F_s W_{\max}) + \TUP(F_s W_{\max}/2)] = 187.0$ Mb/s,
where $\TUP(F_s W_{\max}/2)$ is the channel time to send 100 TCP ACKs, in our case.

\begin{figure}
    \centering
    \includegraphics[width=0.9\linewidth]{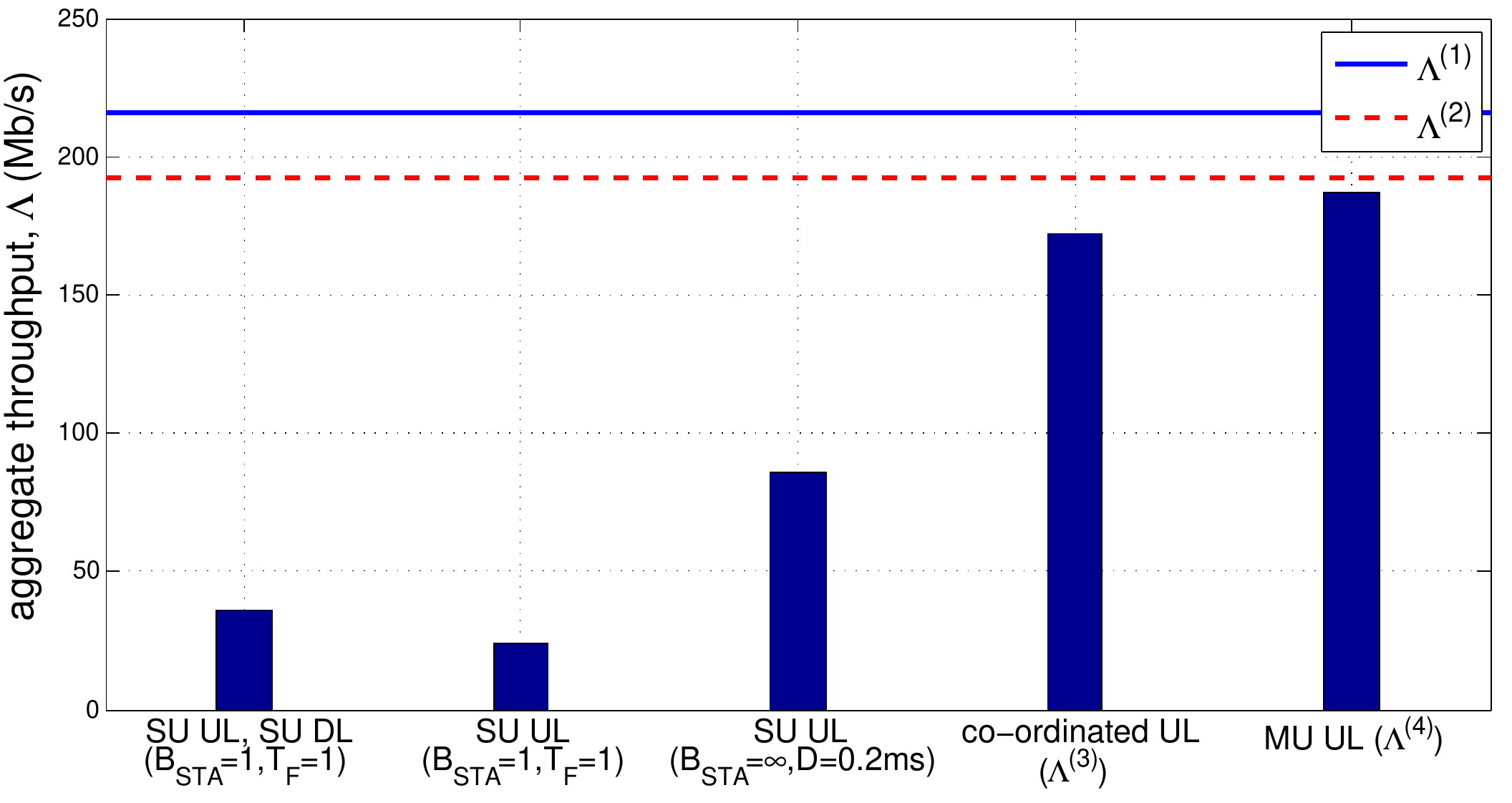}
    \caption{Comparison of throughputs achieved in the reference system, with $D = 0$, $\BAP = \infty$, 
    under different settings and access strategies.}
    \vspace{-0.2cm}
    \label{t_poll}
\end{figure}

Fig. \ref{t_poll} visually compares the throughputs achieved in several  
interesting cases that we have analyzed and discussed so far, 
in our reference system (always with unlimited aggregation by the AP).
{\color{black} The first bar shows that, in the case of $\BSTA = 1$, $T_F = 1$, 
the throughput that we get by using SU DL is actually larger than what we get by 
enabling MU DL (second bar)!}
The third bar (related to the full aggregation regime) 
shows the huge throughput loss (around 50\%) intrinsically
due to random channel contention. The last two bars are related to the
alternative uplink strategies discussed in this section.

\section{Impact of Packet Losses} \label{sec:losses}
To simplify the analysis and capture the key performance factors 
into a parsimonious model, we have assumed so far that packets are never 
lost/corrupted in the network, allowing TCP flows to reach their  
maximum congestion window size. Specifically, TCP flows do not experience
losses because we have considered a scenario in which: i) the capacity 
of the wired portion of the network is overprovisioned; ii) the AP/STAs 
can buffer unlimited amount of data for each flow; iii) the wireless channel
is error-free. What happens when the above assumptions do not hold?   

We can expect that factors preventing MU-MIMO from achieving
its maximum capacity gains under ideal conditions (\textit{i.e.}, in the absence of losses)
persist also under non-ideal conditions (\textit{i.e.}, in the presence of losses),
leading to even worse throughput. Here, we explore this issue in exemplary scenarios as follows, for both the downlink and uplink bottleneck regime.

\begin{figure}
    \centering
    \includegraphics[width=0.9\linewidth]{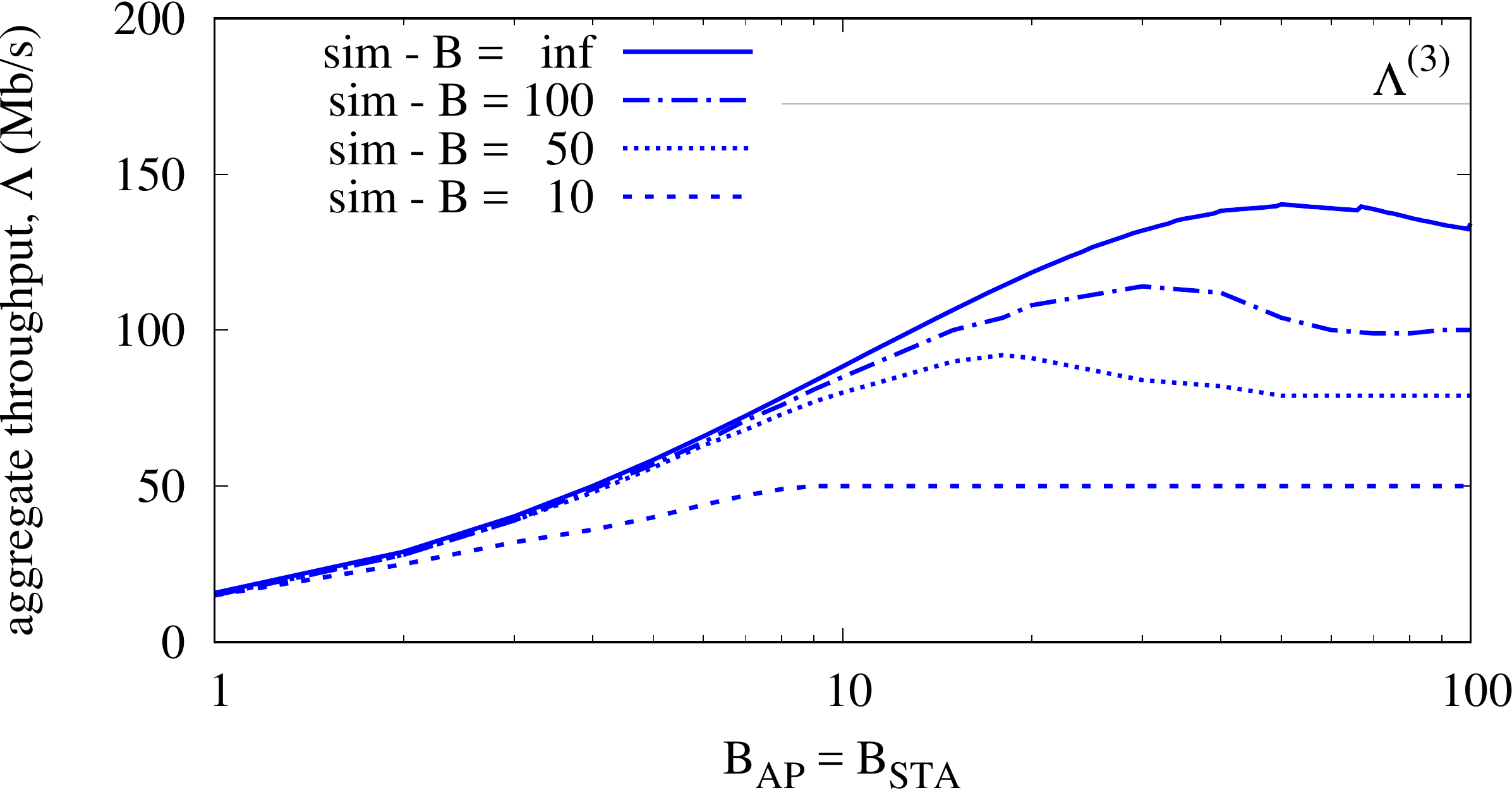}
    \caption{Impact of finite buffer size at the AP, in the downlink bottleneck regime with $\BAP = \BSTA$.}
    \label{plot1bis}
    \vspace{-0.2cm}
\end{figure}

Fig. \ref{plot1bis} shows TCP throughput obtained by simulation
in the reference scenario with $K = 4$ stations, $D = 0$, under the
downlink bottleneck regime (the analogous of Fig. \ref{plot1}). We assume now 
that the AP can buffer a finite number $B$ of data packets
for each flow, and consider $B = 100, 50, 10$. 
Here, TCP flows suffer from losses due to buffer overflow
at the AP, limiting their congestion window size as dictated by the
congestion control algorithm (in our case, NewReno).
Since aggregation of a number of data frames larger that the buffer
size is impossible, curves flatten for $\BAP \geq B$. 
A small overshoot is observed when $\BAP$ approaches $B$ from below.

\begin{figure}
    \centering
    \includegraphics[width=0.9\linewidth]{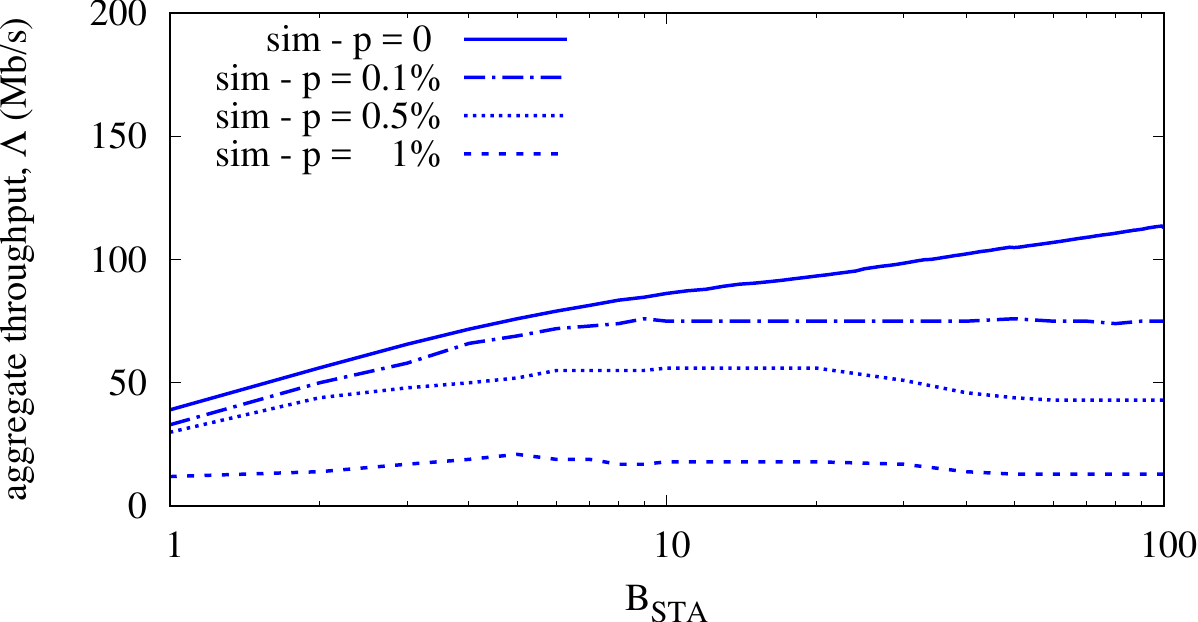}
    \caption{Throughput obtained by simulation in the reference system, with $\BAP = \infty$, $D = 0$, $K = 4$
    as function of $\BSTA$, for different values of packet loss probability.}
    \label{plot3bis}
    \vspace{-0.2cm}
\end{figure}

Fig. \ref{plot3bis} shows TCP throughput obtained by simulation
in the reference scenario with $K = 4$ stations, $D = 0$, under the
same MAC and physical-layer assumptions considered in Fig. \ref{plot3}, where
we observed, in the absence of losses, the onset of the 
uplink bottleneck regime. 
We now ask whether the presence of losses 
can cause the network to not enter the uplink bottleneck regime.
The answer to this question is in the affirmative since, by increasing
the loss probability we make TCP flows operate at smaller and smaller
window sizes, eventually leading to the condition in which
station queues are no longer saturated, and performance gets limited
by TCP dynamics, rather than by wireless channel dynamics.  
In this respect, the (poor) throughput obtained in the absence of losses 
(when we get into the uplink bottleneck regime), acts as an upper bound to the 
throughput achievable with the addition of losses. 
  
To show this fact, we introduce an artificial (Bernoulli) loss probability $p$ 
in the downlink, which could model non-ideal conditions
due to i) congestion in the wired network, ii) residual errors 
on the wireless channel not automatically recovered by the MAC protocol.

With loss probability $0.1\%$, we obtain a curve similar to that obtained
in the uplink bottleneck regime (achieved under zero losses). We also observe
that increasing the aggregation size at the STA beyond about 10 does not provide any benefit,
which can be explained by the fact that frame aggregation performed at the STAs gets
limited by the TCP window size.  
Higher loss probabilities lead to further throughput reductions: when TCP 
flows operate at too small window size, MAC-layer dynamics are no longer the bottleneck 
(actually we are no longer in the uplink bottleneck regime), and
overall performance is essentially determined by TCP dynamics in response
to losses and end-to-end delay.

\chapter{Passive AP-side Estimation and Decomposition of WLAN Latency}\label{uscope_ch}

This chapter presents the design of Uplink Latency Microscope (uScope) which enables a passive AP-side estimation of WLAN uplink latency and its constituent components. In Sec.~\ref{uproblem}, we present the high level problem formulation followed by a description of uScope framework in Sec.~\ref{uscope_framework}. 

\section{High Level Problem Formulation}\label{uproblem}
Let $\tenq$ denote the time at which a particular packet to be analyzed is enqueued at the STA and $\tend$ denote the time when the AP successfully receives the packet. Our goal is to enable the AP to passively compute the mean uplink latency ($\ulatency$), \textit{i.e.}, the average duration between $\tenq$ and $\tend$ for any of its associated STAs. Further for each associated STA, the AP should break down the mean uplink latency into its constituent components. While doing so, the AP cannot perform any active measurements (such as probing), cannot seek any STA side co-operation, does not have any additional hardware infrastructure (\textit{e.g.}, a network of sniffers) for collecting extra measurements and is thus constrained to make an estimate solely based on passive AP side observations. 

For ease of discussion, the associated STA under consideration is hereby referred to as the target STA. Our network scenario comprises multiple BSSs co-existing together. Remaining STAs from the same and neighboring BSSs as well as APs from neighboring BSSs are hereby referred to as non-target STAs. The notations used in this and the following sections are summarized in Table~\ref{tab:notation}.  

To understand the components of $\ulatency$ and the challenges in estimating $\ulatency$ and these components based on passive AP-side observations, consider the journey of a packet from when it gets enqueued at the target STA until it reaches the AP successfully. It is possible that when the packet gets enqueued, the target STA's queue already has a backlog of previously queued packets that are waiting to get serviced. Consequently, prior to reaching the head of the queue at time denoted by $\thead$, the packet will experience  a queuing delay of $\thead - \tenq$. The average queuing delay is denoted by $\qdelay$. If the target STA has multiple flows pushing packets into the layer 2 queue, the average queuing delay $\qdelay$ will have contributions from each of these flows and consequently, $\qdelay = \sum_{i=1}^{N} \bar{\Phi}_{q,i}$  where $\bar{\Phi}_{q,i}$ denotes the average contribution to $\qdelay$ from the $i^{th}$ flow from the target STA. 

\begin{figure}
    \centering
    \includegraphics[width=1\linewidth]{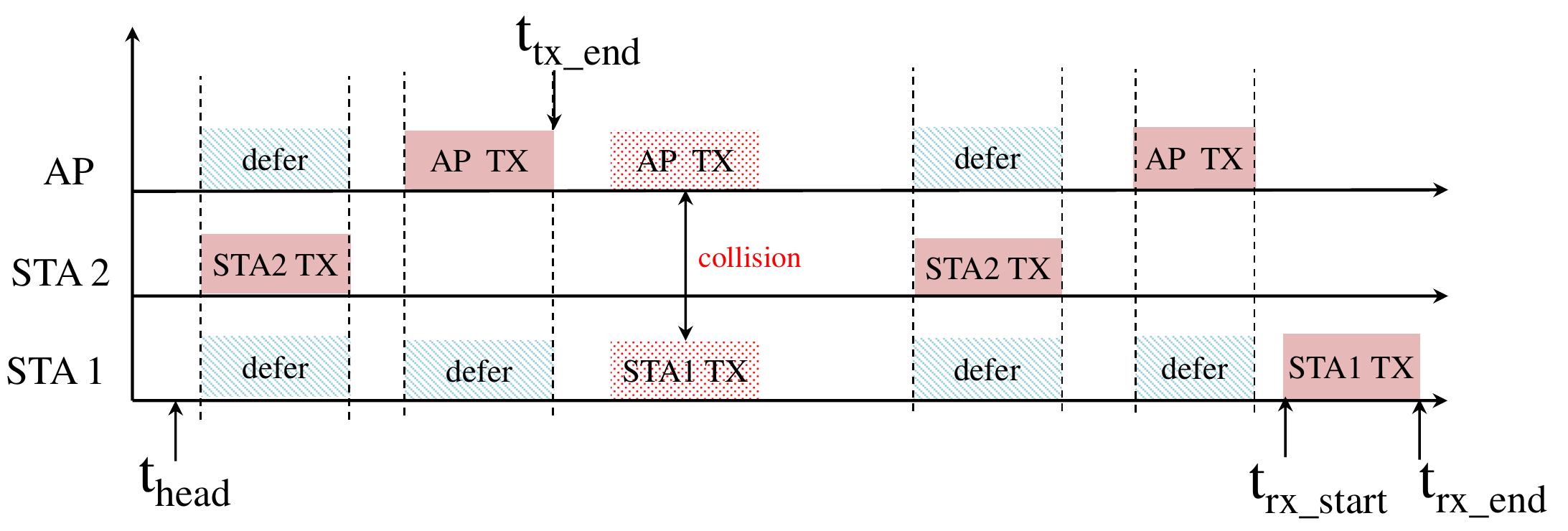}
    \caption{An example timeline to denote $\thead$, $\tsuccess$, $\tend$ and $\ttxend$. Here the target STA (STA 1) makes one retransmission and defers twice each to STA 2 and AP.}
    \label{general_timeline}
\end{figure}

When the packet reaches the head of the queue at $\thead$, the STA chooses a random backoff and begins to contend in accordance with the rules of 802.11 to gain access to the wireless channel. As depicted in Fig.~\ref{general_timeline}, the target STA contending for channel access might have to defer as some non-target STA captures the channel. It is also possible that the target STA captures the channel and makes an unsuccessful transmission (due to collisions, channel errors, etc.) forcing the STA to double its contention window size, choose another random backoff and attempt again. Finally, after several retransmission attempts the AP successfully receives the packet. The start time of this reception is denoted by $\tsuccess$. We define the duration between $\thead$ and $\tsuccess$ as the uplink channel access delay with its mean value denoted by $\adelay$.

Recall that the AP is constrained to estimate $\ulatency$, $\qdelay$, $\bar{\Phi}_{q,i}$ and $\adelay$ based on passive observations. Consequently, the AP encounters a few challenges. First, these parameters are determined by the joint effect of a number of factors such as network topology, interfering links to the target STA (from same BSS as well as neighboring BSSs), user activity, traffic load of interfering nodes, their PHY capabilities, data rates, etc. These factors are not directly observable by the AP. Further, for each packet received on the uplink by the AP, it cannot directly observe $\tenq$ and $\thead$. As a result, the AP cannot estimate $\ulatency$, $\qdelay$ and $\adelay$ based on direct observation of uplink transmissions from the target STA. Since the AP is unaware of $\tenq$ and $\thead$, for a packet received on the uplink from a target STA, the AP cannot directly observe which of the target STA's flows contributed to the queuing delay of the observed packet and the amount of contribution made.\footnote{Under the assumption that the target STA is fully backlogged, $\thead$ for a given packet is indeed the end time of the preceding packet's transmission. However, devices can also have idle times based on user activity.} 

The mean uplink channel access delay ($\adelay$) is affected by two key factors - (i) the number of retransmissions the STA has to make and (ii) the amount of time the STA defers prior to $\tsuccess$. We denote the average number of retransmissions and the mean defer time per packet transmission by $\retrans$ and $\tdefer$ respectively. Passive estimation of these parameters is challenging as the AP is unaware of which non-target STAs the target STA defers to as some of them may be hidden from the AP. This prevents the AP from directly observing the amount of time the channel is kept busy by  hidden non-target STAs. The AP also cannot directly count the number of retransmissions from a target STA. Consequently, based on passive observations, the AP cannot directly compute $\tdefer$ and $\retrans$ for a target STA. 

Finally, as the STA makes a successful transmission, it occupies the channel for a time which includes any MAC layer overhead, interframe spacings, data transmission time and the time to send the MAC layer acknowledgement. The average duration between the start of the successful transmission and its completion is the mean transmission delay ($\tdelay$). 

Thus the mean uplink latency ($\ulatency$), which is the average duration between $\tenq$ and $\tend$, has contributions from three components as follows  

\begin{equation}
\ulatency = \qdelay + \adelay + \tdelay.
\end{equation}

Our objective is to enable the AP to passively estimate $\ulatency$ and decompose it into its  components. Since $\tdelay$ can be directly observed by the AP, we focus only on $\qdelay$ and $\adelay$. As stated previously, the AP should be able to further decompose these two components. Thus, the target STA can have an arbitrary number of flows contributing to $\qdelay$ and the AP should be able to infer their individual contributions ($\Phi_{q,i}$). Further, the AP should also be able to estimate $\retrans$ and $\tdefer$ which affect $\adelay$ for the target STA. Note that $\ulatency$, $\adelay$, $\tdefer$, $\retrans$, $\qdelay$ and $\bar{\Phi}_{q,i}$ can vary from STA to STA depending on their traffic load and the network conditions experienced by the individual STA.

\begin{table}
\begin{center}
\vspace{0mm}
\caption{Definition of parameters used in the estimation and decomposition technique\label{tab:notation}}{%
\scriptsize
\begin{tabular}{||c|p{6.5cm}||}
\hline
\hline
\rule{0pt}{3ex}
$\tenq$  & time when a packet is enqueued at the target STA \\[1ex]
\hline
\rule{0pt}{3ex}
$\thead$  & time when a packet reaches the head of the target STA queue\\[1ex]
\hline
\rule{0pt}{3ex}
$\tsuccess$ & time when the packet reception starts at the AP \\[1ex]
\hline
\rule{0pt}{3ex}
$\tend$  & time when an uplink packet reception is complete \\[1ex]
\hline
\rule{0pt}{3ex}
$\ttxend$ & time when a downlink packet transmission is complete \\[1ex]
\hline
\rule{0pt}{3ex}
$\segend$  & time when TCP segment transmission on the downlink is complete \\[1ex]
\hline
\rule{0pt}{3ex}
$\ackenq$ & time when TCP ACK is enqueued at the target STA \\[1ex]
\hline
\rule{0pt}{3ex}
$\ackhead$ & time when TCP ACK reaches the head of the STA queue \\[1ex]
\hline
\rule{0pt}{3ex}
$\ackstart$ & time when TCP ACK reception starts at the AP \\[1ex]
\hline
\rule{0pt}{3ex}
$\ackend$ & time when TCP ACK reception on the uplink is complete \\[1ex]
\hline
\rule{0pt}{3ex}
$\intkstart$ & time when the reception of the $N^{th}$ intermediate uplink packet from the target STA starts. An intermediate uplink packet is a packet received by the AP from the target STA between $\segend$ and $\ackstart$\\[1ex]
\hline
\rule{0pt}{3ex}
$\intkend$ & time when the reception of the $N^{th}$ intermediate uplink packet from the target STA ends. \\[1ex]
\hline
\rule{0pt}{3ex}
$\qdelay$ & the average queuing delay at the target STA. \\[1ex]
\hline
\rule{0pt}{3ex}
$\bar{\Phi}_{q,i}$ & the average contribution of the $i^{th}$ flow from the target STA to $\qdelay$. \\[1ex]
\hline
\rule{0pt}{3ex}
$\adelay$ & the average access delay for the target STA. \\[1ex]
\hline
\rule{0pt}{3ex}
$\tdelay$ & the average transmit delay. This duration includes time to transmit the packet, any MAC layer overhead and interframe spacings.\\[1ex]
\hline
\rule{0pt}{3ex}
$\ulatency$ & the average uplink latency for a target STA.\\[1ex]
\hline
\rule{0pt}{3ex}
$\retrans$ & the average number of retransmissions faced before a packet is successfully transmitted on the uplink by the target STA.\\[1ex]
\hline
\rule{0pt}{3ex}
$\tdefer$ & the average defer time faced by the target STA for an uplink transmission.\\[1ex]
\hline
\rule{0pt}{3ex}
$\deferk$ & the average defer time faced by the target STA in the $k^{th}$ retransmission attempt.\\[1ex]
\hline
\rule{0pt}{3ex}
$\contnk$ & the average contention time for the $k^{th}$ retransmission attempt.\\[1ex]
\hline
\rule{0pt}{3ex}
$\txk$ & the average amount of time for which the channel is occupied in the $k{th}$ retransmission attempt.\\[1ex]
\hline
\hline
\end{tabular}}
\end{center} 
\end{table}

\section{uScope Framework}\label{uscope_framework}
This section presents \technique, a tool to perform AP-side estimation and decomposition of WLAN uplink latency for each of its associated STAs. We first present virtual probing, the key idea for estimation and decomposition of WLAN uplink latency in Sec.~\ref{vp}. Sec.~\ref{ta} and \ref{rwfit} describe how by leveraging virtual probing, \technique performs the estimation and decomposition. 

\subsection{TCP Handshake as a Virtual Probe}\label{vp}
Consider a layer-4 handshake: The STA receives a TCP segment from the AP on the downlink sent by any arbitrary server on the internet. Since the WLAN is the last hop for this TCP segment, reception of this segment results in the generation of a TCP ACK. This layer-4 handshake exposes three key attributes that are not directly observable to the AP.

\textbf{(i) Uplink enqueue timestamp.} The STA will attempt to return the TCP ACK as fast as possible. In other words, when the handshake occurs, the time when the TCP ACK gets enqueued is approximately the time when the TCP segment is received on the downlink from the AP. Therefore, the  enqueue timestamp for the TCP ACK can be inferred by the AP as $\ackenq \approx \segend$. 

\textbf{(ii) Uplink aggregate layer-2 delay.} After the layer-4 ACK gets enqueued at the STA, it experiences queuing delay as the STA transmits previously backlogged packets, if any. As the ACK reaches the head of the queue, it experiences additional delays from uplink contention, deferral, retransmissions and uplink transmission prior to reaching the AP. Thus, the duration between the transmission of a TCP segment to reception of TCP ACK is a sum of all layer-2 delays that any packet transmitted by the STA would encounter.

\textbf{(iii) Uplink backlog state indicator.} In the time interval between transmission of the TCP segment to reception of the TCP ACK, the STA is guaranteed to be backlogged with at least one packet, the TCP ACK. As a result, until the AP receives the TCP ACK, the STA's queue can be considered to be in a backlogged state. Any other packet received from the STA in this period of time can be considered to be already present in the queue when the TCP ACK gets enqueued.

\technique leverages these three inferences obtained from a layer-4 handshake to estimate total uplink latency and its constituent components as described in the following subsections. Thus, the methodology used in \technique is to perform an assessment of uplink latency by leveraging layer-4 handshakes from TCP flows in which the target STA receives data. Analogous to active probing (e.g, ping) where the AP generates probe requests and the STA sends probe responses, \technique leverages TCP handshakes as virtual probes. However, unlike active probing, this methodology does not increase the traffic load as it leverages layer-4 handshakes that would anyways occur due to TCP. Further, since \technique leverages the fundamental closed loop property of TCP flows, the handshakes can be a part of \textit{any} TCP flow in which the target STA receives data (\textit{e.g.}, a Netflix video stream). 

\subsection{Timing Analysis on Virtual Probes for Delay Estimation}\label{ta}
For every packet received on the uplink, the AP can directly observe the timestamps for reception start ($\tsuccess$) and reception end ($\tend$). Further, for each packet transmitted on the downlink, the AP can observe the timestamp corresponding to the end of the transmission ($\ttxend$). This section describes how virtual probing enables \technique to estimate the total uplink latency, access delay and queuing delay based on these timestamps for TCP segments and ACKs.

\subsubsection{Total Uplink Latency Estimation}
The key challenge faced in estimation of the total uplink latency ($\ulatency$) is that the AP cannot not directly observe the enqueue timestamp ($\tenq$) for a packet received on the uplink from a target STA. Inferring $\tenq$ for all or a subset of packets received on the uplink would enable the AP to use $\tend$ (a directly observable parameter at the AP-side) and $\tenq$ corresponding to such packets to obtain $\ulatency$. However, as stated previously, when a layer-4 handshake occurs, the time at which the TCP ACK gets enqueued is approximately the time at which the TCP segment is received, \textit{i.e.}, $\ackenq \approx \segend$. Therefore, for these TCP ACKs, the enqueue timestamp can be inferred at the AP-side. \technique leverages this key idea to estimate the total uplink latency for a target STA as 

\begin{equation*}
\ulatency = \overline{\big(\ackend - \segend \big)}.
\end{equation*}

\subsubsection{Uplink Access Delay Estimation}
\begin{figure}[]
    \centering
\begin{minipage}[t]{1\linewidth}
    \begin{subfigure}[t]{1\textwidth}
        \centering
		\includegraphics[width=\textwidth]{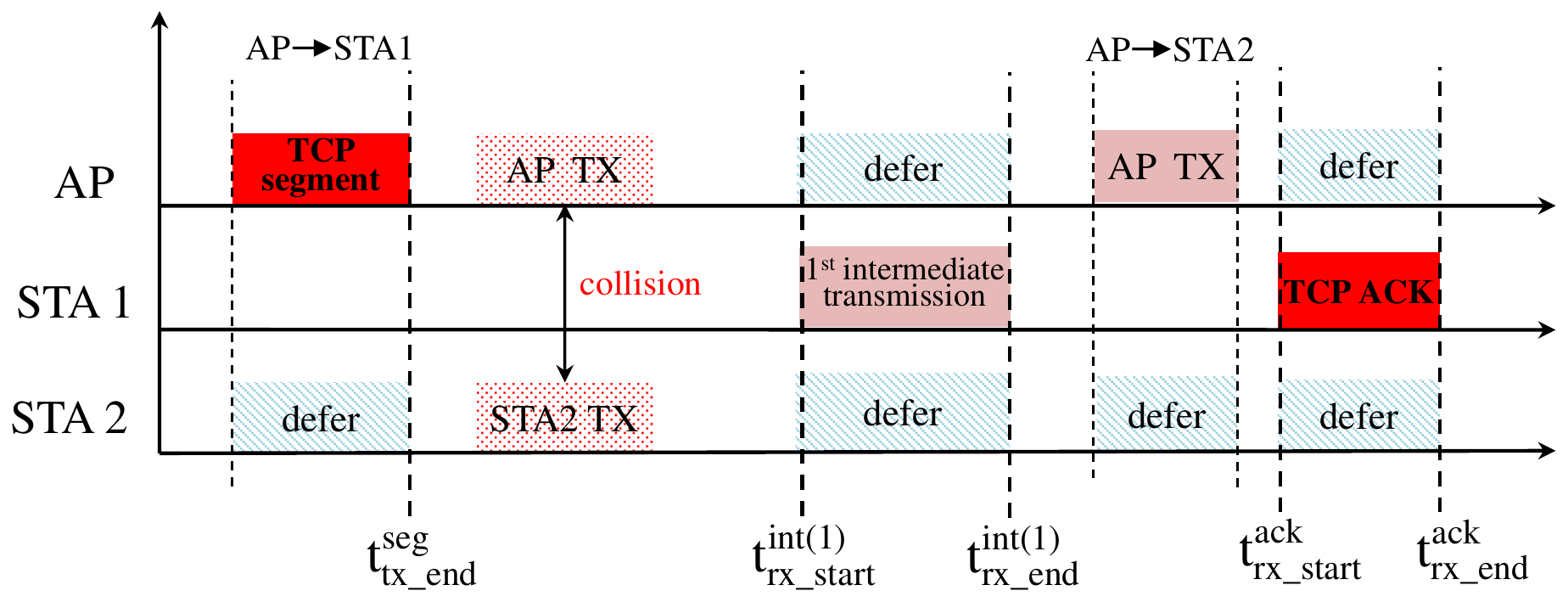}
        \caption{Queued ACK handshake}
    \end{subfigure}
    \begin{subfigure}[t]{1\textwidth}
        \centering
        \includegraphics[width=\textwidth]{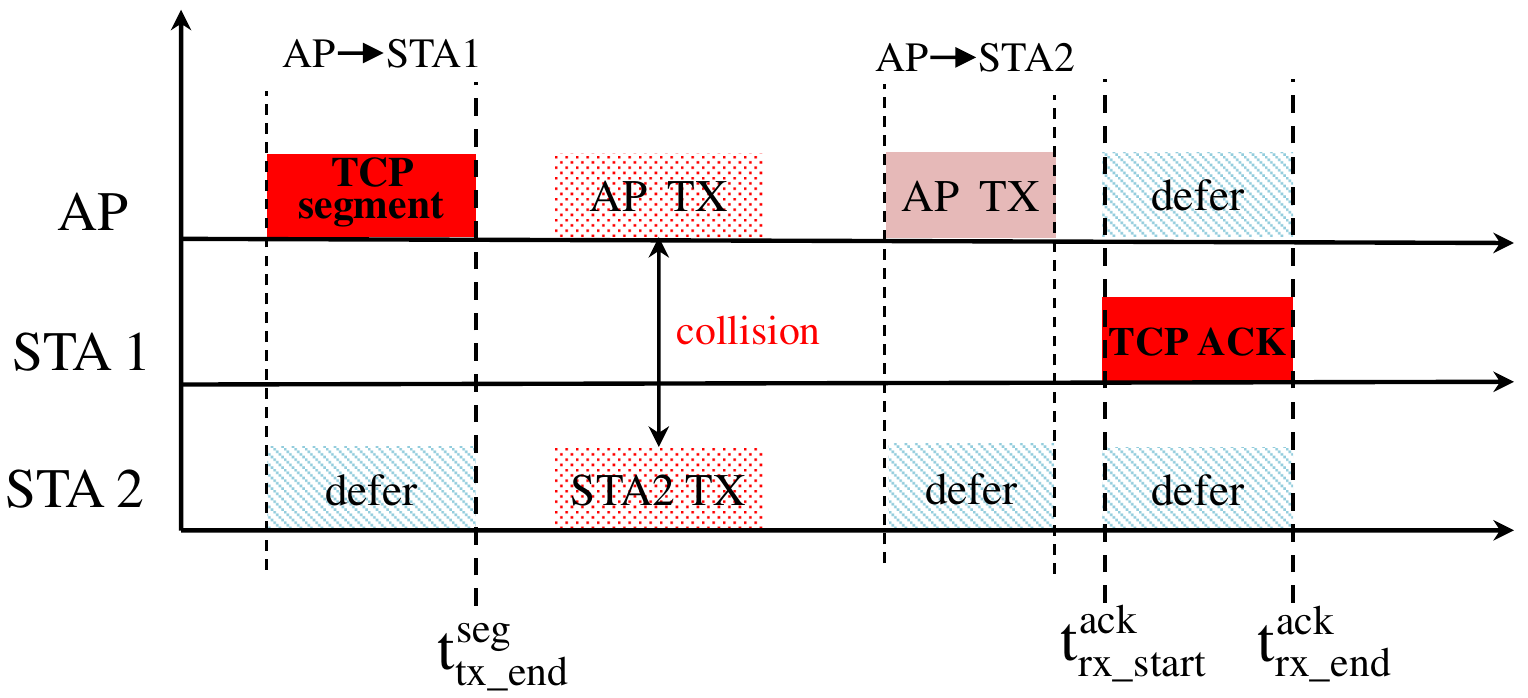}
        \caption{Immediate ACK handshake}
    \end{subfigure}
    \vspace{-0.2cm}
    \caption{Timeline to show queued ACK handshake and immediate ACK handshake. In the illustration, STA 1 is the target STA and STA 2 is the non-target STA. In queued ACK handshake shown in (a), there is one intermediate transmission between $\segend$ and $\ackstart$. In the immediate ACK handshake shown in (b), the AP does not receive any intermediate transmissions from the target STA before receiving the TCP ACK.}
    \label{queuedACK}
    \end{minipage}
\vspace{-0.4cm}
\end{figure} 
To estimate the access delay, the key information missing at the AP is the time when the packet reached the head of the STA queue ($\thead$). Suppose that there is a time interval in which the target STA's queue is backlogged with a few packets. Each packet will reach the head of the queue at the end time of the transmission of the previous packet and thus the AP can infer $\thead$ for each packet received in this duration on the uplink. This inference, combined with the time of reception start ($\tsuccess$), which is directly observable at the AP, can help compute the access delay that the STA encounters. Unfortunately, the AP cannot directly observe time intervals when a STA's queue is backlogged as the STA can have some idle times based on user activity. However, virtual probing provides a unique opportunity to the AP: Recall that the STA is guaranteed to be backlogged until the TCP ACK is received, \textit{i.e.}, in the interval from $\segend$ to $\ackend$. Therefore, by using packets received on the uplink from the target STA in this duration, \technique can estimate the uplink access duration as follows.

Let us say that the AP observes $N$ transmissions from the target STA between $\segend$ and $\ackstart$. We term such transmissions as intermediate transmissions. Handshakes with $N > 0$ are referred to as queued ACK handshakes and those with $N = 0$ are termed as immediate ACK handshakes. Fig.~\ref{queuedACK} shows an illustration of these two handshakes. \technique analyses the timestamps of the packets received from the target STA during these handshakes to obtain instantaneous values of access delay as follows. $\adelay$ is computed by averaging over all these values.  

(a) Queued ACK handshake: In such a case, the STA queue is not empty when the TCP ACK is enqueued. Consequently, the AP witnesses transmission of previously queued packets as  intermediate transmissions prior to reception of the TCP ACK on the uplink. The queued ACK handshake provides instantaneous values of uplink access delay in two ways. One measurement is provided by each of the intermediate transmissions and then one comes from the TCP ACK as follows. For the $j^{th}$ intermediate transmission ($1 < j \leq N$), $t_{\textrm{head}}^{int(j)} = t_{\textrm{rx\_end}}^{int(j-1)}$. Consequently, the instantaneous value of uplink access delay is given by $\big(t_{\textrm{rx\_start}}^{int(j)} - t_{\textrm{rx\_end}}^{int(j-1)}\big)$. The TCP ACK reaches the head of the queue after the $N^{th}$ intermediate transmission. Consequently, for the TCP ACK, the instantaneous value of uplink access delay is given by $\big(\ackstart - \intkend \big)$.

(b) Immediate ACK handshakes: In this case, the STA queue is empty when the TCP ACK gets enqueued and hence the TCP ACK does not wait prior to reaching the head of the queue, \textit{i.e.}, $\ackhead = \ackenq$. Recall that for all handshakes, $\ackenq = \segend$ and hence the instantaneous value of uplink access delay for the immediate ACK handshake is given by $\big(\ackstart - \segend\big)$.

\subsubsection{Queuing Delay Estimation and Decomposition}
Recall that our goal is to estimate the average queuing delay ($\qdelay$) and further decompose it into contributions coming from individual uplink flows. To this end, \technique leverages the two classes of TCP handshakes mentioned above as follows.

Consider the case of immediate ACK handshake. In this case,  the target STA's queue is empty when the TCP ACK gets enqueued. Consequently, the ACK immediately reaches the head of the queue (\textit{i.e.}, $\ackenq = \ackhead$) and the queuing delay is 0.

However, in the case of a queued ACK handshake, the TCP ACK experiences a delay equal to the amount of time required to transmit the $N$ packets queued before it. Consequently, the ACK reaches the head of the queue after the $N^{th}$ intermediate transmission ($\ackhead = \intkend$) and hence the instantaneous value of queueing delay is given by $\big(\intkend - \segend\big)$. As before, $\qdelay$ can be computed by averaging over all the instantaneous values obtained from the two types of handshakes. 

The net queuing delay has contributions from each flow coming on the uplink from the target STA. Each packet delays the TCP ACK from reaching the head of the queue by an amount of time equal to the duration between when the packet reaches the head of the STA queue to when the packet gets transmitted successfully. As shown in Fig.~\ref{queuedACK}a, the contribution to the net queuing delay from packets of the $i^{th}$ flow is given by $\big(t_{\textrm{rx\_end}}^{int(j-1)} - t_{\textrm{rx\_end}}^{int(j)}\big)$ if the $j^{th}$ intermediate packet belonged to the $i^{th}$ flow from the target STA. 

\subsection{Retransmission and Defer Delay Estimation}\label{rwfit}
The value of $\adelay$ is influenced by two key factors, (i) $\tdefer$ which is the amount of time the medium is sensed as busy by the target STA as it attempts to transmit and (ii) $\retrans$ which is the average number of retransmissions attempts per packet that the target STA makes prior to the successful transmission. 

The AP's lack of knowledge of the connectivity links and contention relationships for a target STA prevents it from knowing which nodes a target STA defers to. In the network scenario we consider, due to asymmetry in the contention relationships for the downlink and uplink, the target STA may be deferring to nodes that are hidden from the AP. Likewise, the AP may hear nodes that are hidden from the target STA. Further, the AP cannot directly observe and record each failed transmission of the target STA. As a result, both $\tdefer$ and $\retrans$ can be directly observed only at the STA and are unknown to the AP. This subsection presents a technique that \technique leverages to perform joint estimation of $\tdefer$ and $\retrans$ passively at the AP.

Suppose that the STA has to make $K$ attempts on average prior to a successful transmission ($K$ is not known or observable at the AP). Let $\bar{Z}_{k}$ denote the average duration for the $k^{th}$ attempt ($1 \leq k \leq K$). Each attempt duration consists of contention time, defer time and transmit time. Therefore,

\begin{equation}
Z_{k} = \contnk + \deferk + \txk
\label{zk}
\end{equation}

\noindent where $\contnk$ represents the mean contention time, $\deferk$ denotes the mean defer time and $\txk$ denotes the mean transmission time in the $k^{th}$ attempt. We assume that $\txk$ remains fixed until the successful transmission and hence $\txk \approx \tdelay$. Further, we also assume that $\deferk$'s are iid. This results in an error and we show how \technique compensates for it later. While limited to first order effects, these assumptions lead to a simple methodology to estimate $\tdefer$ and $\retrans$ that nonetheless lead to accurate results (as shown in later sections).

The number of such attempt durations that can fit within the sum of average access delay ($\adelay$) and transmission delay ($\tdelay$) is equal to the average number of transmission attempts or in other words the average number of retransmission attempts ($\retrans$) plus one. The sum of defer time across all these intervals is the average defer delay ($\tdefer$). Therefore, to estimate $\retrans$ and $\tdefer$, \technique must estimate the number of attempt durations that can fit within $\adelay + \tdelay$ (a value that can be computed based on previously described techniques). However, notice that for doing this, \technique must have knowledge of the average duration of each attempt for a given STA. Unfortunately, the average attempt durations are not directly observable at the AP. Therefore, \technique must estimate the average duration for each attempt for a given STA. To this end, \technique uses a protocol based inference methodology coupled with virtual probing to make estimates for each attempt duration. 

\subsubsection{Protocol Based Inference} Under the above formulation, the duration of each attempt differs from the others based on the value of contention time. However, contention time can be estimated by leveraging knowledge of the 802.11 standard. Recall that the standard defines the rules governing the contention process. In each round of attempt, the target STA chooses a random number that is uniformly distributed in $[0,W-1]$ where $W$ is the maximum contention window size. $W = 2^{A}$ where $A$ starts with an initial value of 4 for the first attempt and increments for each round of transmission attempt. Therefore, 

\begin{equation}
\contnk = \frac{(2^{3+k} - 1)*\sigma}{2} 
\label{contn}
\end{equation}

\noindent where $\sigma$ is the slot duration. Based on our assumptions and from Eq. (\ref{contn}) and Eq. (\ref{zk}), we get

\begin{equation}
Z_{k} = \frac{(2^{3+k} - 1)*\sigma}{2} + \deferone + \tdelay
\end{equation}

\noindent where $\deferone$ is the average defer delay faced by the STA during first attempt. Therefore, to obtain an estimate for each $Z_{k}$, the only unknown left is $\deferone$. However, the key challenge to estimate $\deferone$ is that the AP cannot directly measure who the target STA defers to and for how long. Further, the AP also cannot directly observe when each retransmission attempt of the target STA started. As a result, $\deferone$ is not directly known at the AP either. Next, we show how \technique leverages virtual probing to overcome this challenge. 

\subsubsection{Virtual Probing for First Attempt Defer Delay} 
Recall that virtual probing provides instantaneous values of uplink access delays as stated in the previous subsection. We restate that each uplink access delay is a sum of total contention time, defer delays and transmission delays. Consider TCP handshakes whose TCP ACK was successfully transmitted by the STAs in the first attempt. Notice that the average defer delay faced by such ACKs is the average defer delay for the first attempt, \textit{i.e.}, $\deferone$. Since the TCP ACK is transmitted in the first attempt, the average contention time can be computed based on Eq. (\ref{contn}) with $k = 1$. Further recall that transmission delays are directly observable to the AP. By subtracting these two quantities from the average defer delay of such TCP ACKs, \technique can estimate $\deferone$. 

To identify such handshakes, \technique uses the retry bit in the 802.11 MAC header of the packet carrying the TCP ACK. The retry bit indicates if the current packet experienced any retransmissions and is set to 1 for all retransmitted packets and 0 for those that are successfully transmitted in the first attempt. Note that the retry bit does not indicate the number of retransmissions experienced by the packet. Based on such handshakes, \technique estimates $\deferone$ as  

\begin{equation}
\deferone = \overline{\ackend - \ackhead} - (\bar{\theta}_{\textrm{contn,1}} + \bar{\theta}_{\textrm{tx}}). 
\label{fa}
\end{equation} 

Finally, to compensate for any errors arising from the iid assumption made about $\deferk$, \technique adds the residual after subtraction of all possible $Z_{k}$s from $\adelay + \tdelay$ to the $\tdefer$ estimate.
\chapter{Experimental Platforms}

We experimentally validate virtual speed test and \technique using off the shelf Wi-Fi chipsets as well as simulation based frameworks. This section describes the platforms that were used and the framework implementation using those platforms. Section~\ref{nuc} describes the Intel NUC based testbed used for the validation of virtual speed test. This is followed by a description of the implementation of \technique in Sec.~\ref{wrtacm}. Finally, we describe out ns 3 based implementation for both virtual speed test as well as \technique in Sec.~\ref{ns3}. 

\section{Intel NUC Based Testbed}\label{nuc}
\begin{figure}
    \centering
    \includegraphics[width=0.7\linewidth]{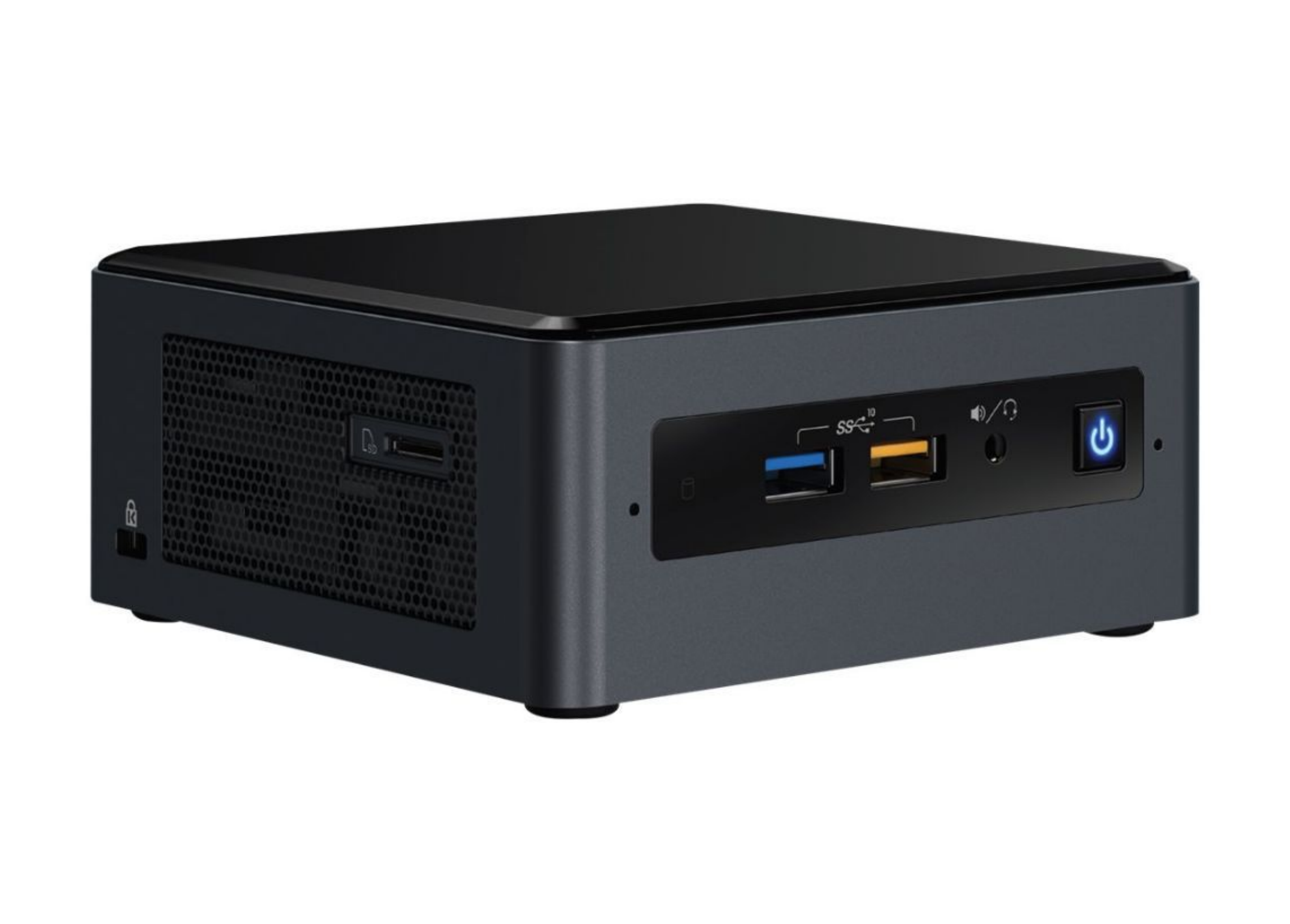}
    \caption{Intel NUC for virtual speed test validation \cite{intel_nuc}.}
    \label{intel_nuc}
\end{figure}

Virtual speed test is evaluated using a commodity hardware based testbed developed by using an Intel NUC shown in fig.~\ref{intel_nuc}. Below we describe the detailed capabilities of the hardware platform. 

\textbf{Platform specifications.} Our AP runs on a Linux operating system and is factory installed with 32 GB DDR4 SO-DIMM RAM, 2.4 GHz dual core CPU with slots for USB, HDMI and a Gigabit LAN port. The AP is equipped with a Ralink RT3070 off-the-shelf WiFi chipset. The radio card supports IEEE 802.11b/g/n utilizing up to 40 MHz bandwidth and a peak PHY rate of 300 Mbps. The STAs are a mix of portable laptops running on either Windows or Linux OS whose network interface card supports 802.11b/g/n as well. This AP is hereby referred to as the VST (virtual speed test) AP. These parameters are summarized in Table \ref{tab:intelNUC}.

\textbf{Per-packet statistics.} We build APIs that enable the acquisition of a number of per packet statistics at the access point as well as the STAs. While a rich raw characterization of each packet is available at all the devices, we only feed the AP side packet timestamps, source and destination IP addresses, frame sizes, and PHY rates into the L2 edge TCP model for throughput estimation. Nonetheless, the remaining statistics enable us to characterize the operating environment as will be described later. As described earlier, the parameter estimation methodologies employ packet timestamps as a part of the computation process. While the absolute value of these timestamps can be infected with system dependent offsets, their post-subtraction residue becomes negligible as they have a low second moment. The timestamps on the VST AP  are available on a nanosecond granularity. However, timestamp resolution has hardware dependency and can vary based on processor architecture, system clock, operating system time stamping policies, etc. Since the packet timeline operates on a scale greater than microsecond, even a microsecond resolution - a capability available on many hardware platforms \cite{guide2011intel}, is sufficient to capture the time domain information between packets necessary to estimate $\uldelay$.

\begin{table}
\begin{center}
\vspace{0mm}
\caption{Intel NUC based testbed specification \label{tab:intelNUC}}{%
\scriptsize
\begin{tabular}{||c|p{6.5cm}||}
\hline
\hline
\rule{0pt}{3ex}
Parameter  & Value \\[1ex]
\hline
\hline
\rule{0pt}{3ex}
Peak datarate  & 300 Mbps\\[1ex]
\hline
\rule{0pt}{3ex}
Maximum bandwidth & 40 MHz \\[1ex]
\hline
\rule{0pt}{3ex}
CPU & 2.4 GHz Dual core \\[1ex]
\hline
\rule{0pt}{3ex}
RAM & 32 GB DDR4 SO-DIMM \\[1ex]
\hline
\rule{0pt}{3ex}
Operating system & Linux \\[1ex]
\hline
\hline
\end{tabular}}
\end{center} 
\end{table}

\section{Linksys WRT3200ACM Based Testbed}\label{wrtacm}
\technique is evaluated using experiments performed on an off-the-shelf Wi-Fi chipset. This section provides details on the software modules and the commodity hardware based testbed implementation.   

\subsection{System Implementation}
The system consists of two main parts, a \textit{stats manager module} which gathers AP side observations and a \textit{stats processor module} which implements the \technique core. 

\textbf{Stats manager.} The stats manager runs on the AP and implements the framework for collection of AP side observation logs. This module employs a libpcap engine \cite{libpcap} to collect logs from the network interface card of the device. The logs consists of aggregate transmission (TX) and reception (RX) statistics which include packet timestamps, 802.11 radio tap headers and packet headers for transmissions and receptions. Further, the logs also contain neighboring BSS's statistics which comprise of timestamps and airtime utilization for transmissions from co-existing BSSs.   

\textbf{Stats processor.} The information gathered by the stats manager is operated upon by an instance of the stats processor. The stats processor is a Python based framework with over \textit{7,000 lines} of code to implement the \technique core. This module implements techniques to parse the AP-side log and separate individual flows. The parsed log is piped into the \technique core which implements the techniques described in the previous section to estimate $\ulatency$, $\adelay$, $\tdefer$, $\retrans$, $\qdelay$ and $\bar{\Phi}_{q,i}$. 

\subsection{Testbed Characterization}
\begin{figure}
    \centering
    \includegraphics[width=0.7\linewidth]{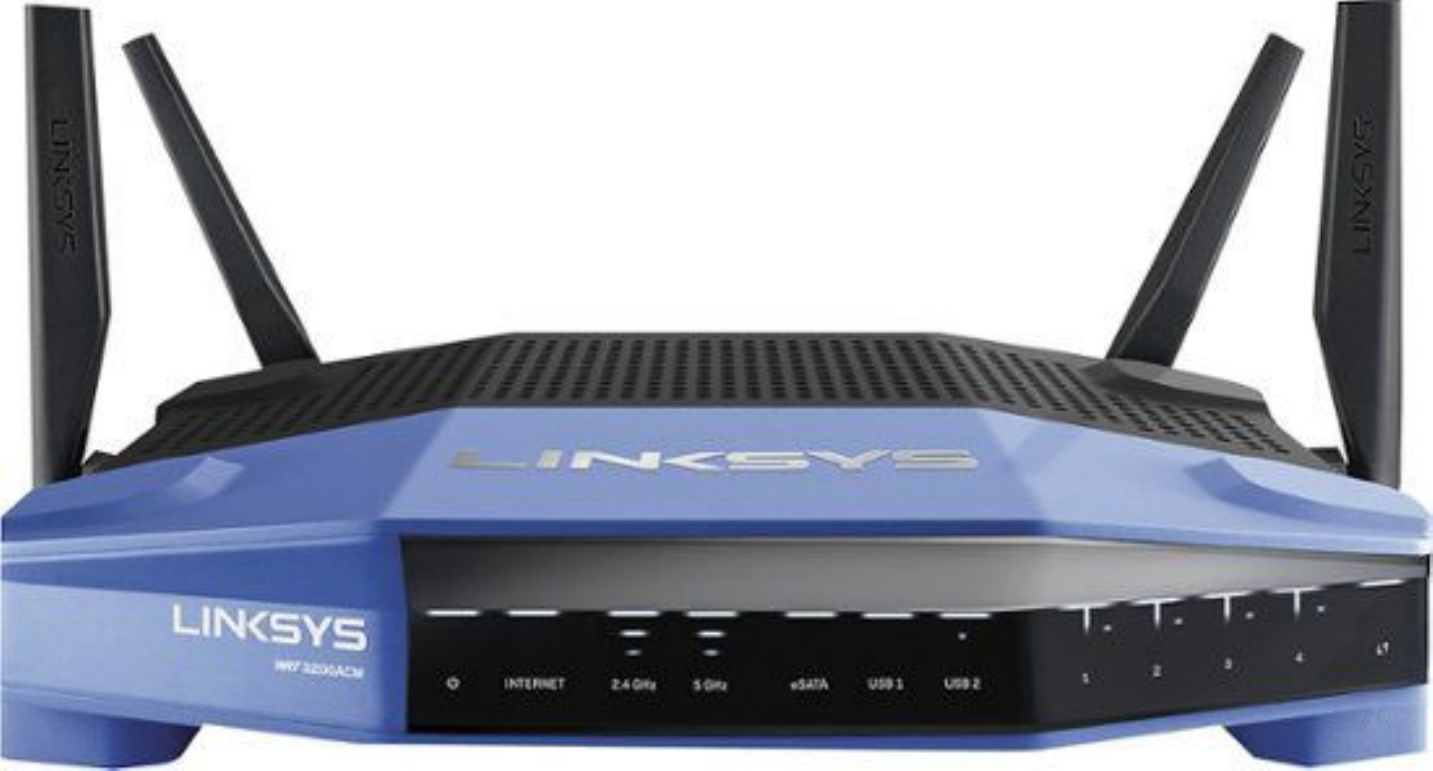}
    \caption{Linksys WRT3200ACM used as an Access Point for uScope validation \cite{linksys_router}.}
    \label{linksys}
\end{figure}

\textbf{Access Point.} We use the Linksys WRT3200ACM as the Access Point as shown in Fig.~\ref{linksys}. The device is an Armada-385 based router running a variant of Linux operating system. The AP has a dual architecture design wherein a general purpose system on chip (SoC) controls the mother board and a peripheral SoC executes the 802.11 PHY and MAC protocols. The dual band radio card has 4 external antennas providing antenna gains of 2.52 dBi in the 2.4 GHz band and 3.81 dBi in the 5 GHz band thereby providing a theoretical range upto 100 m which is far higher compared to those provided by some of the common dualband antenna designs \cite{nayak2013multiband, endluri975low, nayak2012ultrawideband, nayak2013compact, nayak2014novel, endluri975low}. The AP is IEEE 802.11ac compliant and supports up to 3 simultaneous spatial streams and up to 160 MHz bandwidth. The peak data rates supported on the AP are 2.6 Gbps (802.11ac, 5 GHz), 600 Mbps (802.11n, 2.4 GHz) and 54 Mbps (802.11a/g). The AP is equipped with a 1.8 GHz dual core ARM based CPU, 256MB Flash, 521MB DDR3 RAM, 4 Gigabit LAN ports and a Gigabit WAN port. These parameters are specified in Table.~\ref{tab:linksys}.

\begin{table}
\begin{center}
\vspace{0mm}
\caption{Linksys WRT3200ACM based testbed specifications\label{tab:linksys}}{%
\scriptsize
\begin{tabular}{||c|p{6.5cm}||}
\hline
\hline
\rule{0pt}{3ex}
Parameter  & Value \\[1ex]
\hline
\hline
\rule{0pt}{3ex}
CPU & 1.8 GHz dual core ARM based CPU\\[1ex]
\hline
\rule{0pt}{3ex}
Number of antennas & 4 \\[1ex]
\hline
\rule{0pt}{3ex}
Maximum spatial streams & 3 \\[1ex]
\hline
\rule{0pt}{3ex}
Bandwidth & 160 MHz \\[1ex]
\hline
\rule{0pt}{3ex}
Peak datarate (2.4 GHz) & 600 Mbps \\[1ex]
\hline
\rule{0pt}{3ex}
Peak datarate (5 GHz) & 2.6 Mbps \\[1ex]
\hline
\rule{0pt}{3ex}
RAM & 521MB DDR3 RAM \\[1ex]
\hline
\hline
\end{tabular}}
\end{center} 
\end{table}

\begin{figure}
    \centering
    \includegraphics[width=0.9\linewidth]{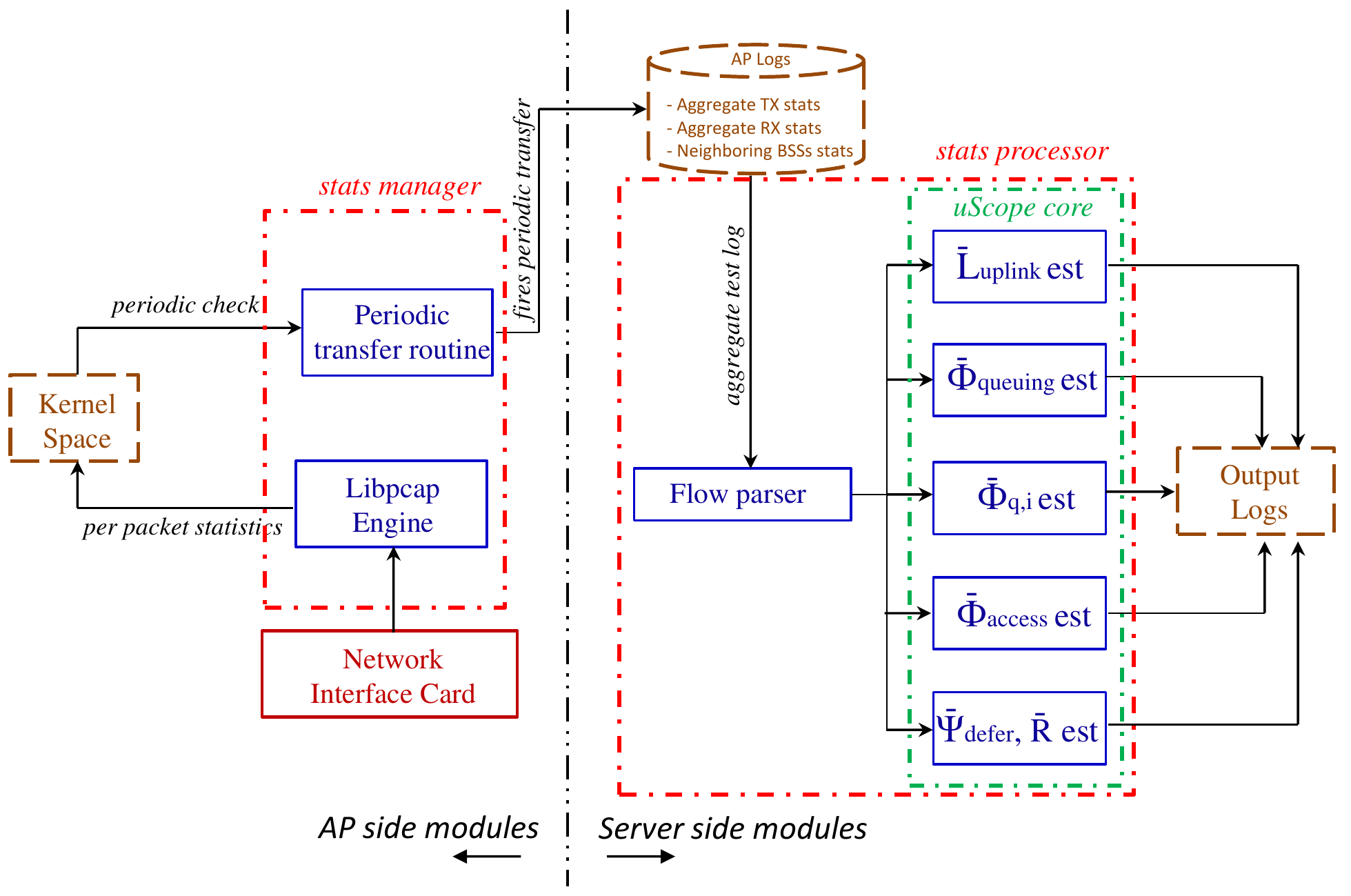}
    \caption{The workflow of commodity hardware based implementation of \technique. The implementation comprises of two key modules: (i) stats manager running on the AP and (ii) stats processor running on a remote server. The stats manager implements key functionalities for recording AP side observations which are sent to the remote server and piped into the \technique core for obtaining the final estimates.}
    \label{sys_arch}
\end{figure}

While the computational capability of the AP is sufficient to support  on board operation of both the stats manager and stats processor, the AP's on board memory becomes a critical bottleneck. Like most commodity hardware platforms, the AP's on board memory is optimized by the vendor to suffice the execution of only legacy functionalities. Consequently, less than 10\% of the memory ($\approx$ 23MB) is available for implementing \technique. To overcome this bottleneck, we offload the stats processor module to a remote server which has sufficient memory for storage of AP logs whereas the stats manager runs on the AP. We further modify the stats manager in the following way. The statistics collected by the libpcap engine are temporarily stored in the kernel space of the AP which is probed by a periodic transfer routine. When a packet log is encountered by the routine, it fires a transfer command which sends the log to the remote server for storage. The estimates from the \technique core are stored in the output log on the remote server. The work flow is illustrated in Fig.~\ref{sys_arch}.

\textbf{STAs.} The STAs are portable laptops running either Windows or Linux operating system. The STAs are upgraded to support IEEE 802.11ac by using an open source IEEE 802.11ac capable Edimax EW-7822ULC Wi-Fi chipset. The radio card has 2 internal antennas and supports communication on both 2.4 and 5GHz bands. The wireless interface has peak rates of 144, 300 and 867 Mbps while using the 20, 40 and 80MHz bandwidth respectively. Web activities on the STAs are performed by using the Mozilla Firefox web browser. 

An instance of the stats manager also runs on the STAs to gather STA side observation log and stores it locally on the device for post-experiment retrieval. While \technique does not require STA side log for making estimations, this information helps in characterization of ground truth which is required for validation of \technique. 

\section{ns-3 Based Simulation Platform}\label{ns3}

ns3 is a discrete event simulator whose source code is open source \cite{ns3rng}. It is one of the most popular tools for performing network simulations \cite{nayak2016performance, nayak2019modeling, ngk} and has rigorously implemented modules for capturing the behavior of the 802.11 MAC and PHY. 

\begin{figure}
    \centering
    \includegraphics[width=0.9\linewidth]{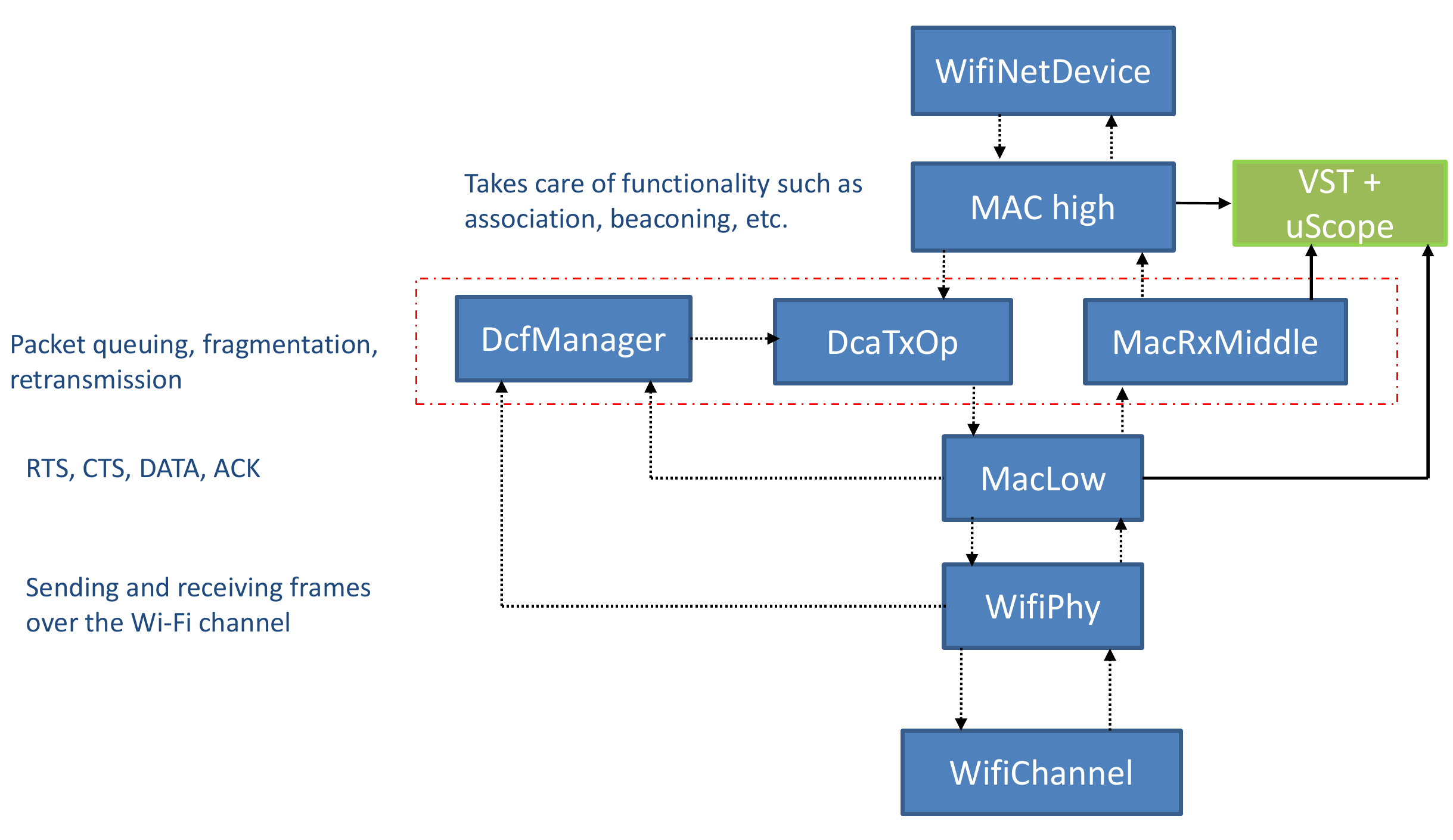}
    \caption{The workflow of the extended ns3 simulator. We extract packet information out of the MAC high, MacRxMiddle and MacLow modules and pass them into the virtual speed test and uScope module.}
    \label{ns3_arch}
\end{figure}

We extend ns3 to include functionalities that enable a validation of both virtual speed test and \technique. Specifically, we extend the MAC high and MAC low modules inside the ns3 core to implement functionalities for virtual speed test and \technique implementation. A workflow of the extended version of the simulator is as shown in Fig.~\ref{ns3_arch}. 

\chapter{Virtual Speed Test Evaluation}

In this chapter, we present the evaluation of virtual speed test. Our experimental methodology is to first evaluate the estimation accuracy of virtual speed test under the variation of one performance factor and then to perform extensive field trials in which a number of these factors co-exist and vary together. In Sec.~\ref{cexp}, we present results from our controlled experiments. This is followed by Sec.~\ref{fieldtrials} which presents results from our in-field trials. 

\section{Controlled Experiments}\label{cexp}

The first factor that we consider for evaluation is network topology. Network topology and especially the presence of hidden terminals has a significant impact on the achievable download and upload speeds. To evaluate the performance of virtual speed test, we set up the following scenario in ns 3. 

\begin{table}
\begin{center}
\vspace{0mm}
\caption{Details of simulation parameters \label{tab:vst_sim}}{%
\scriptsize
\begin{tabular}{||c|p{6.5cm}||}
\hline
\hline
\rule{0pt}{3ex}
Parameter  & Value \\[1ex]
\hline
\hline
\rule{0pt}{3ex}
Wireless standard  & 802.11n\\[1ex]
\hline
\rule{0pt}{3ex}
Rate adapation & Minstrel \\[1ex]
\hline
\rule{0pt}{3ex}
Frequency & 2.4 GHz \\[1ex]
\hline
\rule{0pt}{3ex}
AP antenna count & 1 \\[1ex]
\hline
\rule{0pt}{3ex}
STA antenna count & 1 \\[1ex]
\hline
\hline
\end{tabular}}
\end{center} 
\end{table}

We consider multiple STAs deployed on a circle with a fixed radius around the Access point. Both the AP as well as the STAs have a single antenna. Neither the AP nor the STAs perform any frame aggregation and all nodes are static. In this case, we chose one STA as the target STA and the remaining STAs as non-target STAs. The target STA performs a TCP download from a server with a one way backbone delay of 5 ms to the AP. Using the TCP handshakes from this download, the AP makes an estimate using the L2 Edge TCP model. The non-target STAs have fully backlogged uplink UDP flows. Note that the nature of traffic does not have an impact on the estimation error. Following, the estimate, we re-run the simulation with the same background traffic but with a \textit{speed test-like} flow from a server with a one way latency of 1 ms (low latency server) from the AP. This flow emulates the flooding based mechanism of speed test by having 10 parallel flows with each flow jump started to a congestion window size of 100. The TCP throughput reported by the STA for this flow after the simulation completes is considered as the ground truth. We summarize the simulation parameters used for this and the remaining simulations in Table.~\ref{tab:vst_sim}. The STAs do not perform any frame aggregation in this case. 

\begin{figure}
\centering
\begin{subfigure}[t] {0.50\textwidth}
\centering
\includegraphics[width=0.95\textwidth]{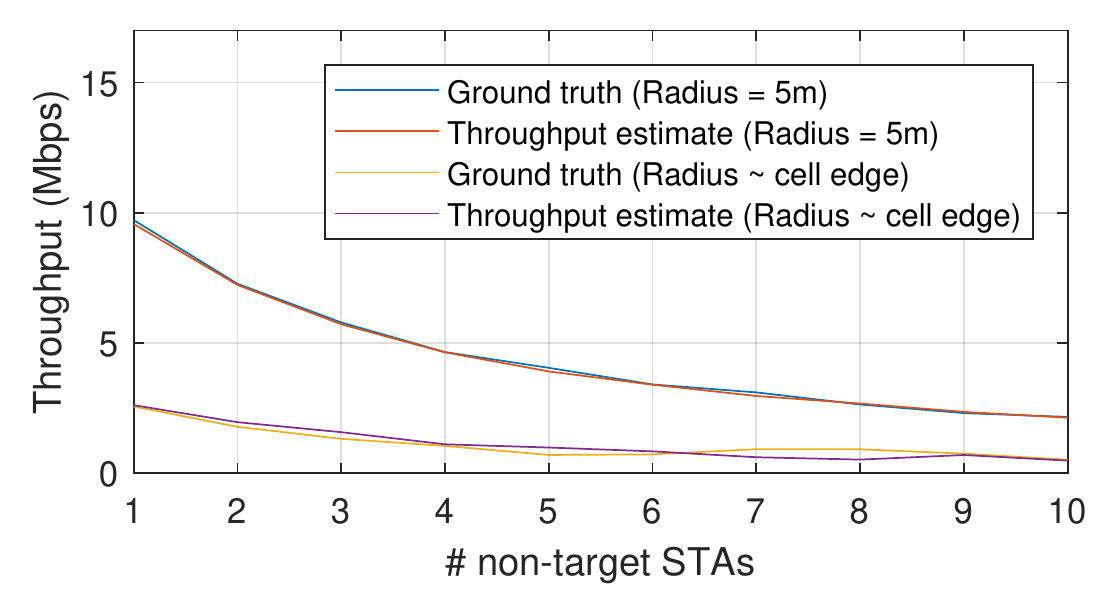}
\caption{Download speed test throughput estimation}
\end{subfigure}
\begin{subfigure}[t] {0.50\textwidth}
\centering
\includegraphics[width=0.95\textwidth]{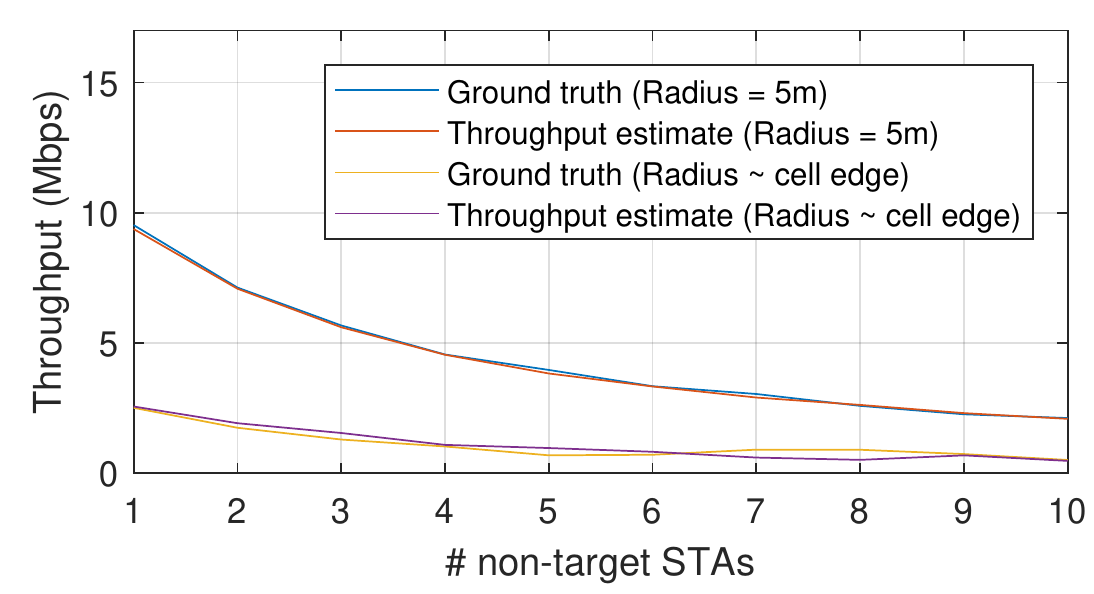}
\caption{Upload speed test throughput estimation}
\end{subfigure}
\caption{Virtual speed test performance in the presence and absence of hidden terminals and with an increasing STA density.}
\label{vst_net_top}
\end{figure}

We first consider a case in which the value of the radius is 5m. In this case, all nodes can carrier sense each other and no hidden terminals are present. Fig.~\ref{vst_net_top} shows the results of the speed test flow in this scenario. As expected, the ground truth value of download and upload speeds decays with an increasing number of STA. Next, we consider the scenario wherein the radius is approximately equal to the cell edge radius. In this case, each additional STA (non-target) added to the network is a hidden terminal. In the case of both download as well as upload, the addition of hidden terminals results in a significant throughput degradation in comparison to the corresponding case without hidden terminals. 

An increase in the number of hidden terminals causes an increase in the collision rate for the STA's transmissions on the uplink. However, recall that the delay caused by retransmissions is accounted for in the uplink channel access time parameter in the model. Consequently, the impact of hidden terminals gets captures in the estimates and the virtual speed test estimates for the download and upload speeds demonstrate a close match with the ground truth values.  This demonstrates the effectiveness of virtual speed test in the presence of hidden terminals as well as with an increasing STA density. 

TCP ACK thinning mechanism enables a TCP receiver to reduce the number of TCP ACKs generated by generating an ACK for every $n^{th}$ TCP segment. Typically, while operating at maximum segment size, $n = 2$ \cite{nayak2016performance, braden1989rfc, nayak2019modeling, ngk}. Note, however, that
massive use of delayed ACK techniques (beyond the standard
TF = 2) has detrimental effects to TCP \cite{miorandi2006queueing}, and would
require sophisticated cross-layer design to be implemented in
a WLAN. 

\begin{figure}
\centering
\begin{subfigure}[t] {0.50\textwidth}
\centering
\includegraphics[width=0.95\textwidth]{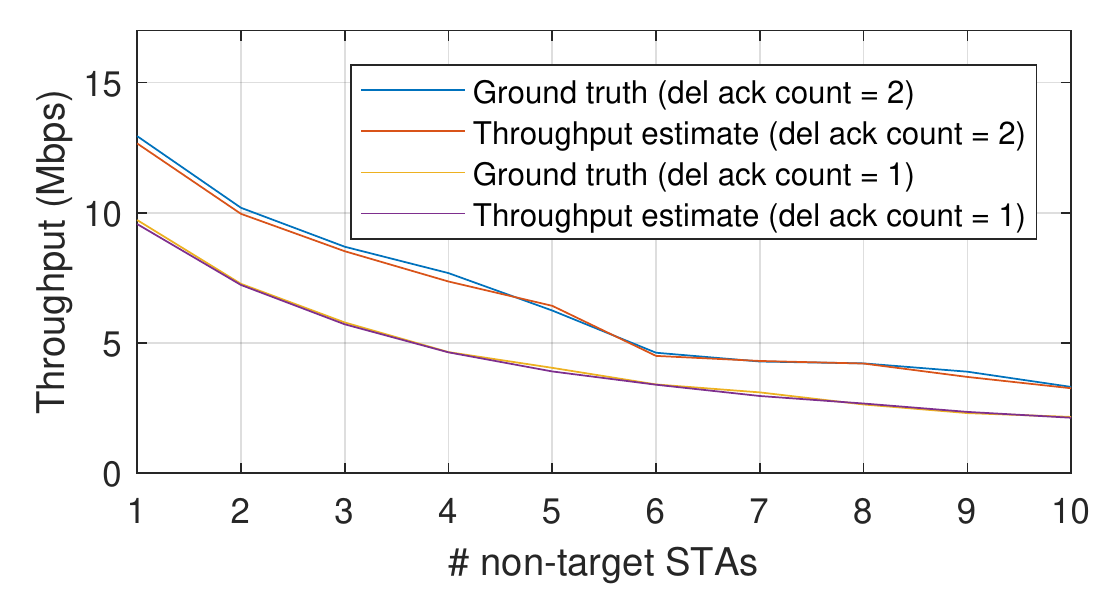}
\caption{Download speed test throughput estimation}
\end{subfigure}
\begin{subfigure}[t] {0.50\textwidth}
\centering
\includegraphics[width=0.95\textwidth]{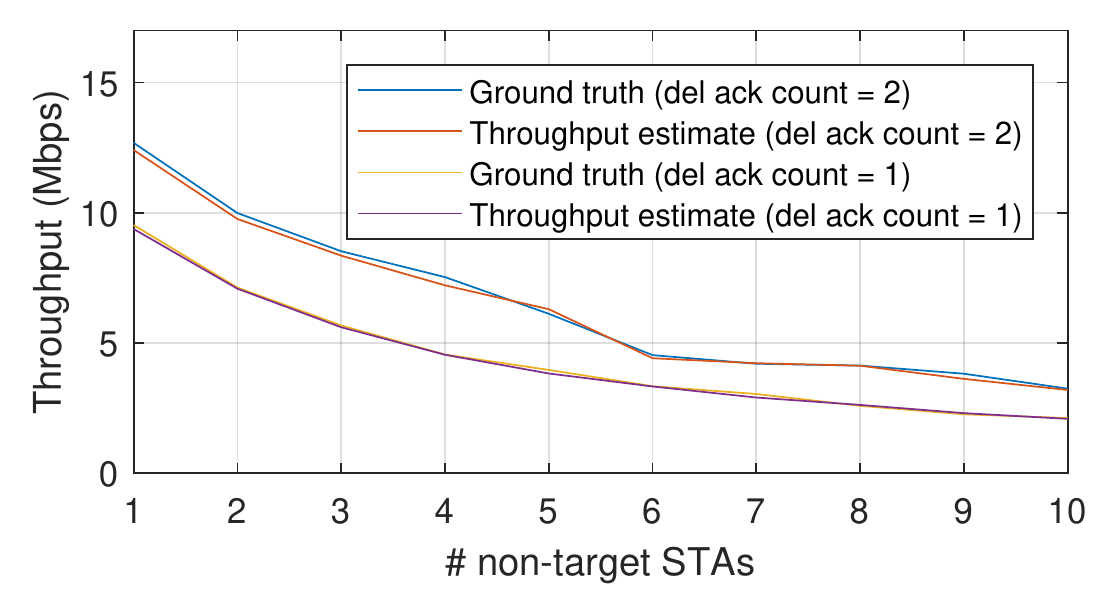}
\caption{Upload speed test throughput estimation}
\end{subfigure}
\caption{Estimation accuracy of virtual speed test in the presence of TCP ACK thinning mechanism.}
\label{del_ack}
\end{figure}

It is important that when TCP uses the TCP ACK thinning feature, virtual speed test can still make estimates accurately. Fig.~\ref{del_ack} shows the estimates and the corresponding ground truth values when the ACK thinning ratio is kept at 1 and 2. Recall from the L2 Edge TCP model described previously that framework accounts for ACK thinning by reducing the number of jobs circulating in the virtual network and instead further inflating the service times for the missing jobs. Consequently, as shown in Fig.~\ref{del_ack}, a variation in the ACK thinning ratio does not affect the estimation accuracy of virtual speed test. 

\begin{figure}
\centering
\begin{subfigure}[t] {0.50\textwidth}
\centering
\includegraphics[width=0.95\textwidth]{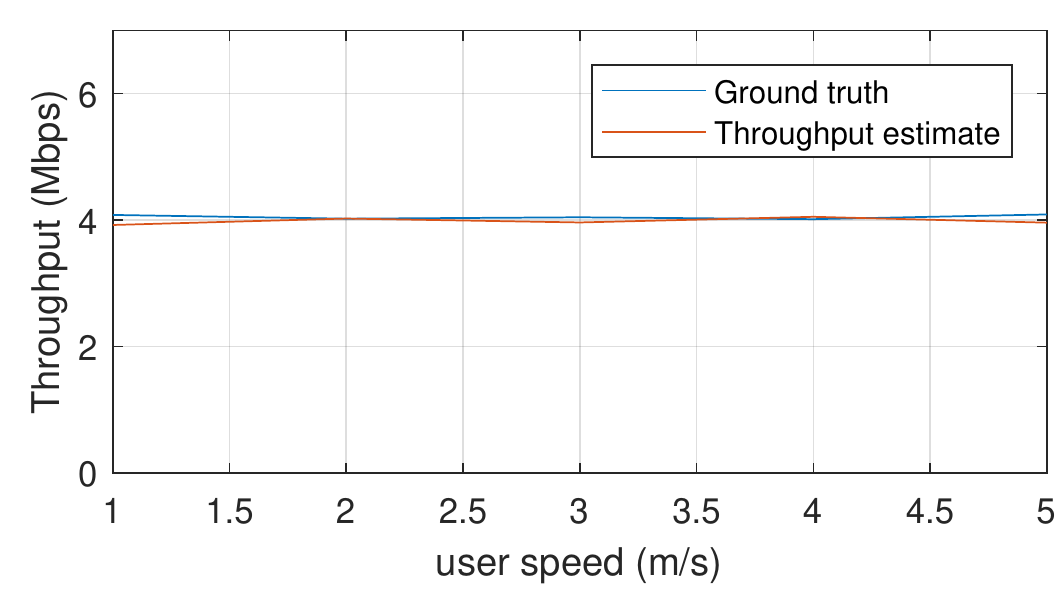}
\caption{Download speed test throughput estimation}
\end{subfigure}
\begin{subfigure}[t] {0.50\textwidth}
\centering
\includegraphics[width=0.95\textwidth]{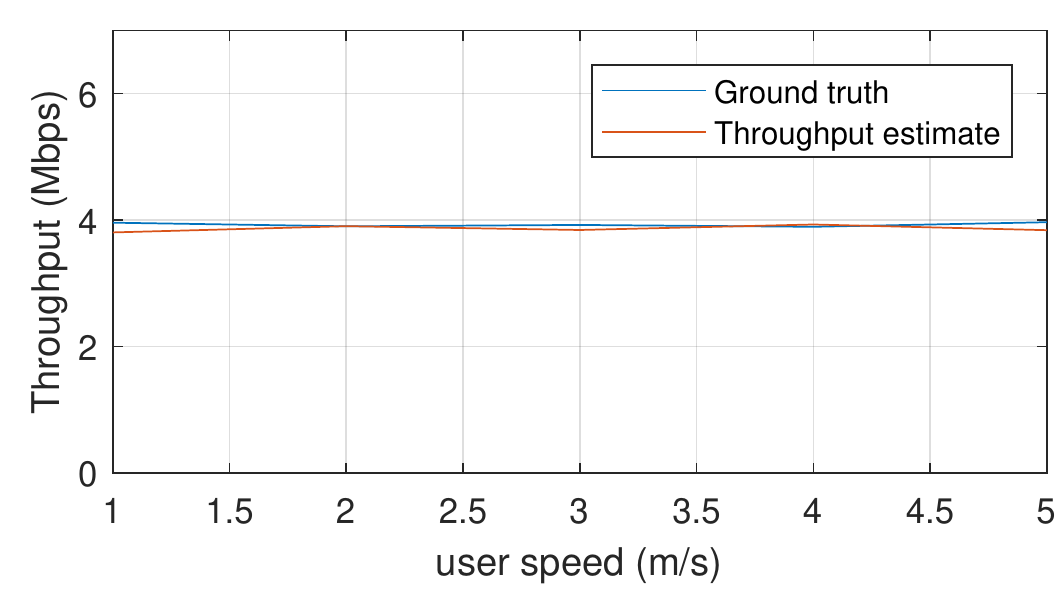}
\caption{Upload speed test throughput estimation}
\end{subfigure}
\caption{Validation of virtual speed test in the presence of node mobility.}
\label{node_mobility}
\end{figure}

Typical WLANs involve presence of mobile users moving at indoor human mobility speeds. Consequently, a speed estimation tool such as virtual speed test much estimate download and upload speeds accurately in the presence of node mobility. Next, we evaluate the performance of virtual speed test in the presence of node mobility. We consider light indoor mobility speeds ranging between walking speeds of 1 m/s to maximum sprinting speeds of up to 5 m/s. Mobility introduces packet loss and forces the devices to perform rate adaptation. Interestingly, we notice from Fig.~\ref{node_mobility}, user mobility has a very negligible effect on the ground truth value of download and upload speeds. Further, virtual speed test demonstrates a high level of accuracy in the estimation for both download as well as upload speed estimations.  

\section{In-field Trials}\label{fieldtrials}
Next, we perform an extensive set of field trials with our commodity hardware based testbed. These field trials are conducted in the residential and office environment. Below, we provide details of our field trials. 

\subsection{Deployment Characterization}
To understand the estimation accuracy of virtual speed test, we deploy the VST AP and STAs in two environments.    

\textbf{Deployment description.} The first deployment is in a 5m x 3m student office located in a 3 storied building on a University campus. In this deployment, the VST AP and the STAs co-exist with a University administered enterprise network and 2 APs deployed in nearby student offices. Here the VST AP has 6 STAs associated with it. We deploy a second network in a 3 storied residential building primarily consisting of apartments with 1 or 2 bedrooms per unit. This residential network consists of 7 STAs. These devices co-exist with 4 APs from neighboring apartments. The VST AP deployed in the office performs measurements for a period of two days whereas the residential scenario measurements are carried out for a period of one week. During the entire duration of deployment, the VST AP observed a total of 113,047 snooped flows. These flows are the result of multiple applications running on end devices such as video streaming (via YouTube), music streaming (via Pandora), pdf downloads (via IEEE Xplore) and sending and receiving emails (via Gmail). The STAs use Mozilla Firefox web browser for performing these online activities. The traffic flows consist of single application traffic where each STA runs only one web application as well as a mix of web applications running concurrently. Aside from the applications mentioned above, some of the STAs have Dropbox installed on them which occasionally adds to the uplink traffic from these devices in addition to that generated by email activities. We do not suppress any control packet transmissions at any of the layers.  

\begin{figure}
\centering
\begin{subfigure}[t] {0.45\textwidth}
\centering
\includegraphics[width=1\textwidth]{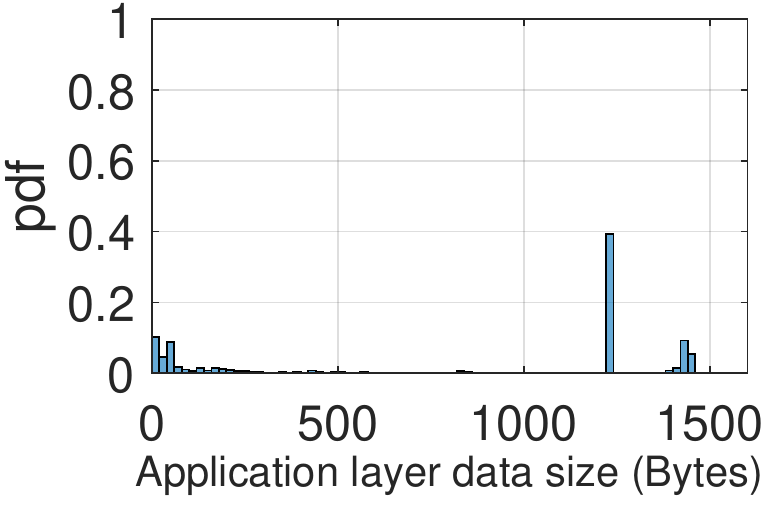}
\caption{Office}
\end{subfigure}
\begin{subfigure}[t] {0.45\textwidth}
\centering
\includegraphics[width=1\textwidth]{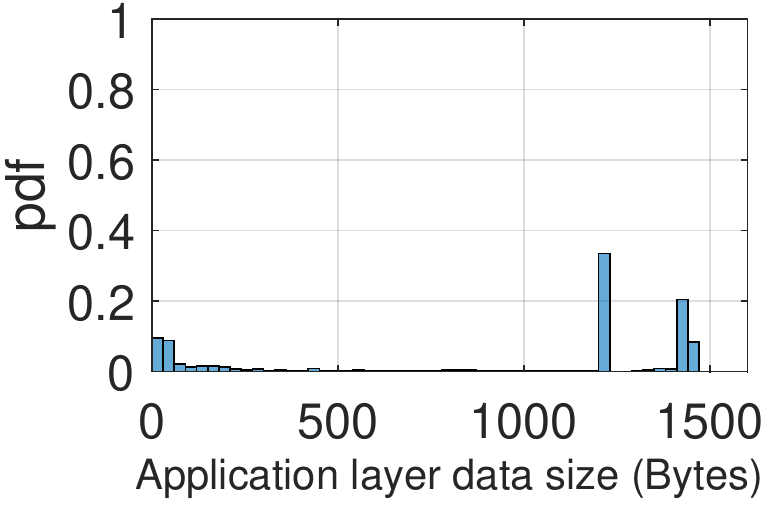}
\caption{Residential}
\end{subfigure}
\caption{pdf of application layer data sizes encountered in the office and residential deployment.}
\label{app_size_dist}
\end{figure}

The ground truth values of the L2 edge TCP model parameters are affected by operating conditions determined by the traffic characteristics, active STA count and the MAC and PHY statistics. It is important that virtual speed test is able to estimate both the upload and download throughputs accurately despite a variation  in these factors.

\begin{figure}
\centering
\begin{subfigure}[t] {0.30\textwidth}
\centering
\includegraphics[width=1\textwidth]{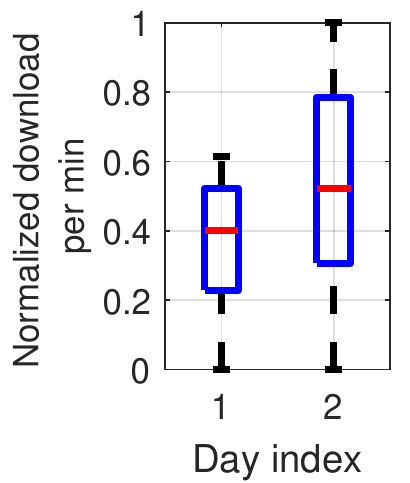}
\caption{Office}
\end{subfigure}
\begin{subfigure}[t] {0.45\textwidth}
\centering
\includegraphics[width=1\textwidth]{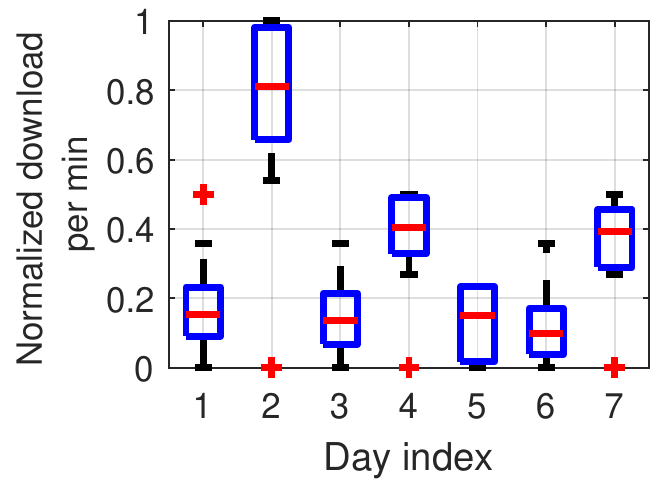}
\caption{Residential}
\end{subfigure}
\caption{Variation in the amount of data downloaded per minute in the office and residential deployment. The values are normalized with respect to the overall maximum encountered in that particular deployment.}
\label{download_rate}
\end{figure}

\textbf{Traffic statistics.} Packet size variation causes a microscopic fluctuation in the air time utilization for non-target STAs and transmit time for the target STAs. Fig.~\ref{app_size_dist} shows the application layer data size distribution for both the scenarios. The minimum size encountered in the trace is 20 bytes whereas the maximum size is 1.4K bytes. In addition to this, a fluctuation in the per minute download statistics as shown in Fig.~\ref{download_rate} causes a variation in the network load affecting queuing delays and idle times for both the target and non-target STA.

\textbf{Active STA count diversity.} A variation in the number of active STAs in the network affects ground truth value of $\usage$ and the `access' parameters. In both deployments, the number of active STAs is varied to cover a total of 36 and 49 combination of STA sets in the office and residential scenario respectively. In these combinations, active STA sets of all possible sizes are covered and the VST AP makes predictions for each device in the set.

\begin{figure}
\centering
\begin{subfigure}[t] {0.50\textwidth}
\centering
\includegraphics[width=1\textwidth]{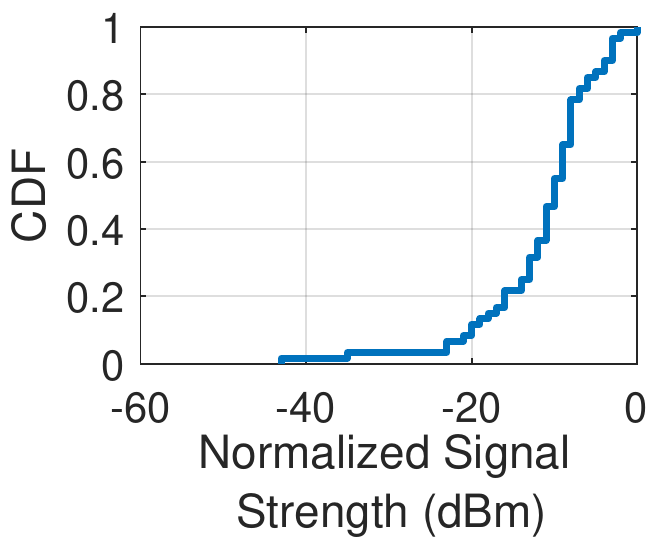}
\caption{Office}
\end{subfigure}
\begin{subfigure}[t] {0.50\textwidth}
\centering
\includegraphics[width=1\textwidth]{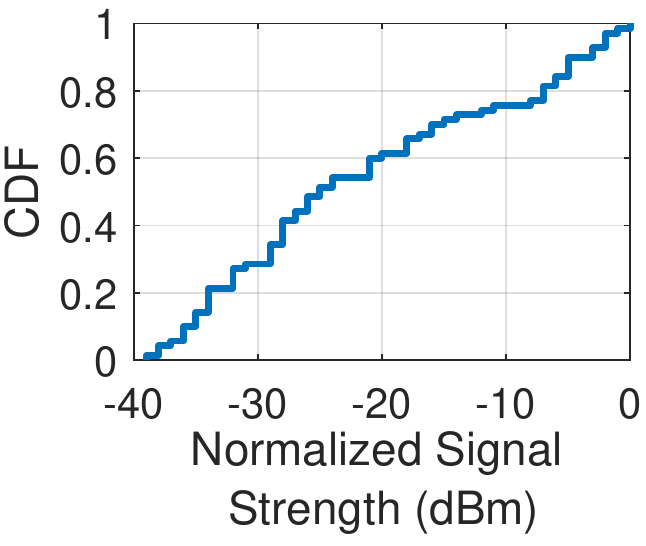}
\caption{Residential}
\end{subfigure}
\caption{Distribution of signal strength of the VST AP across all the STA positions in the office and residential deployments. The values are normalized with respect to the maximum signal strength encountered in that particular deployment.}
\label{ss_dist}
\end{figure}

\textbf{MAC and PHY statistics.} Link diversity in terms of signal propagation characteristics (due to LoS and non-LoS links to the VST AP), overheard transmissions from neighboring BSSs, active STA count variation and traffic statistics mentioned above result in a variation in MAC and PHY statistics across devices.  Fig.~\ref{ss_dist} shows the normalized signal strength distribution of the VST AP across all STA locations in the office and residential scenario. The signal strength distribution is normalized with respect to the maximum VST AP  signal strength encountered in that environment across all STA positions. Due to the small size of the office, the received signal strength of the VST AP across all STA locations is very close to the encountered maximum. In comparison, the residential scenario exhibits a wide distribution of the VST AP's signal strength across the apartment resulting in a variation in supported PHY rates across different STA positions which affects the `tx' parameters. 

\begin{figure}
    \centering
    \includegraphics[width=0.5\linewidth]{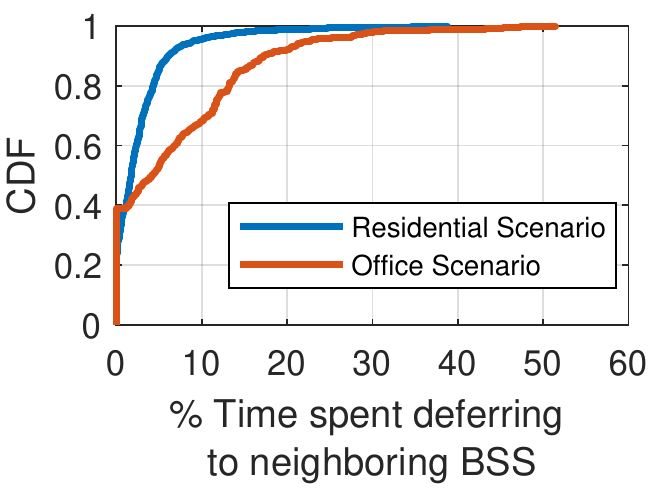}
    \caption{Distribution of fraction of a TCP flow's total duration that a device spends deferring to transmissions from neighboring BSS}
    \label{defer_stats}
    \vspace{-0.9cm}
\end{figure}

Moreover the multi-AP nature of these experiments results in devices in the network deferring to transmissions from co-existing BSS thereby affecting the `access' parameters. Fig.~\ref{defer_stats} depicts a distribution of the fraction of time during a download that a device spends deferring to transmission from neighboring BSS in both the office and residential scenario. The activity in the neighboring BSS varied in both the deployments. For instance, during approximately 30\% of the flows in both the office and the residential scenario, there was no activity on the neighboring BSS. On the other hand, the maximum portion of a TCP flow's duration that a STA spent deferring to neighboring BSS is 38.8\% for the residential scenario and 51.4\% in the office scenario. 

All the above environment characteristics result in a variation in all the model parameters which the VST AP intends to estimate for the purpose of throughput prediction.

\subsection{Ground Truth Procurement}
We compare the outcome of virtual speed test with that obtained from online speed test applications~\cite{ooklaspeedtest,attspeedtest,xfinity} as well as iperf. We focus on iperf for procuring ground truth as online tests are subject to an additional and uncontrolled source of variation due to server selection. Namely, recall that during the setup phase, the server selection process of these speed test applications tries to minimize the backbone delay between the server and the AP with the ideal case being a server in the same LAN as the AP. In practice, a server within the same LAN may not be available in the server pool probed by these applications. Consequently, results obtained from a non-local server may be affected by backbone load and delay. 

\begin{figure}
\centering
\begin{subfigure}[t] {0.50\textwidth}
\centering
\includegraphics[width=1\textwidth]{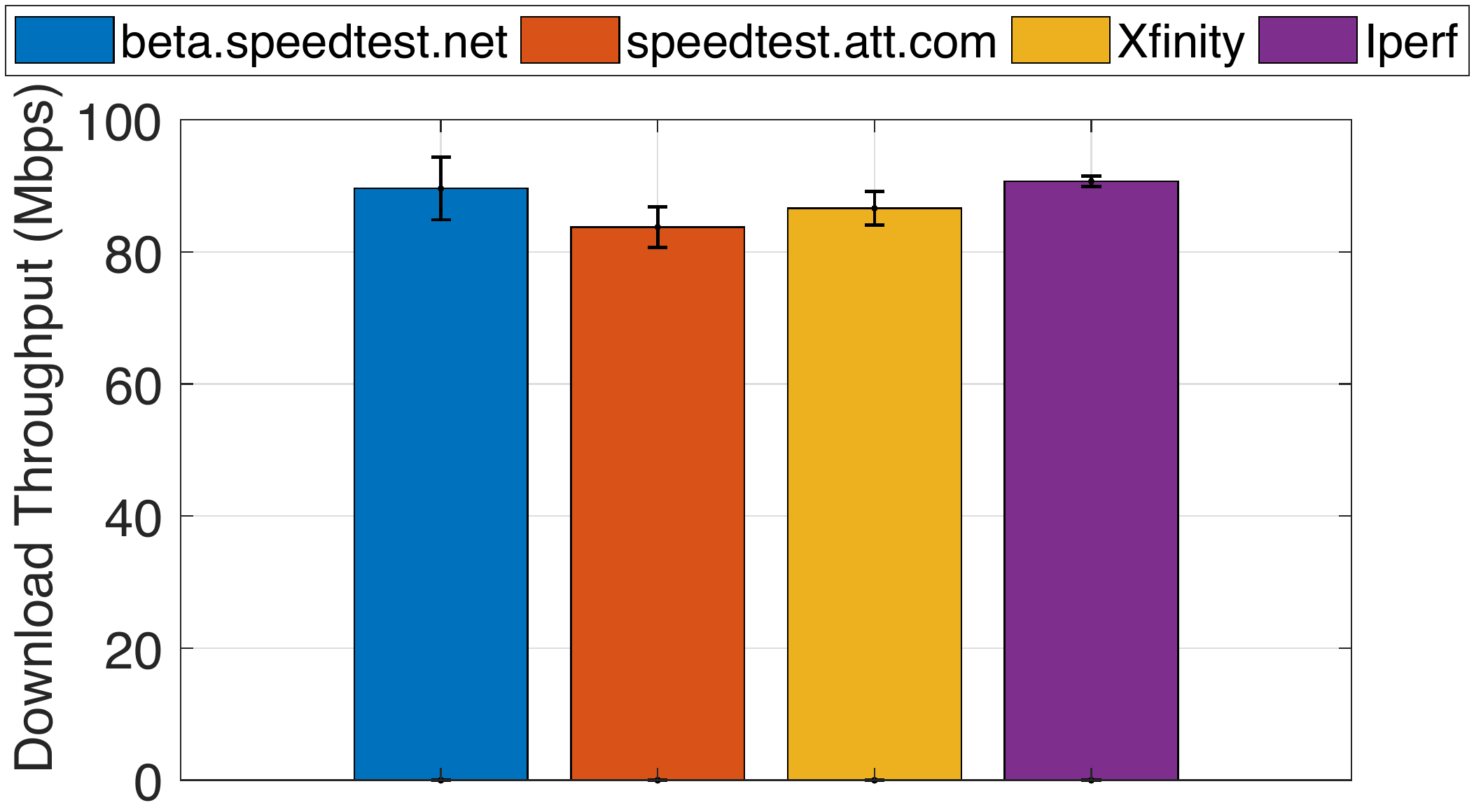}
\caption{Download speed test}
\end{subfigure}
\begin{subfigure}[t] {0.50\textwidth}
\centering
\includegraphics[width=1\textwidth]{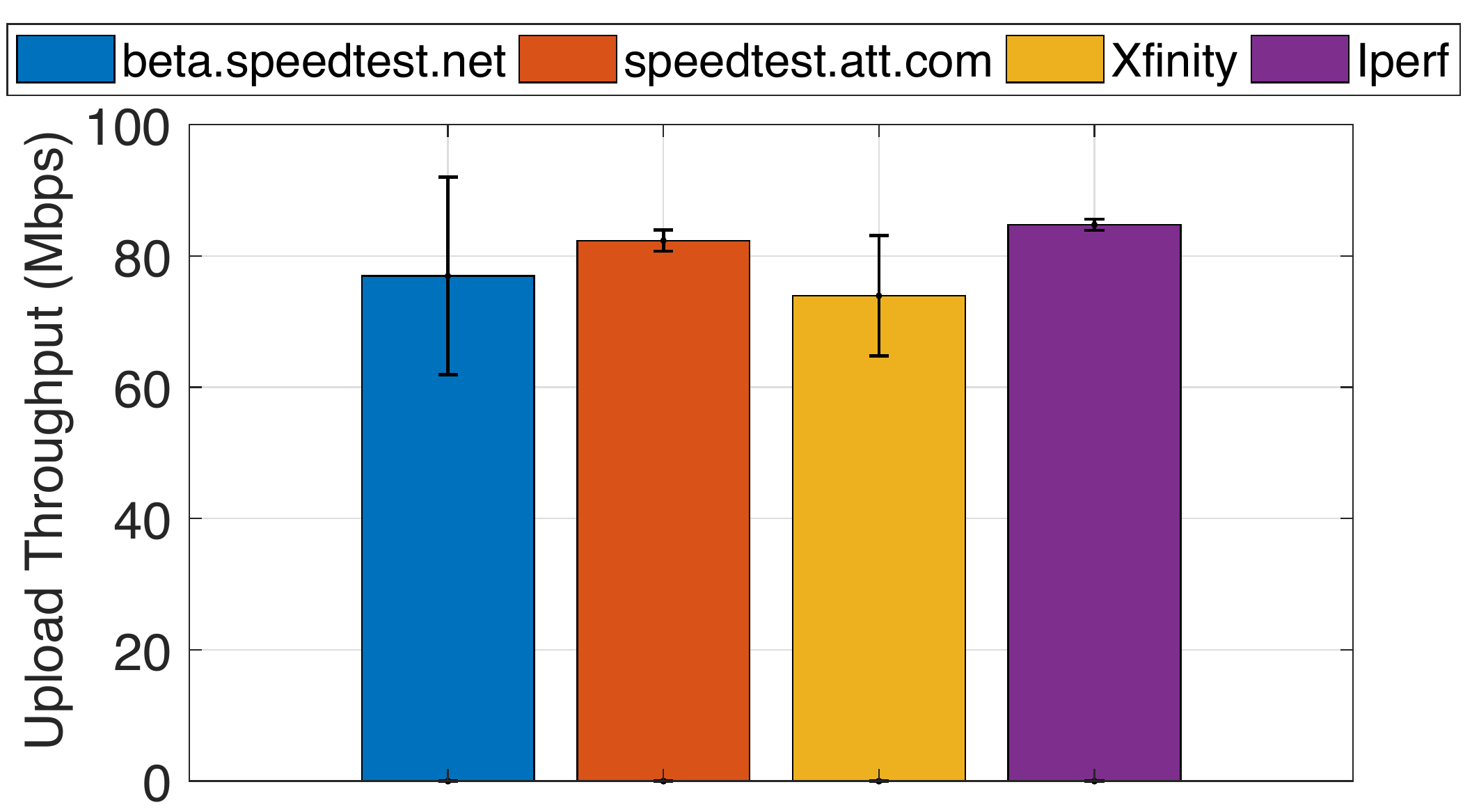}
\caption{Upload speed test}
\end{subfigure}
\caption{Iperf results as compared to exemplary online speed test applications.}
\label{st_comp}
\end{figure}

To explore this effect, we keep one 802.11n laptop associated with the AP and run multiple speed test applications 10 times with a gap of 10 mins between each run. We  attempt to realize static WLAN conditions, \emph{i.e.}, with minimal activity on co-existing BSS, no device or environmental mobility, etc., in which the outcome of speed test applications should remain stable. We compare these results with those obtained by running iperf from a local server to remove backbone and server selection effects. Fig.~\ref{st_comp} shows the values obtained for upload and download throughput from exemplary online speed test applications as well as iperf. Compared to iperf, the speed test applications demonstrate throughput reductions and fluctuations due to backbone and server selection effects. On the other hand, the values obtained from iperf remain stable. Therefore, to obtain stable and more reliable ground truth values in our experiments, we use the iperf tool. 

\subsection{Experimental Methodology}
In our experiments, virtual speed test throughput estimates are obtained solely via AP measurements, \textit{i.e.}, by using AP-snooped TCP data - ACK handshakes from existing network traffic to estimate the parameters required by the L2 edge TCP model. Immediately following each throughput estimation, the ground truth is measured via iperf on instrumented clients. While network traffic  is composed of both download and upload traffic, virtual speed test only leverages download flows for $\uldelay$ estimation.

\subsection{Parameter Estimation Accuracy}
Throughput and virtual service time are impacted by the uplink access time. Recall that this model parameter is not directly observable at the AP. Rather, the VST AP leverages snooped handshakes from TCP flows existing in the network to estimate uplink access time. In our deployment, the VST AP observes cases involving only single application traffic as well as those with mixed application traffic involving concurrent TCP flows with a mix of downloads and uploads. To correctly estimate the uplink access time, the VST AP needs to first identify TCP flows based on the address and size signatures and then accurately identify the occurrence of the ACK queuing and ACK immediate regimes. We first explore the efficacy of the $\uldelay$ estimation method by comparing the estimates made by the VST  AP using snooped flows with the ground truth values obtained from the STA-side. 

\begin{figure}[]
    \centering
\begin{minipage}[t]{1\linewidth}
    \begin{subfigure}[t]{0.4\textwidth}
        \centering
		\includegraphics[width=\textwidth]{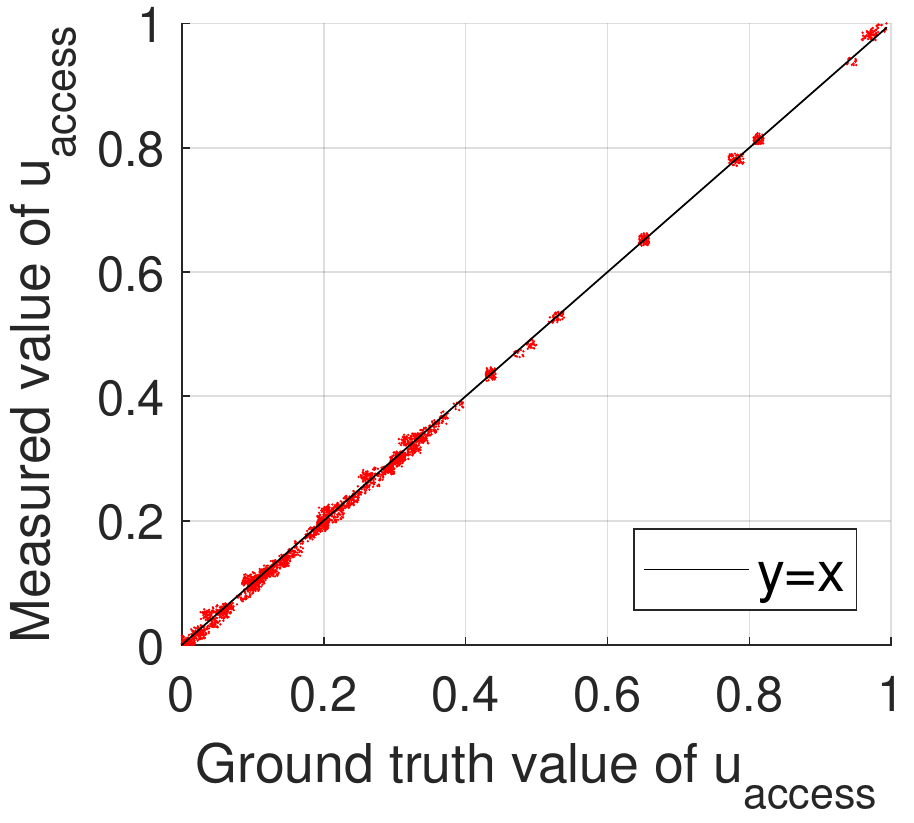}
        \caption{Office Scenario}
    \end{subfigure}
    \hspace{1cm}
    \begin{subfigure}[t]{0.4\textwidth}
        \centering
        \includegraphics[width=\textwidth]{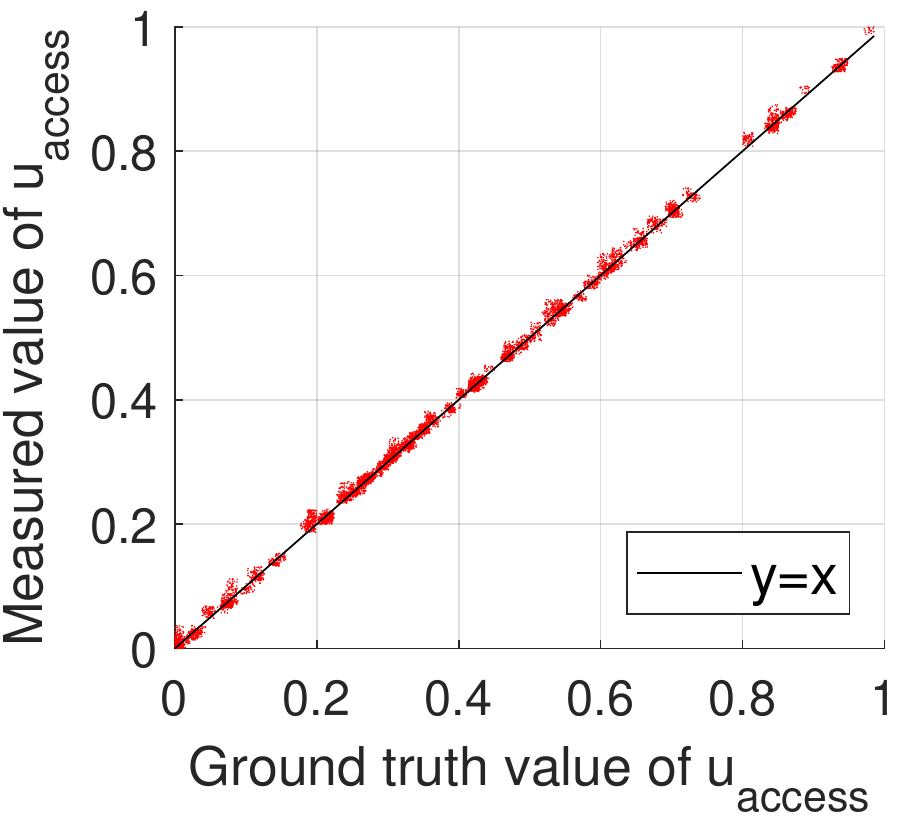}
        \caption{Residential Scenario}
    \end{subfigure}
    \vspace{-0.2cm}
    \caption{Scatter plot of $\uldelay$ measured and ground truth value in the office and residential deployments. In each sub-figure, both the axis are normalized with respect to the maximum ground truth value in that particular deployment.}
    \label{u_access}
    \end{minipage}
\vspace{-0.4cm}
\end{figure}

Fig.~\ref{u_access} shows a scatter plot of measured values of $\uldelay$ vs. the ground truth value. For each deployment scenario, both of these values are normalized with respect to the maximum ground truth value in that particular scenario. Fig.~\ref{u_access} shows that as the ground truth values of $\uldelay$ varies due to numerous factors such as background traffic characteristics, active STA count, etc., the points in the scatter plot continue to be closely located to the identity line. Fig.~\ref{u_access} also demonstrates the existence of homoscedasticity between estimates obtained by VST AP and the ground truth value of $\uldelay$ obtained from the STA-side. In essence, this shows that as the ground truth value of $\uldelay$ varies due to the numerous factors mentioned previously, the estimated value of $\uldelay$ retains its accuracy.

\subsection{Throughput Estimation Accuracy}
The ultimate goal of virtual speed test is to estimate $\sdl$ and $\sul$. To evaluate throughput estimation accuracy, we calculate the percent estimation error between the ground truth values obtained from iperf and the estimates of virtual speed test as $\%error = \frac{|(\theta_{\textrm{\tiny gt}} - \theta_{\textrm{\tiny est}})|*100}{\theta_{\textrm{\tiny gt}}}$ where $\theta_{\textrm{\tiny gt}}$ denotes the ground truth value of throughput and $\theta_{\textrm{\tiny est}}$ denotes the estimated value from virtual speed test. We remark that since the $\%error$ is calculated after weighing by $\theta_{\textrm{\tiny gt}}$, it is sensitive to the value of $\theta_{\textrm{\tiny gt}}$, \emph{i.e.}, the same absolute error at a smaller ground truth value is bound to yield higher estimation error. 

\begin{figure}[]
    \centering
\begin{minipage}[t]{1\linewidth}
    \begin{subfigure}[t]{0.4\textwidth}
        \centering
		\includegraphics[width=\textwidth]{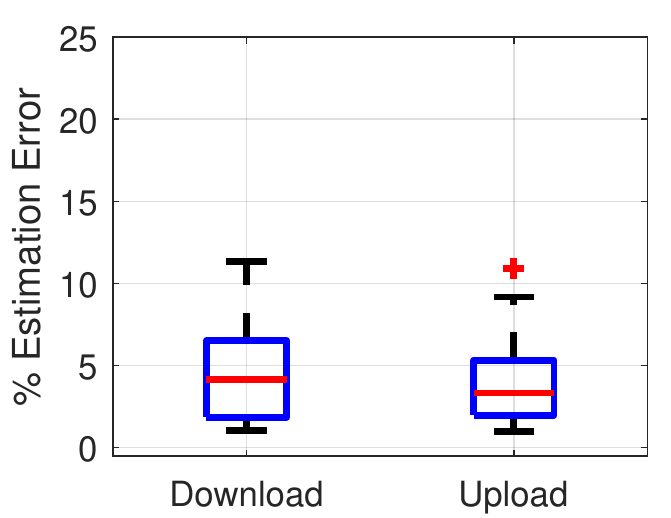}
        \caption{Office Scenario}
    \end{subfigure}
    \hspace{1cm}    
    \begin{subfigure}[t]{0.4\textwidth}
        \centering
        \includegraphics[width=\textwidth]{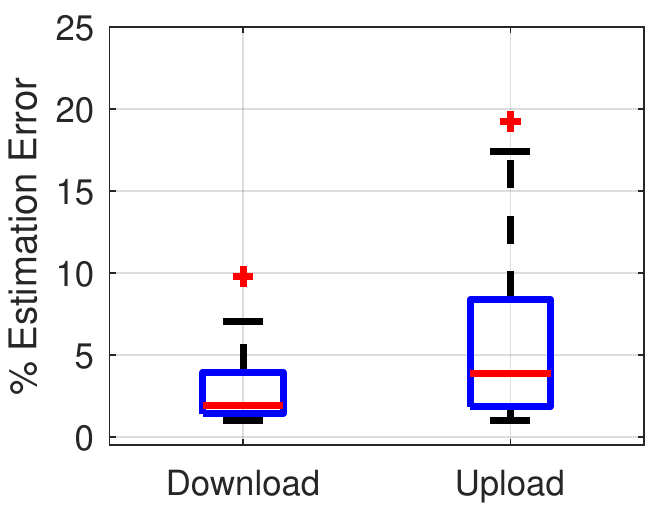}
        \caption{Residential Scenario}
    \end{subfigure}
    \vspace{-0.2cm}
    \caption{Percent throughput estimation error statistics in the office and residential deployments. In the table, GT denotes the ground truth value and Est denotes the throughput estimated by virtual speed test.}
    \label{throughput_error}
    \end{minipage}
\vspace{-0.4cm}
\end{figure}

Fig.~\ref{throughput_error} summarizes the percent throughput estimation error statistics in both the office and residential deployments. Overall virtual speed test shows a good match against ground truth values with a mean percentage error of 4.09\% and 4.3\% for upload and download speeds respectively in the office scenario. In the residential scenario, these values are 5.51\% and 2.9\% respectively. 

\subsection{Observation Set Size Characterization}
\begin{figure}[]
    \centering
\begin{minipage}[t]{1\linewidth}
    \begin{subfigure}[t]{0.49\textwidth}
        \centering
		\includegraphics[width=\textwidth]{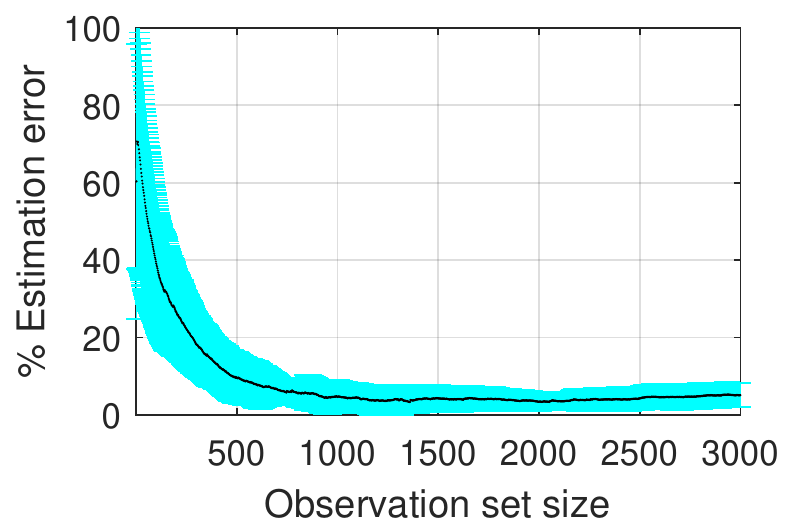}
        \caption{Office: Download}
    \end{subfigure}
    \begin{subfigure}[t]{0.49\textwidth}
        \centering
        \includegraphics[width=\textwidth]{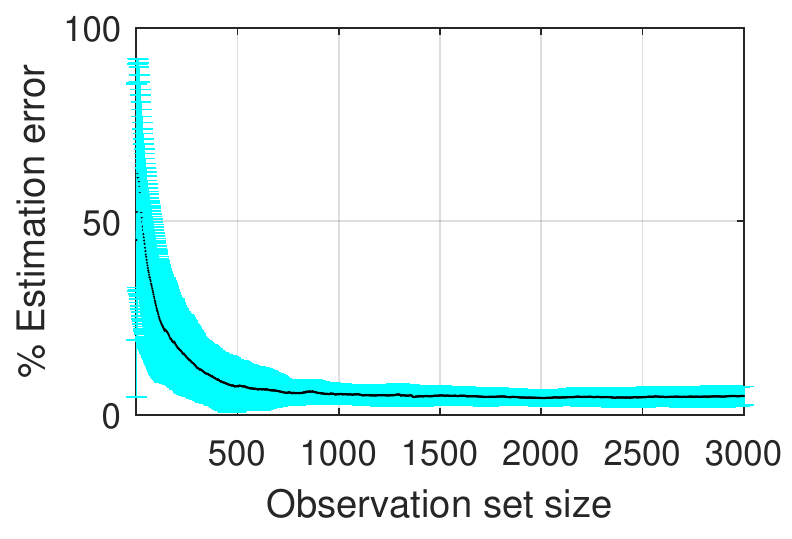}
        \caption{Office: Upload}
    \end{subfigure}
    \vspace{-0.3cm}
    \begin{subfigure}[t]{0.49\textwidth}
        \centering
        \includegraphics[width=\textwidth]{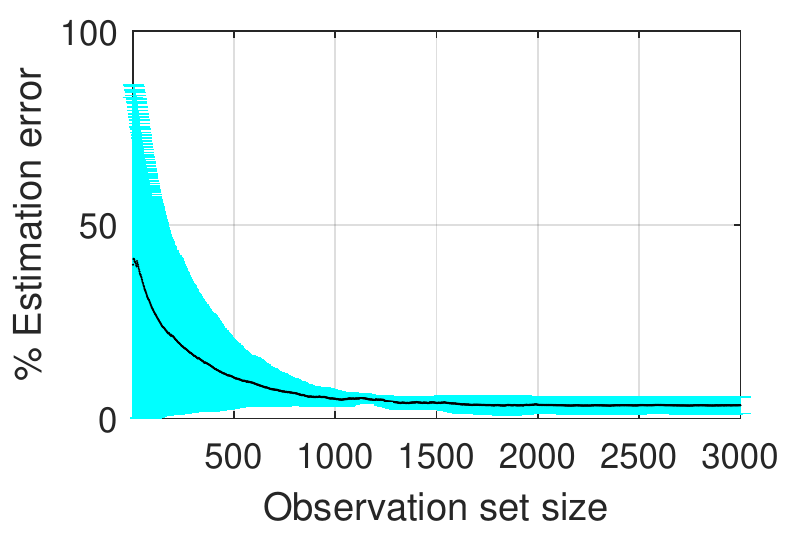}
        \caption{Residential: Download}
    \end{subfigure}
    \begin{subfigure}[t]{0.49\textwidth}
        \centering
        \includegraphics[width=\textwidth]{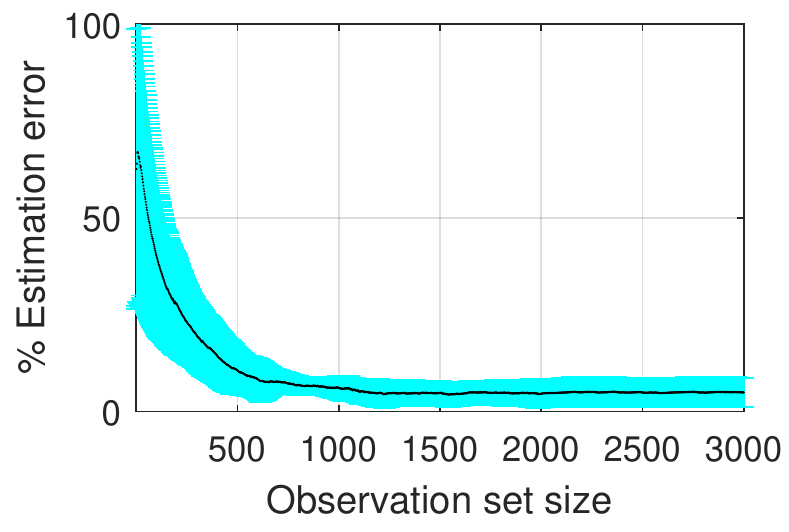}
        \caption{Residential: Upload}
    \end{subfigure}
    \caption{Throughput estimation error as a function of the observation set size available at the AP}
    \label{download_size}
    \end{minipage}
\vspace{-0.6cm}
\end{figure}

Virtual speed test estimates the target STA's throughput by collecting measurements from its TCP traffic. Therefore, a key factor that determines the estimation accuracy of virtual speed test is the number of measurements that the VST AP can obtain from the target STA's TCP traffic which in turn depends on the target STA's online activity. On one extreme, a large number of observations obtained from either a single large file download or a series of consecutive small file download results in refined parameter estimation. We experimentally investigate the observation set size required to avoid a deterioration in estimation accuracy as follows. Since the backbone delay between the snooped flow's server and the VST AP is inconsequential, we initiate a single TCP flow from a local server to the target STA and treat it as a snooped flow. We control the observation set size available at the VST AP by changing the download file size and computing the percent throughput estimation error as before.  

Fig.~\ref{download_size} depicts the mean estimation error as a function of the size of the observation set available at the VST AP. The mean estimation error demonstrates an exponential decrease with an increasing set size. When the sample set size is on the order of a few 10s of samples, the estimation error is as high as 70\%. The variance in this case is also very high. This is due to the fact that the number of measurement samples are insufficient to accurately characterize the model parameters. With an increasing sample set size, the precision in the model estimation increases as expected. Fig.~\ref{download_size} reveals that even when the observation set contains a few 1000s of samples (which could easily be obtained from observing the download of a research paper from IEEE Xplore), the mean estimation error of virtual speed test is under 5\% in both the deployments.

\chapter{uScope Evaluation}

In this chapter, we show the evaluation of \technique under the impact of numerous factors such as network topology, device mobility, environmental mobility, etc. First we show controlled experiments in Sec.~\ref{control} followed by in-field trials experiments in Sec.~\ref{field}.

\section{Controlled Experiments}\label{control}
The goal of \technique is to estimation uplink WLAN latency and decompose it into its constituent components. Specifically, \technique enables an AP-side estimation of queuing delays, channel access delays, defer delays and the retransmission rate for the STA on the uplink. Our experimental approach is to validate the estimation accuracy of \technique in the presence of factors that cause a variation in these parameters.

We begin by evaluating the estimation accuracy of \technique in the presence of a variation in queuing delays. In this experiment, we keep one STA associated with the AP. The STA performs a TCP download from a server with a one way latency of 10 ms. It is important to note that the value of the backbone latency does not affect the ground truth or the parameter estimate for the uplink. In addition to this flow, the STA has one additional uplink UDP flow which pushes packets into the STA queue. The packet generation rate of this uplink UDP flow is controlled in the experiment.  

\begin{figure}
    \centering
    \includegraphics[width=0.5\linewidth]{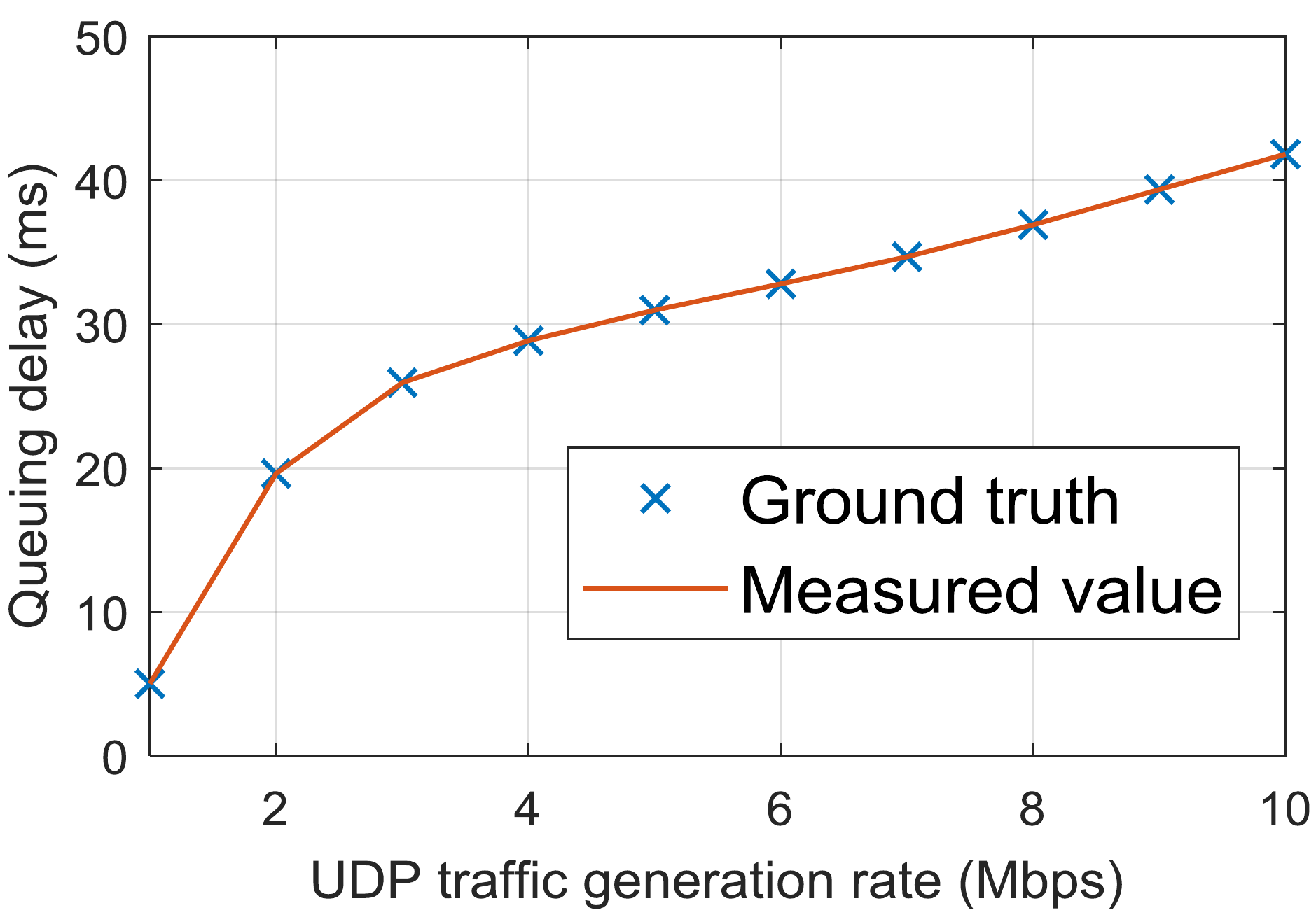}
    \caption{Estimation accuracy of \technique in the presence of varying queuing delay on the STA.}
    \label{q_delay}
\end{figure}

In this experiment, the AP observes the handshakes from the TCP download of the STA and makes an estimate for the queuing delay at the STA-side on the uplink. The ground truth is reported based on the knowledge of the timestamps corresponding to when the packet gets enqueued and when it reaches the head of the STA queue. 

As shown in Fig.~\ref{q_delay}, a variation in the traffic generation rate of the uplink UDP flow causes an increase in the amount of queuing delay that is experienced on the STA side. However, as the queuing delay varies from a minimum of 1 Mbps to 10 Mbps, the estimate made by the AP shows a close match with the ground truth value. 

Next, we validate the accuracy of \technique in estimating the amount of time a STA defers to transmit on the uplink. Here we consider to types of topologies and validate the performance for an increasing number of nodes for each case. First, we consider a scenario in which all nodes are within carrier sensing range of each other. In this scenario, each additional node added to the experiment makes the target STA or in other words the STA for which the AP is making an estimate defer. Next, we consider a scenario in which every non-target STA is hidden from the target STA. Consequently, an increasing number of non-target STAs results in an increasing number of hidden terminals in the network for the target STA.

\begin{figure}
\centering
\begin{subfigure}[t] {0.45\textwidth}
\centering
\includegraphics[width=0.95\textwidth]{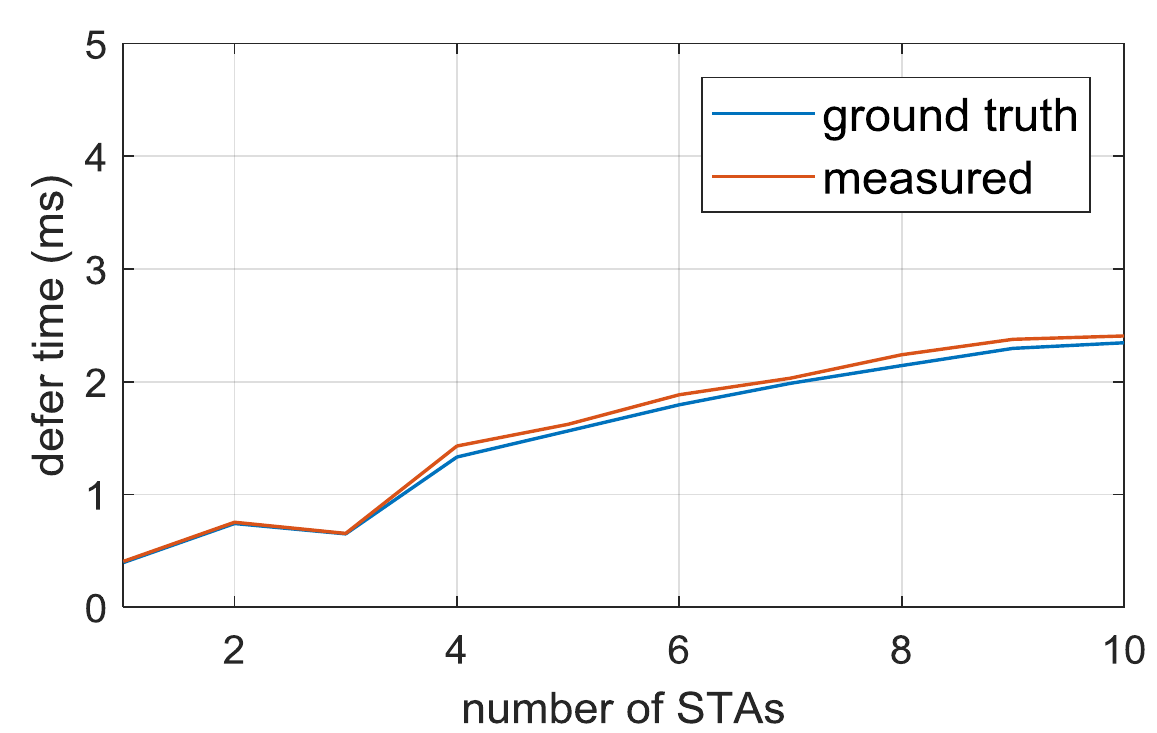}
\caption{Absence of hidden terminals}
\end{subfigure}
\begin{subfigure}[t] {0.45\textwidth}
\centering
\includegraphics[width=0.95\textwidth]{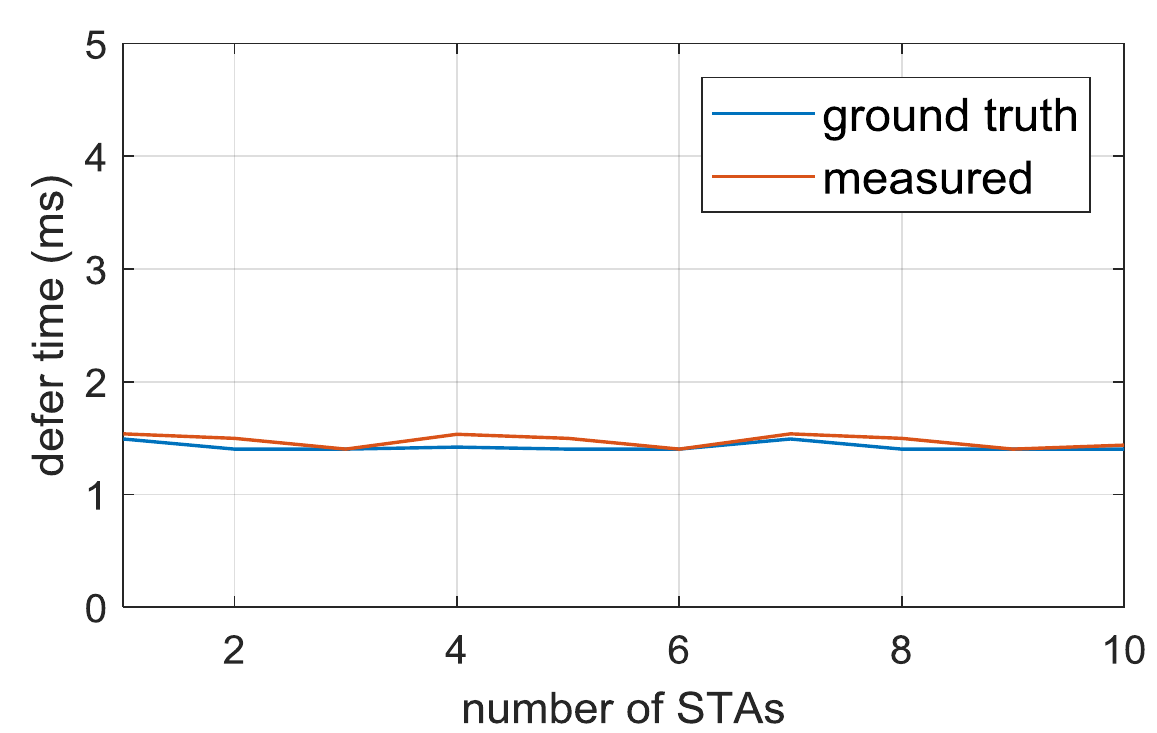}
\caption{Presence of hidden terminals}
\end{subfigure}
\caption{Defer delay estimation accuracy of \technique in the presence and absence of hidden terminals.}
\label{defer_delay}
\end{figure}

Fig.~\ref{defer_delay} shows the defer delay estimate made by the AP for the target STA. As depicted, in the case where all nodes are within carrier sensing range of each other, an increasing number of non-target STAs results in an increasing defer delay for the target STA. However, in the case where no hidden terminals are present, an increasing number of non-target STAs does not result in an increasing defer delay. This is because these non-target STAs are hidden from the target STA. Consequently, the target STA does not defer to them. In fact, the target STA defers to only one node which is the AP. Therefore, an increasing number of non-target STAs does not result in an increasing defer delay for the target STA. It is important to note that if the AP had solely made an estimate based on the defer delay for the downlink, this case would have caused a false positive. However, \technique can estimate the defer delay in both the cases with a high level of accuracy. 

\begin{figure}
\centering
\begin{subfigure}[t] {0.45\textwidth}
\centering
\includegraphics[width=0.95\textwidth]{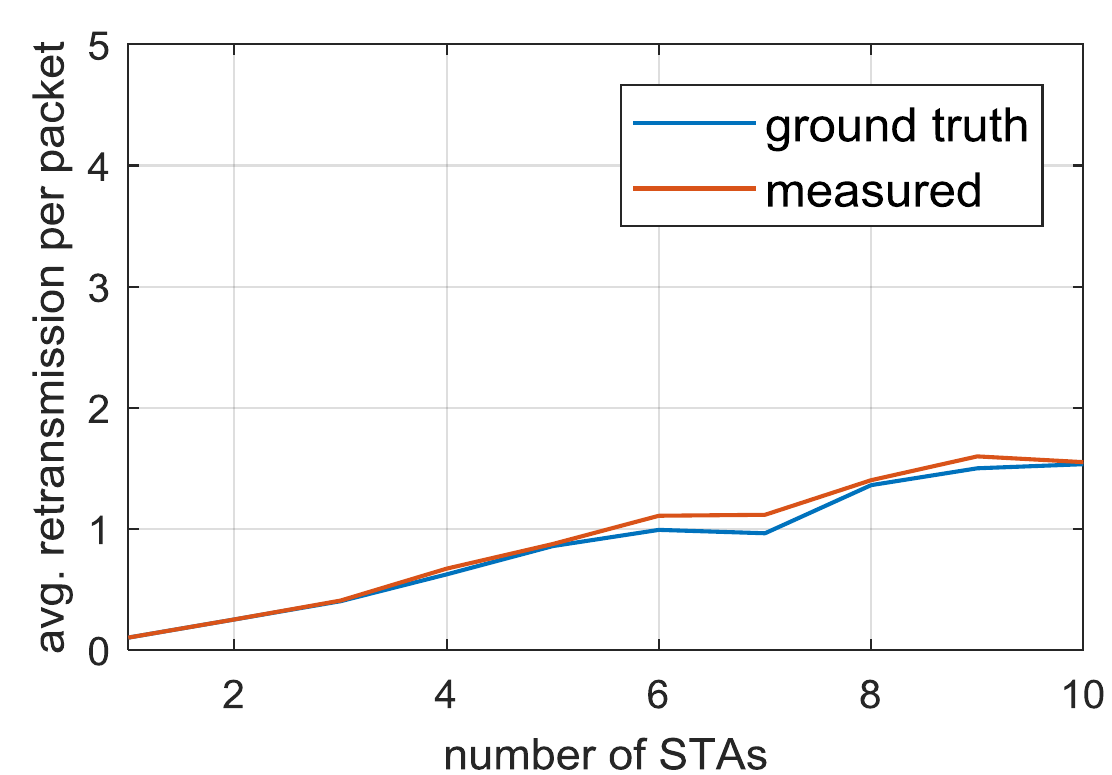}
\caption{Absence of hidden terminals}
\end{subfigure}
\begin{subfigure}[t] {0.45\textwidth}
\centering
\includegraphics[width=0.95\textwidth]{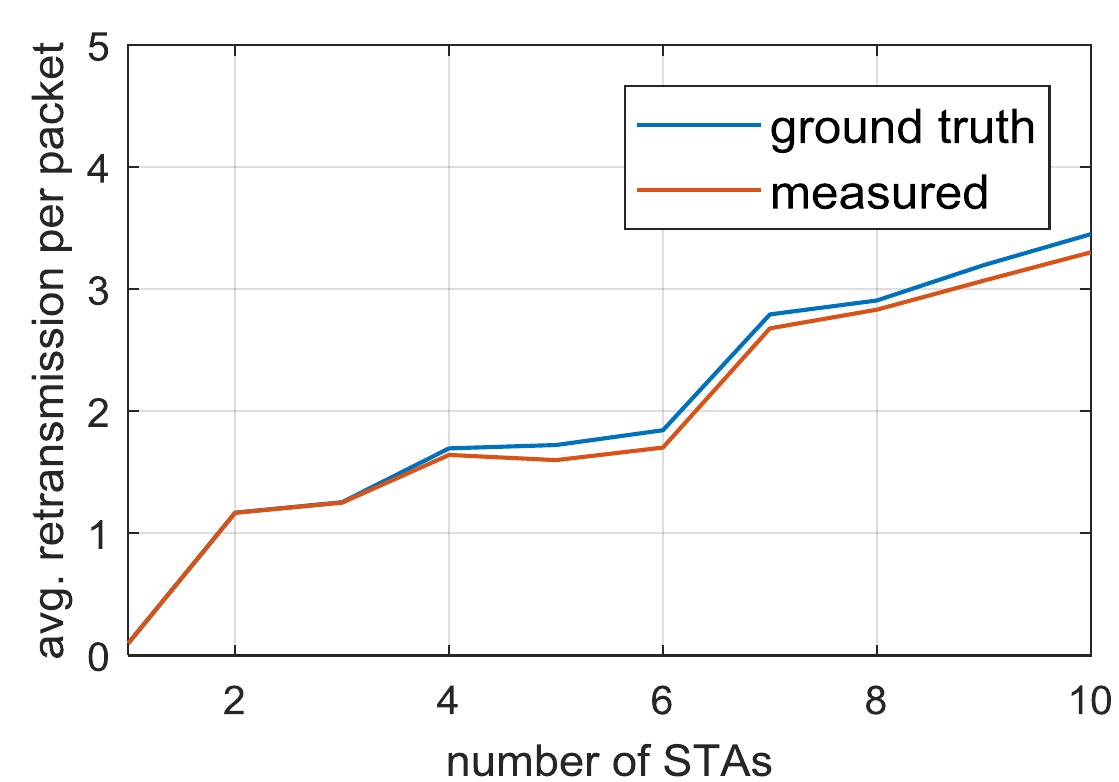}
\caption{Presence of hidden terminals}
\end{subfigure}
\caption{Retransmission rate estimation accuracy of \technique in the presence and absence of hidden terminals.}
\label{retrans}
\end{figure}

Next, we evaluate the estimation accuracy of \technique in estimating the retransmission rate. We consider the same scenarios as before and evaluate the accuracy of \technique in both the scenarios. Fig.~\ref{retrans} shows the retransmission rate estimation accuracy as a function of increasing number of non-target STAs. In the first scenario, an increasing number of non-target STAs increases the retransmission rate due to a rise in collision probability as the number of contending STAs increases. However, in the second case, the rise is even faster as each non-target STA added to the simulation is hidden from the target STA and consequently results in a rapid increase in the collision probability. Fig.~\ref{retrans} reveals that in both the scenarios, \technique estimates the retransmission rate with a high accuracy which is demonstrated by a close match between the ground truth value and the estimated values.\footnote{Since, we already show the estimation accuracy for uplink access delay in the previous chapter, we do not repeat it hear again.} 

\section{In-field Trials}\label{field}
Next we perform extensive field trials in which a number of factors co-exist together thereby enabling us to validate the performance of \technique in complex scenarios representative of a typical multi-BSS environment. 

\subsection{Deployment Characterization}
The empirical values of total uplink latency and its constituent components are affected by a number of factors such as traffic statistics, network topology as well as MAC and PHY statistics. It is important that as these factors vary, \technique is able to accurately estimate total uplink latency as well as decompose it into is constituent components. This section characterizes the diverse operating conditions encountered in these tests. 

\textbf{Deployment overview.} We deploy our AP and STAs in two locations. The first location is a university campus. Here the AP is deployed in a 3m x 5m office located in a 3 story building. The STAs are deployed both inside the office as well as outside the office at different locations on the same floor. The second deployment is an apartment located in a residential complex which comprises 1 or 2 bedroom units. Both of these environments comprise multiple BSSs co-existing together. In the university deployment, our devices co-exist with a university administered enterprise WLAN and 4 student deployed APs in nearby offices. Whereas in the residential environment, our devices co-exist with 12 APs from neighboring apartments. In each of these deployments, our AP has a total of 10 STAs associated with it. Each of these scenarios are characterized by a diversity in links (\textit{i.e.}, LoS and non-LoS paths), light human and environmental mobility as well as device mobility. We validate \technique via an extensive number of tests conducted in these deployments. Specifically, the total number of tests run were \textit{1,296,000} of which 576,000 were run in the university deployment whereas 720,000 were run in the residential deployment.
\begin{figure}
\centering
\begin{subfigure}[t] {0.45\textwidth}
\centering
\includegraphics[width=0.95\textwidth]{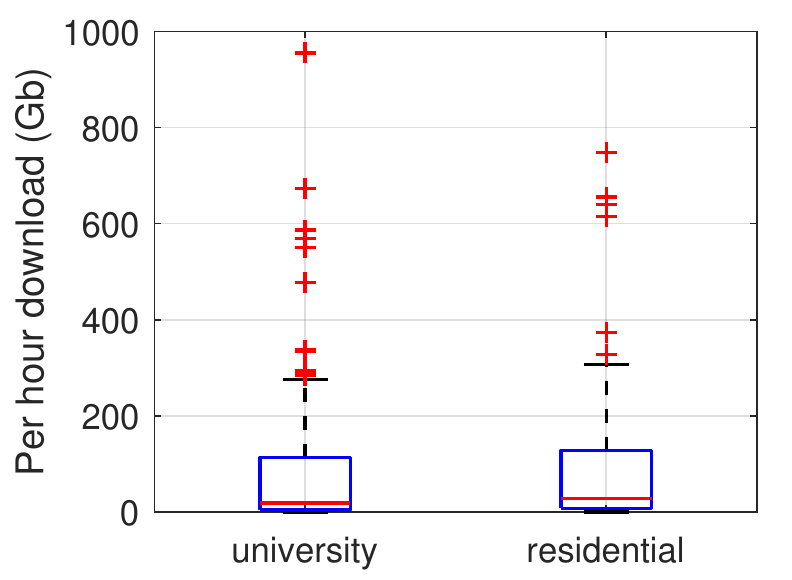}
\caption{}
\label{fig:ullatency}
\end{subfigure}
\begin{subfigure}[t] {0.45\textwidth}
\centering
\includegraphics[width=0.95\textwidth]{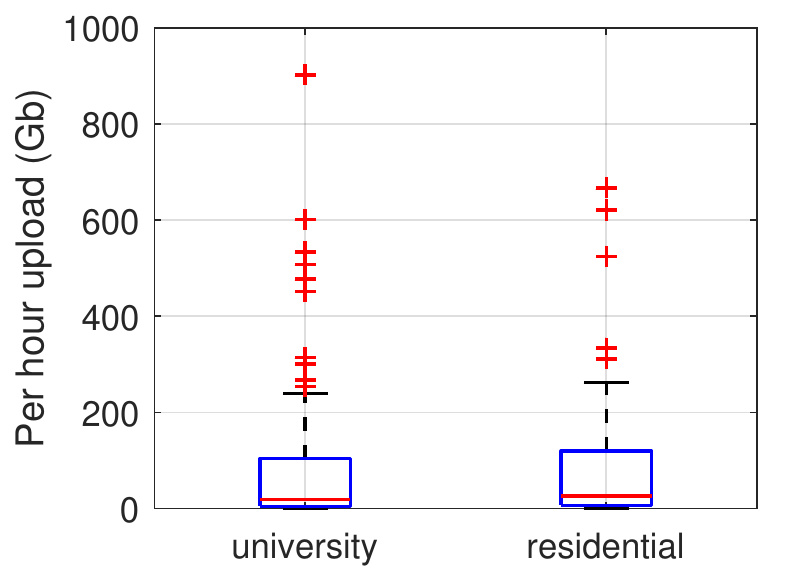}
\caption{}
\label{fig:adelay}
\end{subfigure}
\caption{Aggregate per hour download and upload statistics in the university and residential deployments (a) Per hour download distribution, (b) Per hour upload distribution.}
\label{per_hr}
\end{figure}

\begin{figure}
\centering
\begin{subfigure}[t] {0.45\textwidth}
\centering
\includegraphics[width=0.95\textwidth]{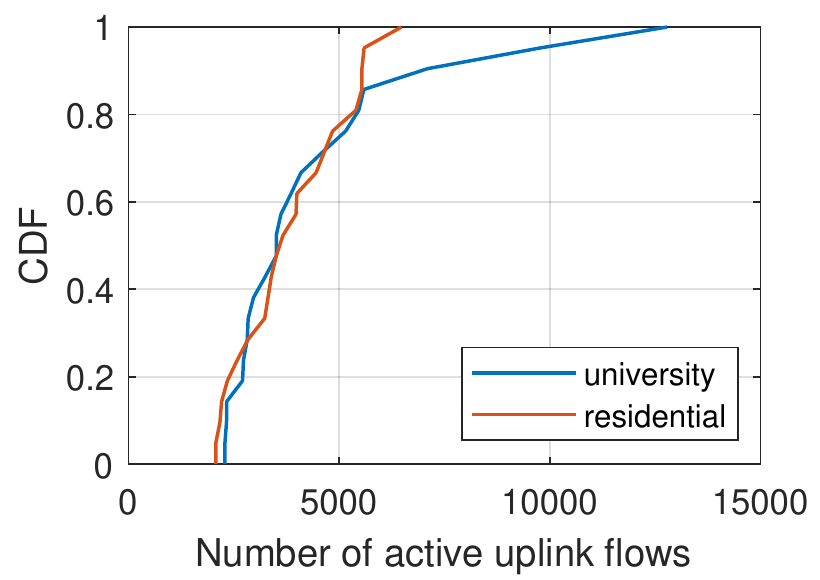}
\caption{}
\label{fig:ullatency}
\end{subfigure}
\begin{subfigure}[t] {0.45\textwidth}
\centering
\includegraphics[width=0.95\textwidth]{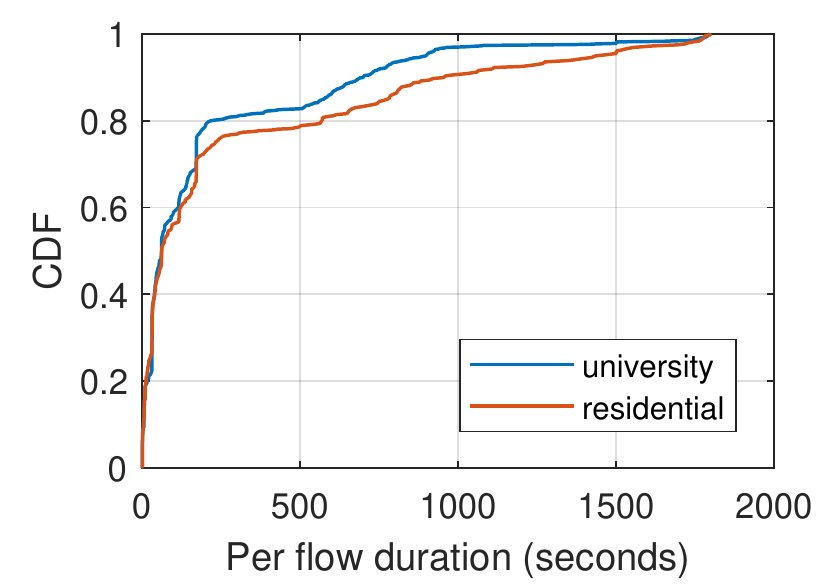}
\caption{}
\label{fig:adelay}
\end{subfigure}
\caption{Aggregate uplink flow statistics in the university and residential deployments. (a) Distribution of number of active uplink flows per hour (b) Distribution of the duration of an active flow.}
\label{uplink_flows}
\end{figure}

\textbf{Traffic statistics.} The STAs in our deployments run online internet applications that perform video streaming (using YouTube and Amazon Prime Video), music streaming (via Pandora), pdf downloads (from IEEE Xplore), email activities (using Gmail) and Gigabit file downloads and uploads to Dropbox and Google Drive. The tests involve cases with both single application traffic where only one of the above applications is run at a time as well as mixed application traffic were a number of these applications run in parallel. These internet applications are run using Mozilla Firefox web browser. In addition, some of the STAs also performed UDP downloads and uploads to local servers. A distribution of the amount of per hour download and upload is shown in Fig.~\ref{per_hr} and a distribution of the number of active uplink flows and their duration is shown in Fig.~\ref{uplink_flows}. This variation causes a fluctuation in the queuing delays experienced by the STAs as well as a variation in the per flow queuing delay. 

\begin{figure}
\centering
\begin{subfigure}[t] {0.45\textwidth}
\centering
\includegraphics[width=1\textwidth]{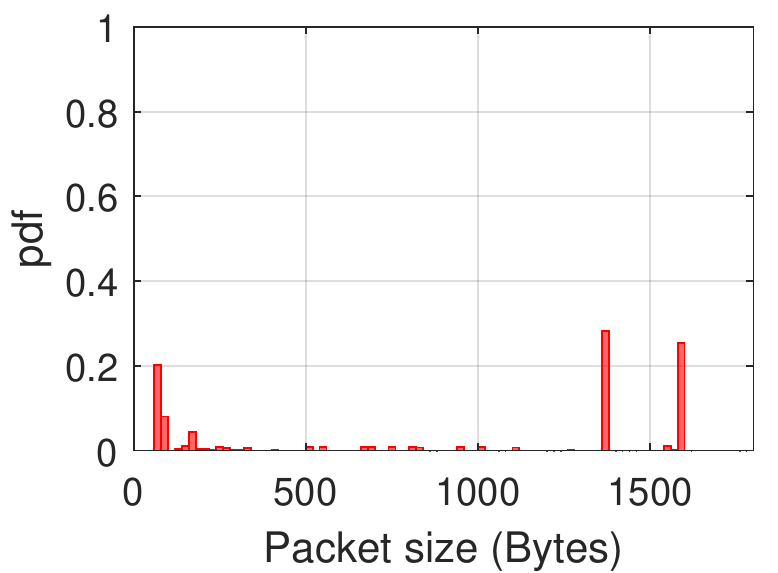}
\caption{University deployment}
\label{fig:ullatency}
\end{subfigure}
\begin{subfigure}[t] {0.45\textwidth}
\centering
\includegraphics[width=1\textwidth]{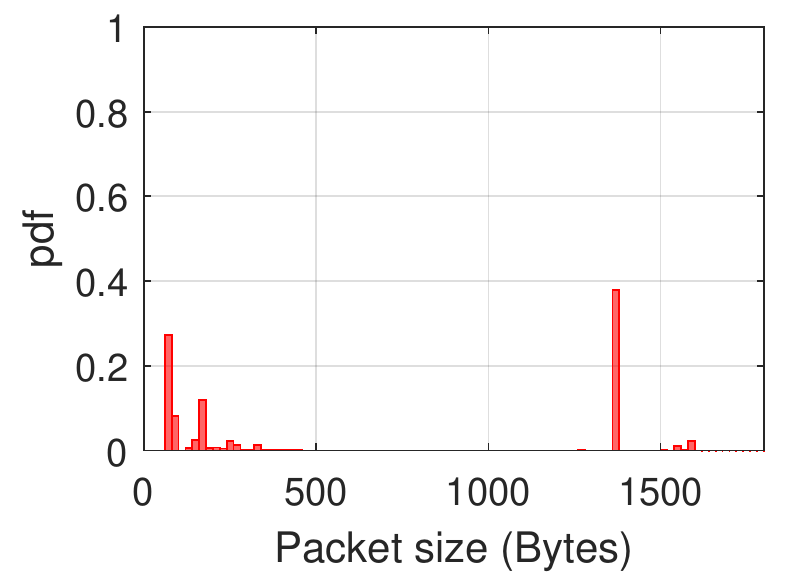}
\caption{Residential deployment}
\label{fig:adelay}
\end{subfigure}
\caption{Packet size distribution for the university and residential deployments. In each of these deployments, the minimum packet size encountered is around 200 bytes and the maximum packet size is around 1.6KB.}
\label{flow_dist}
\end{figure}

The packet size distributions encountered in each of these scenarios are shown in Fig.~\ref{flow_dist}. The minimum packet size is around 200 bytes and the maximum is around 1600 bytes. A variation in a STA's packet sizes affects the amount of air time utilization thereby causing a microscopic fluctuation in the defer delays experienced by other STAs. This in turn affects uplink access delays. Further, packet size variation also affects transmission delays. 

\textbf{Network topology.} The network topology determines the number of interfering nodes which affects the retransmission rate and defer delays. These in turn affect the uplink access delay. The residential and university scenarios cover a total of \textit{250} and \textit{200} number of topologies respectively. These topologies arise from a combination of weak links, strong links, near and far away nodes, hidden terminals, etc. Further, the tests cover cases with active STA sets of all possible sizes and the AP makes an estimate for each associated STA.  

\begin{figure}
\centering
\begin{subfigure}[t] {0.45\textwidth}
\centering
\includegraphics[width=1\textwidth]{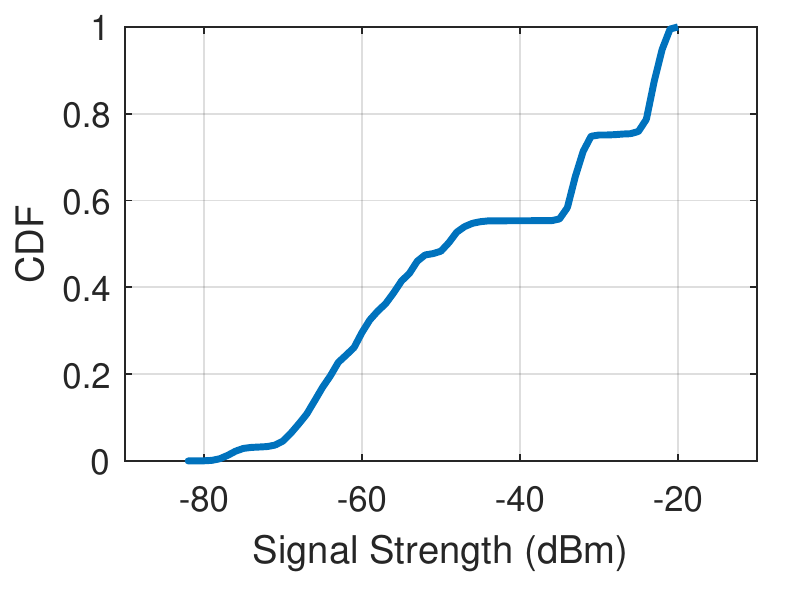}
\caption{University deployment}
\label{fig:ullatency}
\end{subfigure}
\begin{subfigure}[t] {0.45\textwidth}
\centering
\includegraphics[width=1\textwidth]{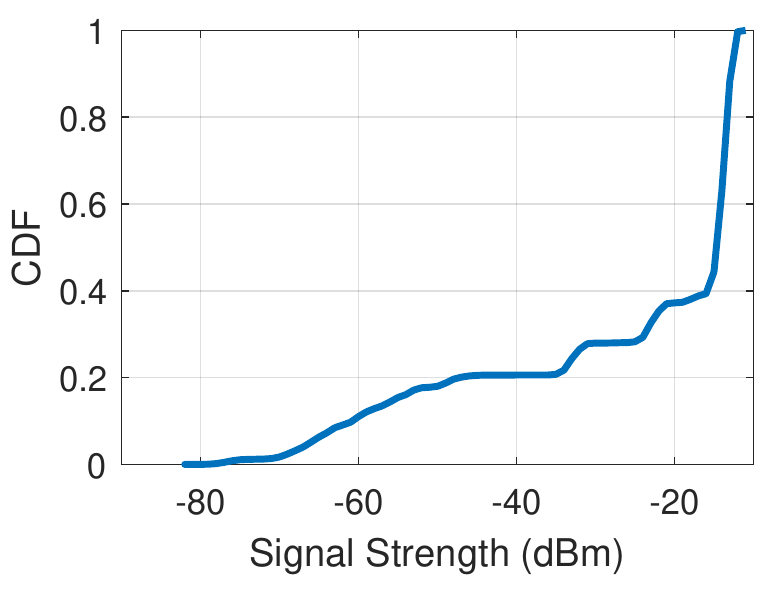}
\caption{Residential deployment}
\label{fig:adelay}
\end{subfigure}
\caption{Distribution of STA's signal strength in the university and residential deployments.}
\label{rssi_dist}
\end{figure}

\textbf{MAC and PHY statistics.} A diversity in both the network topology as well as the traffic characteristics of the STAs result in a variation in the MAC and PHY statistics of the devices in our network. This variation affects the total uplink latency as well as its constituent components. Fig.~\ref{rssi_dist} shows the distribution of rssi values for the STAs during the experiments. Rssi affects selected data rates which affects transmission delays and hence total uplink latency. Transmission delays of non-target STAs affects the amount of time that the target STA defers thereby affecting its defer delays and access delays. 

\begin{figure}
\centering
\begin{subfigure}[t] {0.45\textwidth}
\centering
\includegraphics[width=1\textwidth]{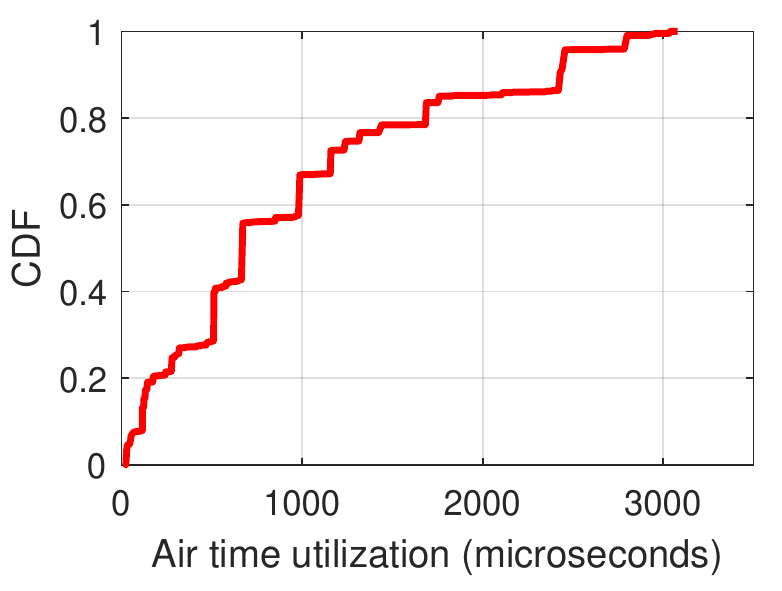}
\caption{University deployment}
\label{fig:ullatency}
\end{subfigure}
\begin{subfigure}[t] {0.45\textwidth}
\centering
\includegraphics[width=1\textwidth]{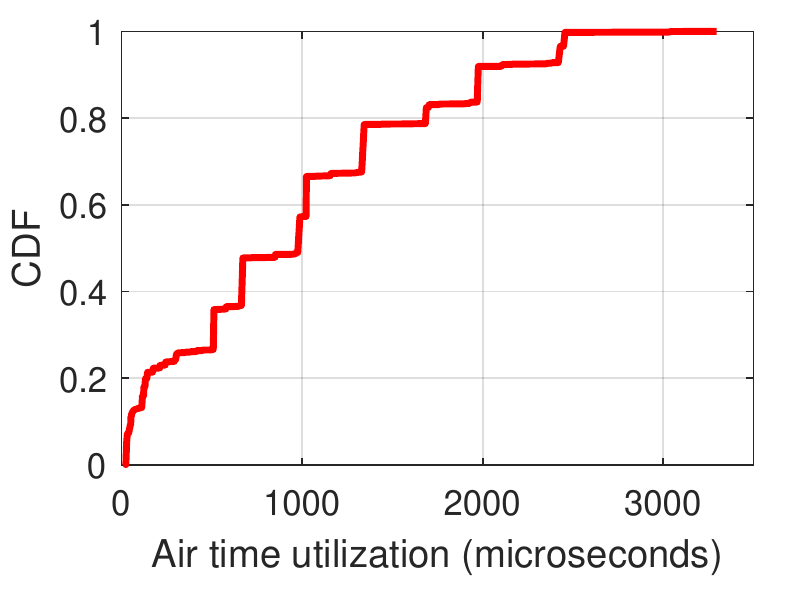}
\caption{Residential deployment}
\label{fig:adelay}
\end{subfigure}
\caption{Airtime utilization of devices in co-existing BSSs in the university and residential deployments.}
\label{air_time_util}
\end{figure}

Both deployment scenarios are characterized by co-existing BSSs. As a result, the STAs in our network defer to these nodes depending on their air time utilization which affects their defer delays and access delays. Fig.~\ref{air_time_util} shows the distribution of air time utilization of devices in the co-existing BSSs.

\subsection{Parameter Estimation Accuracy}
In this section, we investigate the performance of \technique in the two deployment scenarios. 

\begin{figure*}[!]
\centering
\begin{subfigure} {0.3\textwidth}
\centering
\includegraphics[width=1\textwidth]{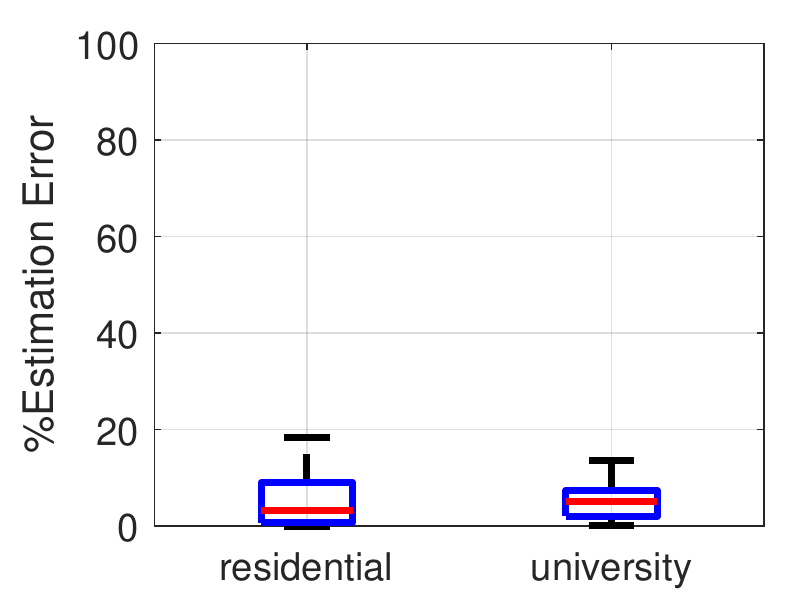}
\caption{Total uplink latency estimation error}
\label{fig:ullatency}
\end{subfigure}
\begin{subfigure} {0.3\textwidth}
\centering
\includegraphics[width=1\textwidth]{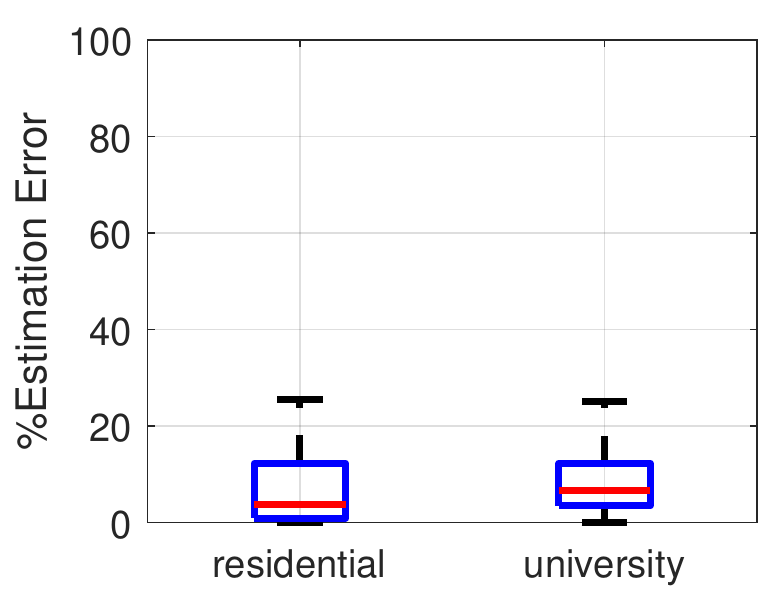}
\caption{Access delay estimation error}
\label{fig:adelay}
\end{subfigure}
\begin{subfigure} {0.3\textwidth}
\centering
\includegraphics[width=1\textwidth]{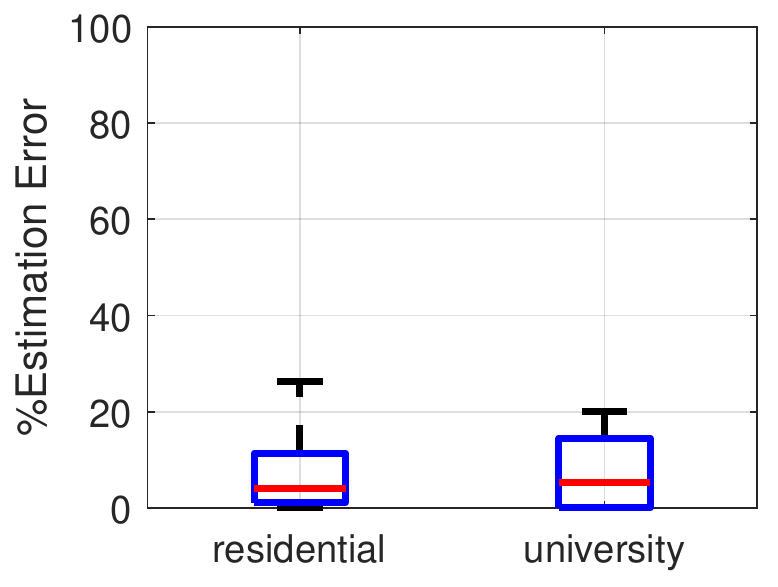}
\caption{Queuing delay estimation error}
\label{fig:qdelay}
\end{subfigure}
\begin{subfigure} {0.3\textwidth}
\centering
\includegraphics[width=1\textwidth]{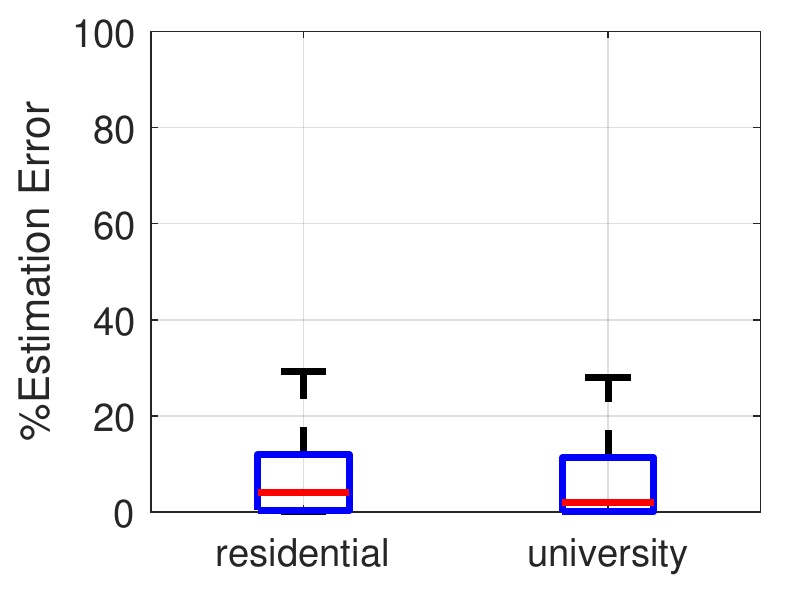}
\caption{Per flow queuing delay estimation error}
\label{fig:pfqueuing}
\end{subfigure}
\begin{subfigure} {0.3\textwidth}
\centering
\includegraphics[width=1\textwidth]{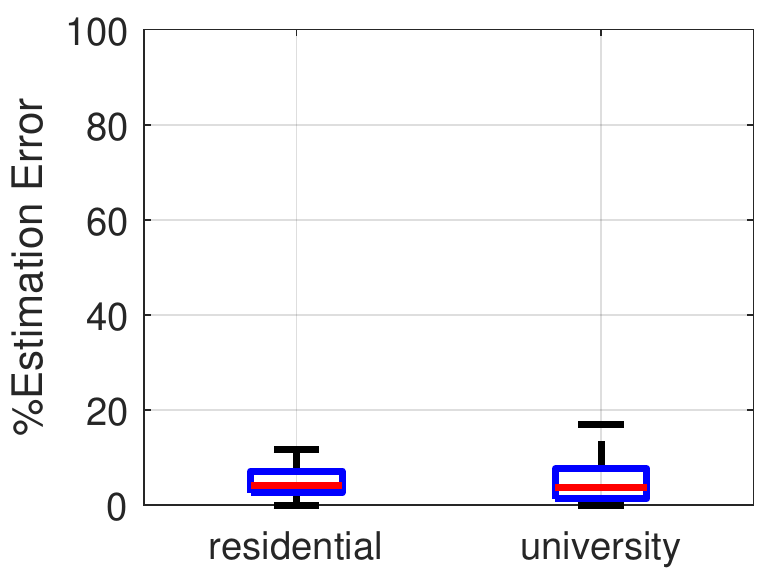}
\caption{Retransmission rate estimation error}
\label{fig:retransmission}
\end{subfigure}
\begin{subfigure} {0.3\textwidth}
\centering
\includegraphics[width=1\textwidth]{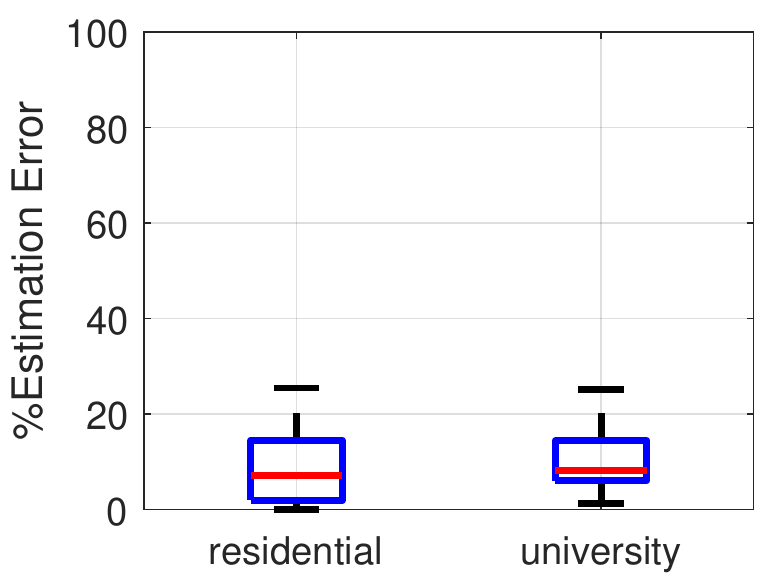}
\caption{Defer delay estimation error}
\label{fig:defer}
\end{subfigure}
\caption{Parameter estimation accuracy of \technique for the university and residential deployments. The average estimation error across all the parameters is less than 10\%.}
\label{fig:para_est}
\end{figure*}

First we experimentally evaluate the parameter estimation accuracy of \technique. Recall that the key idea in \technique is to treat TCP handshakes as virtual probes and use them to drive measurement based estimation of total uplink latency and its constituent components. In our deployment, the system estimates total uplink latency and its components for each associated STA every 30 seconds. Therefore, in this test duration, the system gathers AP logs and pipes them into the \technique core to obtain estimates. These estimates are time-stamped and stored on the remote server. Recall that the instance of stats manager running on the STAs locally stores the STA log. Similar to the AP, this log is collected every 30 seconds. This log captures the ground truth observed by the STAs. In the deployment scenarios, the STA uses each packet transmitted on the uplink as a recording of ground truth. However, the AP only uses the TCP handshake coupled with the techniques described in the Ch.~\ref{uscope_ch} to estimate  the parameters.

We compare the estimates for $\ulatency$, $\adelay$, $\qdelay$, $\bar{\Phi}_{q,i}$, $\retrans$ and $\tdefer$ made by the \technique system with those obtained from the STA side. To evaluate the parameter estimation accuracy, we compute the percent error in the estimate calculated as $\frac{|\beta_{est} - \beta_{gt}|*100}{\beta_{gt}}$ where $\beta_{est}$ is the estimate for a parameter obtained from the \technique system and $\beta_{gt}$ is the ground truth for the parameter obtained from the STA side. 

Fig.~\ref{fig:para_est} summarizes the estimation error statistics for both the university and the residential scenario. Overall,  \technique demonstrates a mean estimation error below 10\% across all  parameters. However, Fig.~\ref{fig:para_est} reveals that in some cases, the worst case estimation errors are much larger compared to the average. This primarily occurs due to the following reason. Recall that \technique is driven by measurements collected from TCP handshakes at the AP side. As a result, for an accurate estimation, the AP must observe a sufficiently large number of TCP handshakes. In our deployments, the number of handshakes observed by the AP in any given test duration is dependent on the activity of the internet application running on the STA. Consequently, in test intervals where the AP observes an insufficient number of TCP handshakes, the estimate made by the AP demonstrates a high error. We further explore how estimation error varies as a function of number of TCP handshakes in the next subsection.

\subsection{Estimation Error Characterization}
\begin{figure*}[t]
\centering
\begin{subfigure}[t] {0.3\textwidth}
\centering
\includegraphics[width=1\textwidth]{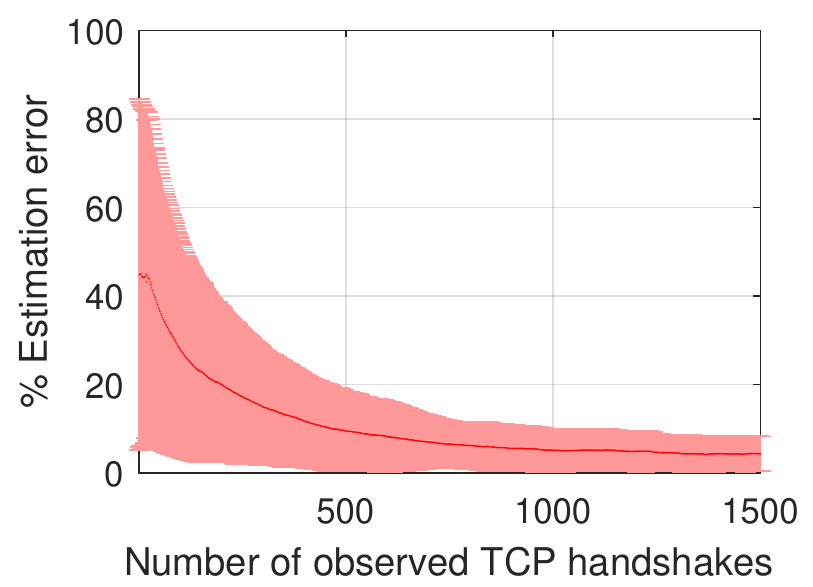}
\caption{Total uplink latency estimation error}
\label{fig:ullatency}
\end{subfigure}
\begin{subfigure}[t] {0.3\textwidth}
\centering
\includegraphics[width=1\textwidth]{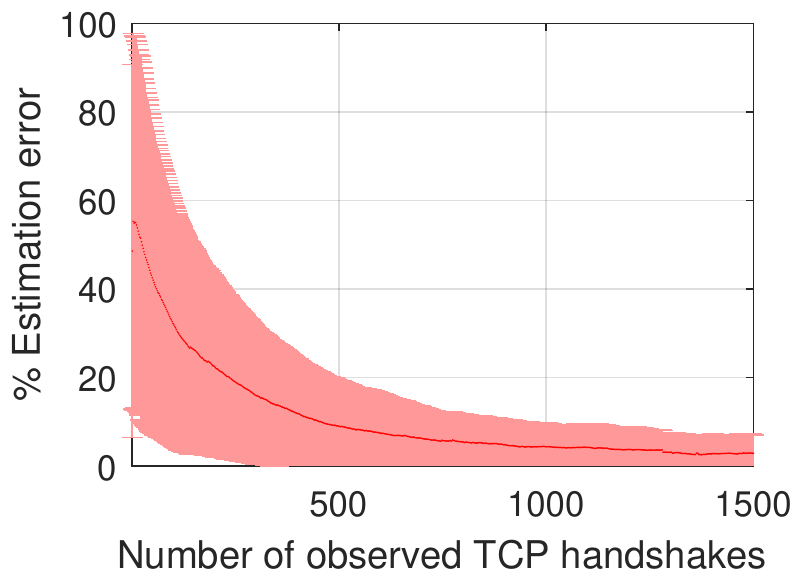}
\caption{Access delay estimation error}
\label{fig:adelay}
\end{subfigure}
\begin{subfigure}[t] {0.3\textwidth}
\centering
\includegraphics[width=1\textwidth]{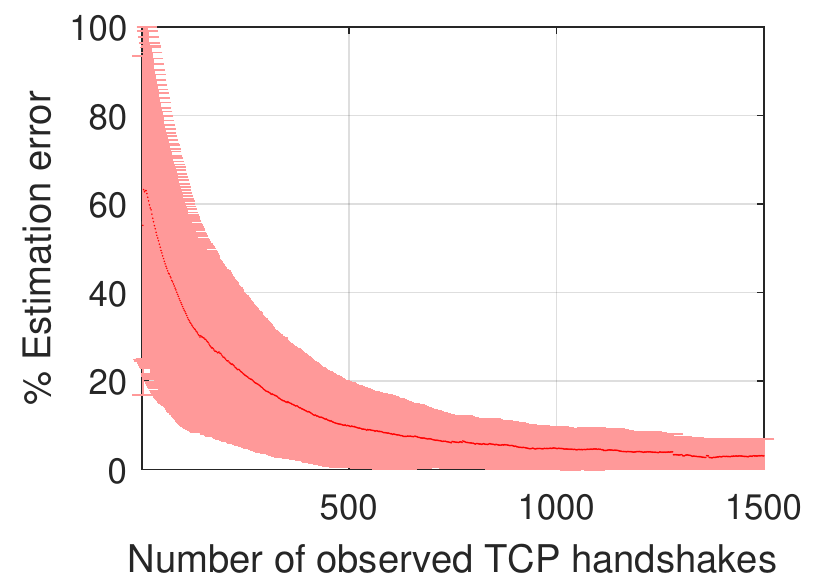}
\caption{Queuing delay estimation error}
\label{fig:qdelay}
\end{subfigure}
\begin{subfigure}[t] {0.3\textwidth}
\centering
\includegraphics[width=1\textwidth]{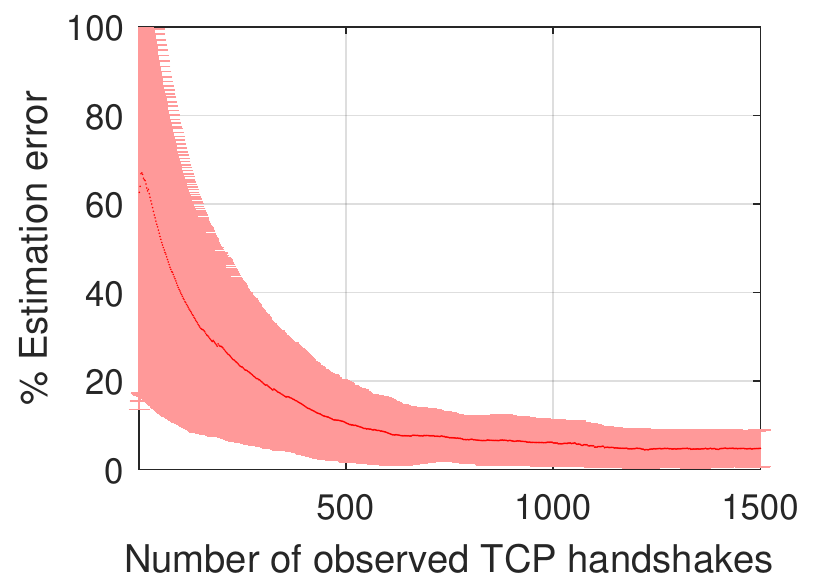}
\caption{Per flow queuing delay estimation error}
\label{fig:pfqueuing}
\end{subfigure}
\begin{subfigure}[t] {0.3\textwidth}
\centering
\includegraphics[width=1\textwidth]{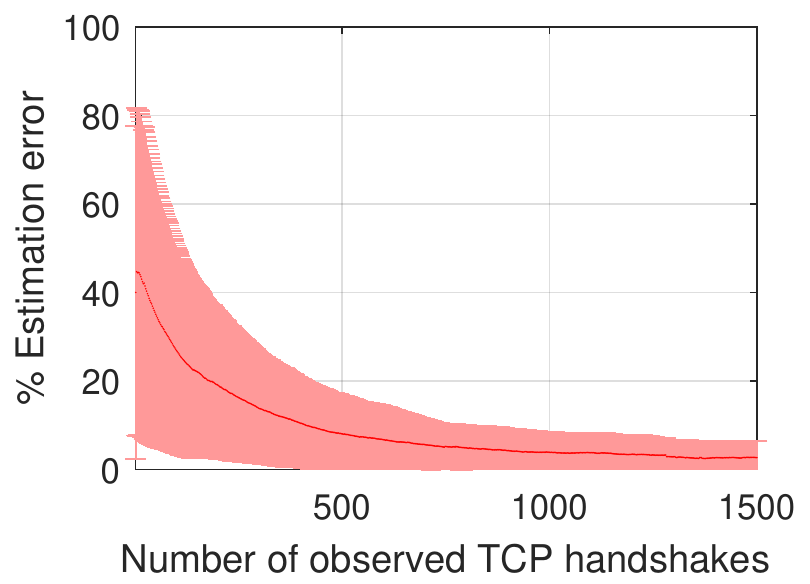}
\caption{Retransmission rate estimation error}
\label{fig:retransmission}
\end{subfigure}
\begin{subfigure}[t] {0.3\textwidth}
\centering
\includegraphics[width=1\textwidth]{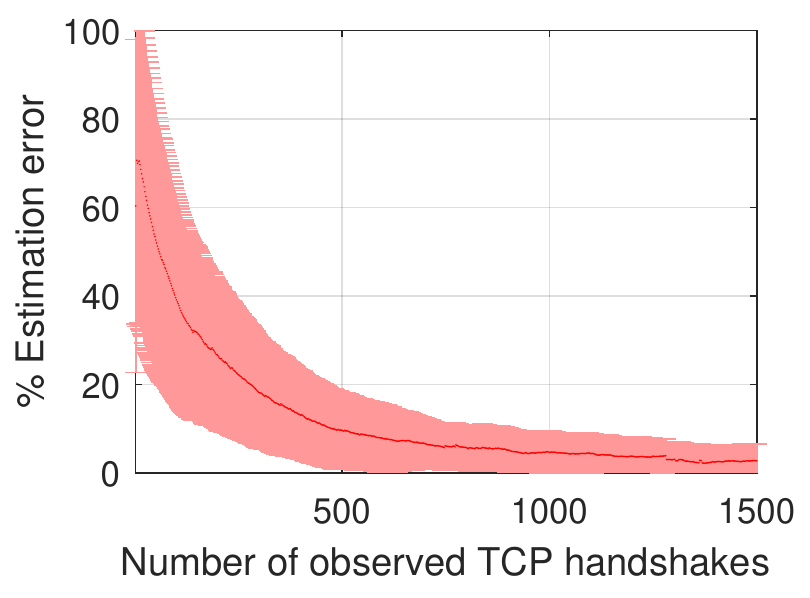}
\caption{Defer delay estimation error}
\label{fig:defer}
\end{subfigure}
\caption{Mean estimation error for total uplink latency and its constituent components as a function of the number of TCP handshake measurements available to the AP while making an estimate in the residential deployment.}
\label{fig:est_error_res}
\end{figure*}

\begin{figure*}[t]
\centering
\begin{subfigure}[t] {0.3\textwidth}
\centering
\includegraphics[width=1\textwidth]{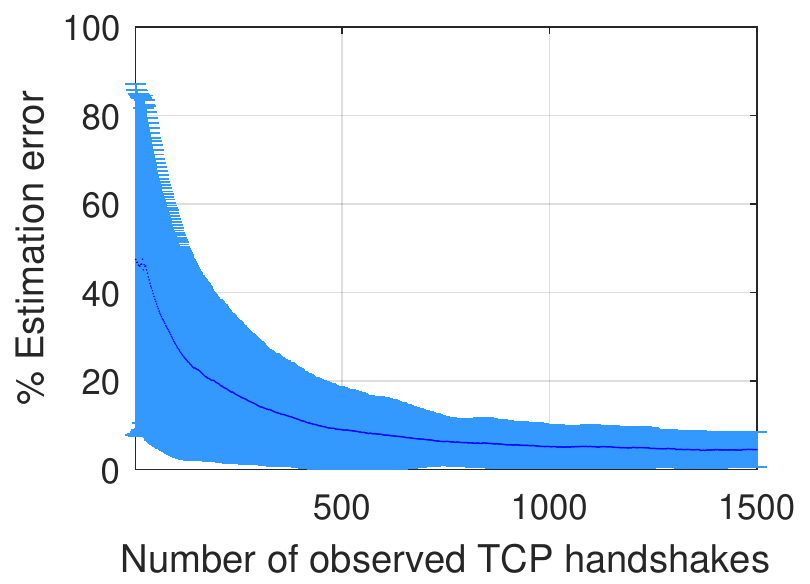}
\caption{Total uplink latency estimation error}
\label{fig:ullatency}
\end{subfigure}
\begin{subfigure}[t] {0.3\textwidth}
\centering
\includegraphics[width=1\textwidth]{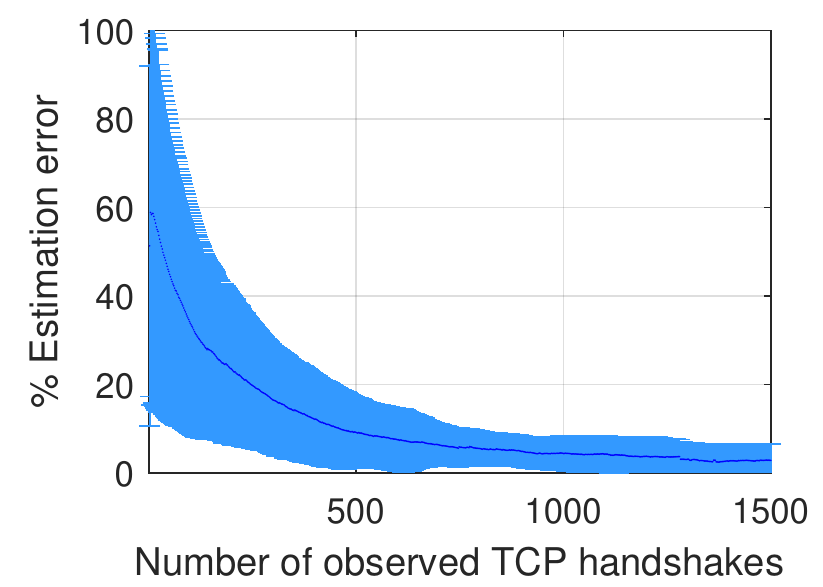}
\caption{Access delay estimation error}
\label{fig:adelay}
\end{subfigure}
\begin{subfigure}[t] {0.3\textwidth}
\centering
\includegraphics[width=1\textwidth]{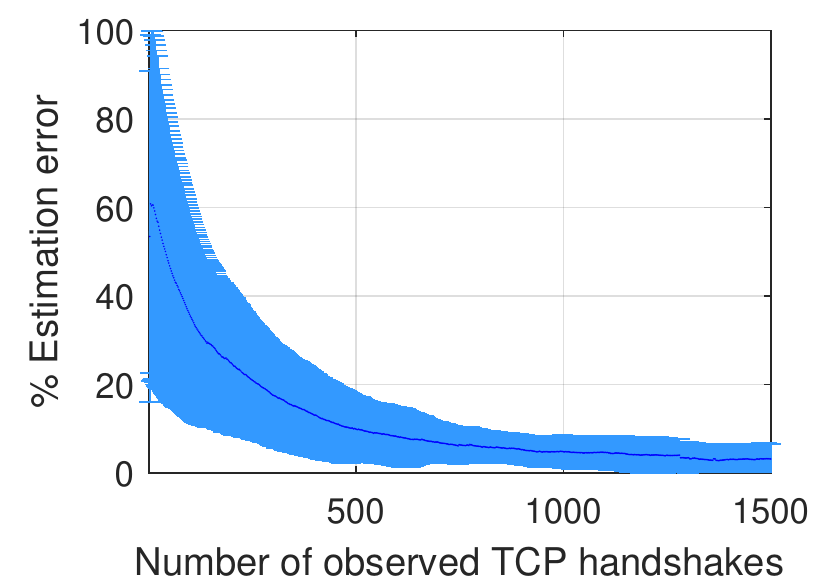}
\caption{Queuing delay estimation error}
\label{fig:qdelay}
\end{subfigure}
\begin{subfigure}[t] {0.3\textwidth}
\centering
\includegraphics[width=1\textwidth]{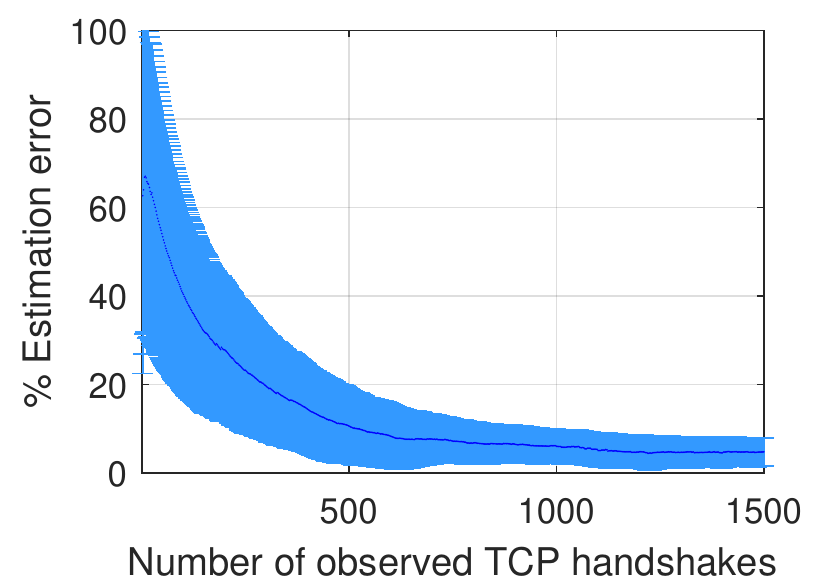}
\caption{Per flow queuing delay estimation error}
\label{fig:pfqueuing}
\end{subfigure}
\begin{subfigure}[t] {0.3\textwidth}
\centering
\includegraphics[width=1\textwidth]{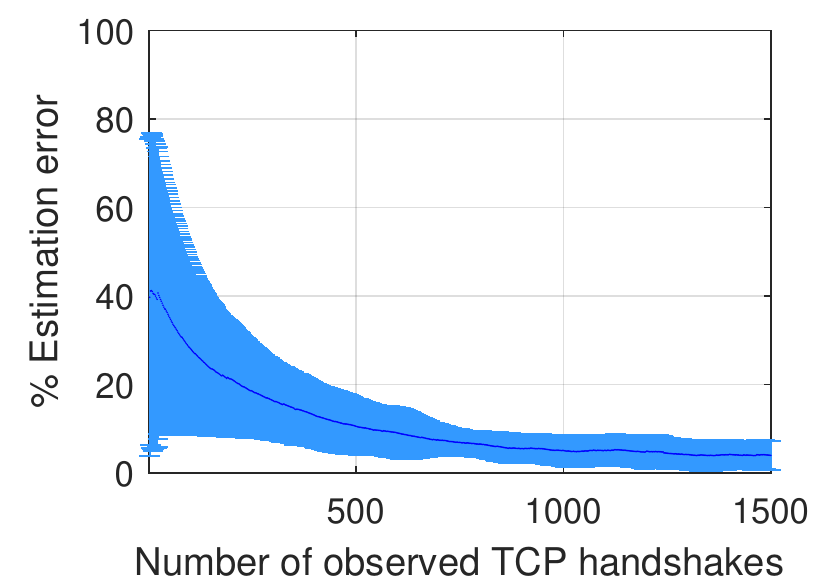}
\caption{Retransmission rate estimation error}
\label{fig:retransmission}
\end{subfigure}
\begin{subfigure}[t] {0.3\textwidth}
\centering
\includegraphics[width=1\textwidth]{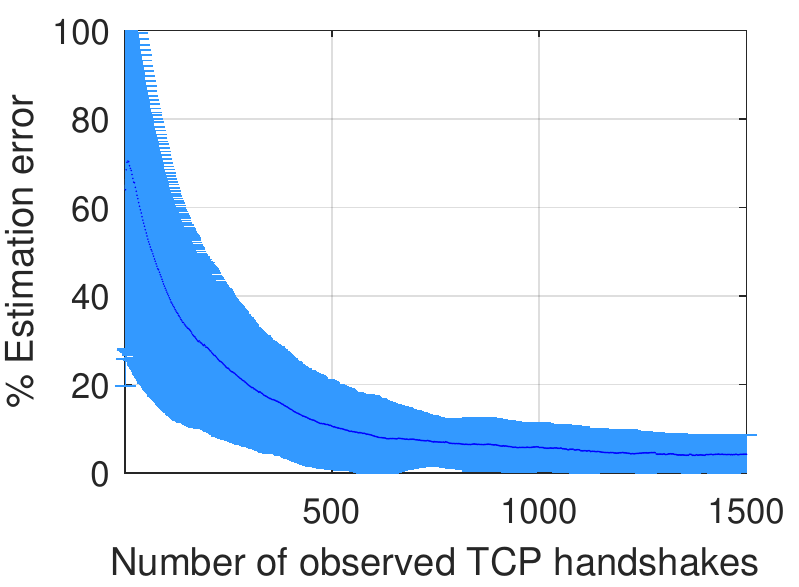}
\caption{Defer delay estimation error}
\label{fig:defer}
\end{subfigure}
\caption{Mean estimation error for total uplink latency and its constituent components as a function of the number of TCP handshake measurements available to the AP while making an estimate in the university deployment.}
\label{fig:est_error_univ}
\end{figure*}

\technique is a measurement driven framework. These measurements are collected by observing TCP handshakes from TCP flows of a STA. Consequently, the parameter estimation error is a function of the number of TCP handshakes that the AP observes. Here we experimentally investigate the relationship between the estimation error and the number of TCP handshakes observed by the AP. 

In order to do so, we leverage the logs collected from the AP and all the STAs in the field trials. We modify the stats processor to iterate through the log and pass a specified number of consecutive TCP handshakes to the \technique core. As a result, \technique is constrained to make an estimate solely based on these samples. This number is varied to understand the relationship between estimation error and number of observed TCP handshakes.\footnote{To ensure that the AP sees a non-retransmitted TCP ACK for defer delay and retransmission rate estimation, we start the test with a minimum of 100 measurement samples.} 

Fig.~\ref{fig:est_error_res} and \ref{fig:est_error_univ} depict the estimation error as a function of the number of observed handshakes in the residential and the university scenario respectively. The estimation error demonstrates a similar trend across all the parameters. When the number of TCP handshakes observed by the AP are on the order of a few 100s, the mean estimation error is extremely high. This is because the number of measurements is not sufficient to characterize the total uplink latency and its constituent components. In this regime, the variance in the error is also high. Both the estimation error and the variance demonstrate a rapid decay with an increase in the number of observed TCP handshake. Fig.~\ref{fig:est_error_res} and \ref{fig:est_error_univ} reveal that even with over 1,000 observed TCP handshakes, the estimation error of \technique across all the parameters is less than 10\%.

\chapter{Related Work}
\label{sec:relatedwork}

\section{Analytical Models for TCP Throughput Estimation}
Several analytical modeling approaches are available in literature that lead to closed form expression for TCP throughput. Classic models for TCP focus on capturing the macroscopic behavior of TCP over a network path with a given round trip time (RTT) and a loss probability \cite{padhye1998modeling, padhye1999stochastic, parvez2010analytic, he2005predictability, kim2011modeling, sikdar2001integrated}. Extensions of these approach have been applied to capture the performance of TCP over advanced physical layer technologies such as MIMO \cite{oh2007mimo, chaurasia2013dynamic, toledo2006cross, toledo2006tcp}. 

On the other hand, powerful models for Wi-Fi focus on capturing the microscopic packet dynamics including queuing behavior, traffic burstiness, etc. Classic models target predictions of layer-2 throughput by taking into account the backoff and contention process of 802.11 \cite{bianchi2000performance, bianchi1998ieee, bianchi2005remarks, bianchi1996performance}. Several papers further extend these models by incorporating various MAC and PHY layer aspects of Wi-Fi \cite{ngk, dae, nayak2019modeling}. 

Unfortunately, the models are not applicable or extensible to our goal of passive AP-side TCP throughput estimation as they require AP-side knowledge of network topology, interfering nodes including those from neighboring BSS, their traffic patterns, PHY capabilities, data rates, etc. Obtaining this information requires STA-side co-operation and regular reporting. On the other hand, the L2 edge TCP model that we propose enables throughput estimation with zero STA-side co-operation and no reporting. 

\section{Active Probing Based Tools}
Active probing techniques involve use of special purpose probing packets. The classic work on packet pair leverages the dispersion of packet pairs to estimate network parameters \cite{liu2005signals, dovrolis2001packet, pasztor2002packet, keshav1995packet}. Probing techniques such as  such as \cite{Probegap, CapProbe} involve usage of probing packets to estimate bandwidth. 

In the context of latency measurements, probing methods available in literature can be further divided into AP-initiated probing and STA-initiated probing. Techniques such as  \cite{sui2016characterizing, cangialosi2015ting, baranasuriya2015qprobe, jackson2015ping, huang2011mobiperf1, huang2011mobiperf, WBest}  make use of AP initiated probing can be used to measure and decompose WLAN uplink latency. Likewise, tools such as \cite{das2017informed, mickulicz2016zephyr, kim2014mot, kim2014wislow, speedtest, rosen2014mcnet} can be used to collect STA side information via user initiated measurements and reporting to analyze uplink latency. 

Unfortunately, these tools involve active measurements which increases the traffic load on the network. Consequently, their usage for periodic monitoring can potentially disrupt user traffic thereby worsening both throughput and latencies for other users and draining the battery of mobile devices. Further, STA-side tools such as \cite{kim2014wislow} also require special purpose software installation on the STAs for data collection and reporting which end users may be unwilling to install. In contrast, our frameworks are completely passive, do not require any special purpose STA side software and make estimates solely based on AP side observations.

\section{Passive Measurements Approaches}
Deploying a network of sniffers can enable collection of packet traces which could facilitate a passive inference of the inter-node connectivity in the network and activity of interfering nodes for a given STA \cite{mishra2015usage, zeng2014delay, paul2013passive, dtr, cbb, kashyap2010deconstructing}. Such insights can enable estimation and decomposition of WLAN uplink latency for all STAs that the sniffers can sense. Unfortunately, such passive approaches involve deploying an additional hardware infrastructure which adds to the cost of network deployment and maintenance. On the other hand, our works do not require any additional hardware infrastructure and can make an estimate solely based on passive observations from a single AP.

\section{Learning Based Approaches}
In this approach, followed by \cite{pgb, rattaro2010throughput, mirza2009accuracy}, network clients  store empirical throughput of all TCP sessions and report them to the AP to build a database of TCP throughputs. This coupled with AP-side records of wireless conditions during the TCP session (\textit{e.g.}, the session's MCS, busy air time, and collision rate) enable the AP to predict throughput by correlating the current conditions with historical averages corresponding to similar conditions. 

Machine learning tools can also be employed to estimate WLAN uplink latency and its components \cite{zhou2017mining, syrigos2019employment, khan2016accurate, croce2018learning}. With an accurately trained model, these approaches can achieve an accuracy over 95\%. 

Unfortunately, these tools require ground truth collection and reporting from each device in the network for the purpose of model training. In contrast, the tools presented in this thesis do not seek any information from the STAs and can make estimates solely based on passive AP-side observations.

\section{TCP Based Network Analysis}
TCP flows have been used to collect information that is valuable for network analysis, performance monitoring as well as detection of security threats. Tools like \cite{tstat, favicftstat} analyze TCP headers to provide several IP and TCP statistics such as segment reordering, duplication, etc. which aids network measurement research whereas tools like \cite{nsk} leverage observation of TCP flows in the network to  predict achievable TCP throughput for all STAs in the network. On the other hand, \cite{chen2007tcp, zhou2006p2p} utilizes TCP flow characteristics to identify security threats. In contrast, the frameworks presented by this thesis leverages TCP's layer 4 handshake to estimate download and upload speeds, WLAN latency and its constituent components.

\section{Conflict Graph Based Estimation} 
Techniques to infer conflict relationships amongst wireless nodes are available in literature \cite{li2017conflict, zhou2015practical, reis2006measurement}. Such techniques augmented with knowledge of conflicting nodes in terms of of their traffic pattern, PHY capabilities, transmission rates, etc. can indeed enable an estimation of TCP throughput, latency and breakdowns. Unfortunately, the AP lacks information about these parameters and would require either active measurements or co-operation among all co-existing BSS nodes to obtain this additional data. On the other hand, our frameworks can make an estimation solely based on passive AP-side observations and does not need any additional data collection mechanism.  
\chapter{Conclusion}
\label{sec:conclusion}

This thesis overcomes the challenges faced in performing AP-side WLAN analytics. This is achieved through the design, implementation and experimental evaluation of two key frameworks that enable passive monitoring of user experience metrics at the AP-side.

The first framework is called virtual speed test. Virtual speed test enables a passive AP-side estimation of the download and upload speeds that a STA can achieve over its current wireless connection. While doing so, virtual speed test does not require any active measurements (\textit{e.g.}, internet speed tests), no client side co-operation (\textit{e.g.}, special purpose software installations) and no additional hardware information (\textit{e.g.}, sniffer deployment) to collect more data. Virtual speed test can make an estimation solely based on passive AP-side observations. Virtual speed test employs a novel data-driven throughput estimation model called L2 Edge TCP model which enables speed test estimation based on passive AP-side observations. 

We implement virtual speed test on a commodity hardware platform and perform extensive field trials in an office environment and in an apartment complex. In both the environments, virtual speed test demonstrates a high estimation accuracy with estimation errors less than 10\%.

The second framework is uplink latency microscope (\technique) which enables an AP-side estimation of uplink WLAN latency and decomposition into its constituent components. Similar to virtual speed test, \technique does not require any active probing, no client side co-operation and no additional hardware installations for collection of more information and can make estimations solely based on passive AP-side observations. The key idea in $\technique$ is to use layer-4 handshakes of TCP as virtual probes to drive a measurement based estimation of uplink WLAN latency, queuing delays, defer delays and retransmission rate on the uplink for any associated STA. 

We implement \technique on a commodity hardware platform and run extensive field trials on a university campus and in a residential apartment complex. In these field trials, \technique demonstrates a high estimation accuracy with estimation errors under 10\% for both the deployments.



\begin{singlespace}

\cleardoublepage
\addcontentsline{toc}{chapter}{\bibsecname}
\renewcommand{\bibname}{\bibsecname}
\bibliographystyle{IEEEtran}
\bibliography{IEEEabrv,thesis}
\end{singlespace}

\begin{appendices}
\chapter{Bounds on $\wmax^{*}$}
\label{Proof}
We break the proof into two parts. (a) In the first part, we show that $\wmax^{*}$ is upper bounded by $4$. (b) In the second part we show this maximum value occurs when $S_{vf} = S_{vr}$.

\noindent(a) Suppose that $S_{vf} = max(S_{vf}, S_{vr})$. Consequently, 

\begin{equation}
\wmax^{*} = \frac{2*(S_{vf} + S_{vr})}{S_{vf}} \leq 4
\end{equation}

\noindent The same can be shown when $S_{vr} = max(S_{vf},S_{vr})$.

\noindent(b) It can be further stated that 

\begin{equation}
\wmax^{*} = \frac{4*(S_{vf} + S_{vr})}{\frac{(S_{vf} + S_{vr} + |S_{vf} - S_{vr}|)}{2}}
\end{equation}

Suppose that $S_{vf} \geq S_{vr}$. When $\wmax^{*}$ reaches its upper bound, it follows that

\begin{equation}
\frac{2*(S_{vf} + S_{vr})}{S_{vf}} = 4
\label{final_eq}
\end{equation}

It follows that $S_{vf} = S_{vr}$. Again, the same can be shown when $S_{vf} \leq S_{vr}$. In the absence of the equality condition, Eq.~\ref{final_eq} will not be solvable. 
\chapter{Computation of Joint Distribution $P(h,b)$}\label{app:joint}
We can limit ourselves to the case in which stations send just one effective TCP ACK
in each channel access ($\SSTA = 1$). The extension to the case in which stations send 
$\SSTA > 1$ effective ACKs in each channel access is trivial, since it just requires
to scale the distribution obtained for $\SSTA = 1$ accordingly.  
 
To obtain an exact expression of $P(h,b)$ in the case of zero-delay
backbone, we separately account for the impact of the 
initial deterministic ACK at the beginning of a cycle (see
Fig. \ref{fig:embed2}). So, let us first consider the distribution
produced by uplink transmissions following the first one. 
For them, we actually compute the more detailed joint 
pdf $\hat{P}(h_1,h_2,b)$ where: $b$ is the maximum queue length;
$h_1 \geq 1$ is the number of queues having exactly length $b$;
$h_2 \geq 0$ is the number of queues having length strictly less 
than $b$.
By conditioning on the backoff value $x$ extracted by the AP, we can write:\vspace{-1mm}
\begin{eqnarray}\label{eq:ph1h2}
\hat{P}(h_1,h_2,b)  =  \int_0^\infty \binom{K}{h_1} \left[ \frac{(\mu x)^b}{b!}e^{-\mu x}\right]^{h_1} \cdot \nonumber && \\
 \binom{K-h_1}{h_2}\!\! \left( \sum_{j=1}^{b-1} \frac{(\mu x)^j}{j!}e^{-\mu x}\right)^{h_2}
\!\!\!\!\! e^{-\mu x (K-h_1-h_2)} \mu e^{-\mu x} \diff x &&
\end{eqnarray}

Despite their ugly look, integrals of the form \equaref{ph1h2} have a closed-form 
expression, obtained by expanding them into a sum of contributions, each 
leading to an analytical solution. Just as an example, in the case of $K=4$, \vspace{-1mm}
$$ \hat{P}(2,1,3) = \binom{4}{2}\binom{2}{1}\frac{1}{(3!)^2}\left(\frac{7!}{1! 5^8}+\frac{8!}{2! 5^9}\right)  = \frac{3024}{390625}$$
Note that $\hat{P}(h_1,h_2,b)$ are some \lq universal' numbers
that depend only on $K$, and that can be computed once and forall
and made available through, \textit{e.g.}, a table lookup.

To derive the final joint pdf $P(h,b)$ we have to add the contribution of the first deterministic ACK:\vspace{-1mm}
\begin{eqnarray}\label{eq:ph}
P(h,b) = \sum_{h_1+h_2 = h-1} P(h_1,h_2,b)\frac{K-h+1}{K} + \nonumber \\
\sum_{h_1+h_2 = h} P(h_1,h_2,b-1)\frac{h_1}{K} + 
\sum_{h_1+h_2 = h} P(h_1,h_2,b)\frac{h_2}{K}
\end{eqnarray}

In the above expression, the first summation corresponds to the case in which the first ACK
increases the user-diversity (and not $b$); the second summation corresponds to the case in which the first ACK
does not increase the user-diversity (but it increases $b$); the third summation corresponds to the case 
in which the first ACK does not increase neither the user-diversity nor $b$.
\chapter{Full Aggregation Case With Non-negligible Delay}\label{app:markov}
We compute the transition probabilities of the Markov
Chain by considering the number of batches that can arrive from the backbone between 
two consecutive transmissions by the AP, and the fact that
the AP, if it receives at least one batch, will transmit
again after a number of transmissions by the stations uniformly 
distributed in $0,1,\ldots,m_1+m_2$.

Let $p_{[m_1,m_2 \rightarrow m_1',m_2']}$ denote the transition
probability from state $(m_1,m_2)$ to state $(m_1',m_2')$.
To avoid unnecessary complications, we assume that batches 
flying in the backbone, in number $m_3 = K - m_1 -m_2$, can arrive at the AP
during an interval of duration
$$V(m_1,m_2) = T(m_1,F_s W_{\max}) + \sum_{j=1}^{m_1+m_2}\left(\frac{1}{\mu j} + \TUP \right)$$   
comprising the AP transmission at the end of previous cycle
plus the time required by stations to send up all of their batches
by competing only among themselves.
Let $\lambda = 1/D$ be the arrival rate of a batch from the backbone.
With probability $e^{-\lambda m_3 V}$ no batch arrives, and we get the (set of) transitions
\begin{equation}\label{eq:nob}
p_{[m_1,m_2 \rightarrow 1,0]} = e^{-\lambda (K-m_1-m_2) V(m_1,m_2)} 
\end{equation}
The reason for transiting to state $(1,0)$ is that, eventually,
one batch will arrive at the AP, while all of the others will be still flying in the backbone.

With probability $$\rho(k,m_1,m_2) = \binom{m_3}{k}(1-e^{-\lambda V})^k e^{-\lambda (m_3-k) V}$$
we have, instead, $k$ batch arrivals during $V$ ($1 \leq k \leq m_3$).
Assuming for simplicity that all these arrivals occur before
nodes start to contend again on the channel, the AP will transmit
the above $k$ batches after a number of transmissions by the stations
uniformly distributed in $0,\ldots,m_1+m_2$, leading to the transition probability:
$$ p_{[m_1,m_2 \rightarrow k,m_1+m_2-j]} = \frac{\rho(k,m_1,m_2)}{m_1+m_2+1} \qquad 0 \leq j \leq m_1+m_2$$
Note that one of these transitions, specifically the one with $k=1$, $j=m_1+m_2$,
should be added (\textit{i.e.}, contribute) to the one introduced before in \equaref{nob}
(we have preferred to split the two contributions, with abuse of notation, for better readability).
 
With the above transition probabilities, one can then solve the Markov Chain
and obtain the stationary probability distribution $\pi_{m_1,m_2}$.

It remains to explain how this distribution can be used to compute
the numerator and the denominator of throughput formula \equaref{ren}.
The numerator (average number of packets sent in a cycle) is simply:
$$\sum_{m_1,m_2} \pi_{m_1,m_2} m_1 F_s W_{\max} $$ 
The denominator (average cycle duration) can be expressed instead as an average \lq reward'
over all possible transitions:
$$\sum_{m_1,m_2,m_1',m_2'}  \pi_{m_1,m_2} \cdot p_{[m_1,m_2 \rightarrow m_1',m_2']} \cdot r(m_1,m_2,m_1',m_2') $$
where $r(m_1,m_2,m_1',m_2')$ is the reward associated to the transition
from state $(m_1,m_2)$ to state $(m_1',m_2')$. 
Hence, we only need to specify the rewards associated to the possible state transitions.
The reward associated to the zero arrival transition \equaref{nob}
is $V(m_1,m_2) + \frac{D}{K} + \frac{1}{\mu}$. Indeed, after the interval of duration $V$ all batches will be
flying in the backbone, and the first will arrive on average $D/K$ later, and 
will be sent by the AP after an average backoff $1/\mu$. 

The reward associated to the transition with probability
$p_{[m_1,m_2 \rightarrow k,m_1+m_2-j]}$ instead, is equal to
$$ T(m_1,F_s W_{\max}) + \sum_{i=0}^{j} \frac{1}{\mu (m_1+m_2+1-i)} + j \TUP $$
since, after the transmission by the AP, we have $j$ transmissions by the stations,
each preceded by a contention phase of proper average duration.
\end{appendices}


\end{document}